\documentclass[twocol,%
 nofootinbib
]{ametsocV6.1}

\bibpunct{(}{)}{;}{a}{}{,}

\usepackage{graphicx}
\usepackage{dcolumn}
\usepackage{lastpage}
\usepackage{ulem}
\usepackage{epsf}
\usepackage{multirow}
\usepackage{caption}
\usepackage{subcaption}
\usepackage{url}
\usepackage{natbib}
\usepackage{setspace}
\usepackage{enumerate}
\usepackage{epstopdf}
\usepackage{wrapfig}
\usepackage{morefloats}
\usepackage{float}
\usepackage{color}
\usepackage{xcolor}
\usepackage{scalerel,stackengine}
\usepackage[export]{adjustbox}

\stackMath
\newcommand\reallywidehat[1]{%
\savestack{\tmpbox}{\stretchto{%
  \scaleto{%
    \scalerel*[\widthof{\ensuremath{#1}}]{\kern-.6pt\bigwedge\kern-.6pt}%
    {\rule[-\textheight/2]{1ex}{\textheight}}
  }{\textheight}%
}{0.5ex}}%
\stackon[1pt]{#1}{\tmpbox}%
}

\def \imag{\textrm{\fontfamily{cmtt}\selectfont i}}

\def\mathunderline#1#2{\color{#1}\underline{{\color{black}#2}}\color{black}}

\def\Jmath{\mathbf{j}}

\renewcommand{\mathsf}[1]{\textrm{\textsf{#1}}}

\allowdisplaybreaks

\begin{document}

\title{Probing the Nonlinear Interactions of Supertidal Internal Waves using a High-Resolution Regional Ocean Model}

\authors{Joseph Skitka\correspondingauthor{Joseph Skitka, jskitka@umich.edu}\aff{a}, Brian K. Arbic\aff{a}, Ritabrata Thakur\aff{a}, Dimitris Menemenlis\aff{b}, William R. Peltier\aff{c}, Yulin Pan\aff{d}, Kayhan Momeni\aff{c} and Yuchen Ma\aff{c}}
\affiliation{\aff{a}Department of Earth and Environmental Sciences, University of Michigan, Ann Arbor, Michigan \\ \aff{b}Jet Propulsion Laboratory, California Institute of Technology, Pasadena, California \\ \aff{c}Department of Physics, University of Toronto, Toronto, Ontario, Canada \\ \aff{d}Department of Naval Architecture and Marine Engineering, University of Michigan, Ann Arbor, Michigan}


\abstract{The internal-wave (IW) continuum of a regional ocean model is studied in terms of the vertical spectral kinetic-energy (KE) fluxes and transfers at high vertical wavenumbers. Previous work has shown that this model permits a partial representation of the IW cascade.  In this work, vertical spectral KE flux is decomposed into catalyst, source, and destination frequency bands of nonlinear scattering, a framework that allows for the discernment of different types of nonlinear interactions involving both waves and eddies.  Energy transfer within the supertidal IW continuum is found to be strongly dependent on horizontal resolution.  Specifically, at a horizontal grid spacing of 1/48$^\circ$, the vast majority of KE in the supertidal continuum arrives there from lower frequency modes through a single nonlinear interaction, while at 1/384$^\circ$ KE transfers within the supertidal IW continuum are comparable in size to KE transfer from lower-frequency modes.  Additionally, comparisons are made with existing theoretical and observational work on energy pathways in the IW continuum.  Induced diffusion (ID) is found to be associated with a weak forward frequency transfer within the supertidal IW continuum.  Spectrally local interactions are found to play an insignificant role within the model evolution.  At the same time, ID-like processes involving high vertical-wavenumber near-inertial and tidal waves as well as low-vertical-wavenumber eddy fields are substantial, suggesting that the processes giving rise to a Garrett-Munk-like spectra in the present numerical simulation and perhaps the real ocean may be more varied than in idealized or wave-only frameworks. }
\maketitle


\section{Introduction}
\label{sec:intro}

\subsection{Motivation}

Breaking internal gravity waves (IWs) are one of the most important processes underlying diapycnal mixing, which is thought to be an important mechanism for closing and regulating the meridional overturning circulation \citep{wunsch04}; in turn, the overturning circulation is a key regulator of the Earth-system’s climate \citep{ipcc14}.   IW energy spectra are readily observable along one dimension in the ocean via dropped, floated, or dragged CTD probes and moored profilers \citep{alford17}.  Phenomenological models of energy spectra associated with these observations were developed in the 1970's by Garrett and Munk (GM) \citep{garrett72,garrett75,garrett79} and theoretical models of the detailed dynamics that give rise to such spectra have been the subject of research in the ensuing decades \citep{mccomas77b, muller86, dematteis22}. However, verification of such theoretical models via observations is difficult due to the intractability of rapidly measuring detailed two- or three-dimensional flow fields.  

More recently, global \citep[e.g.][]{arbic10, arbic18, muller15, rocha16, arbic22} and regional \citep[e.g.][]{nelson20, pan20, thakur22} ocean models have begun to permit detailed representations of the largest scales of the IW continuum due to a combination of tidal forcing, coupling to atmospheric wind and buoyancy forcing fields, and higher horizontal and vertical resolution \citep{arbic22}.  The highest-resolution regional models in particular are able to permit detailed 4-dimensional sampling of realistic internal wave fields that is not possible from observations, potentially a watershed breakthrough for IW modeling and theoretical innovation.  

Outside of the specific context of IW continua, spectral-energy-flux methods have been utilized to study the direction of energy cascades across scales in geostrophic eddies in ocean models and observations \citep{scott07,arbic13}.  \cite{barkan21} used the related coarse-graining technique with a frequency decomposition to isolate spectral energy transfers between mesoscale, submesoscale, and wavelike motions.  The coarse-graining technique allows for spatial resolution of such spectral exchanges, but the process introduces uncertainty in separating large and small scales\footnote{In order to look at spatial dependencies of a spectral quantity, the coarse-graining method imposes Galilean invariance of the local contribution of the spectral integral at the expense of having a clean partition of the fine- and coarse-scale kinetic energies \citep{aluie18}. The computed energy transfer is not strictly a spectral flux between Fourier-defined scales and the uncertainty associated with this becomes larger as the coarse-grain filter becomes more spatially localized and the filter kernel becomes less spectrally localized.}.  Others, such as \cite{wagner16}, have devised simplified models with the explicit inclusion of different eddy and wave components that can be used to infer (rather than compute directly) the exchange of energy between the components.   The aforementioned methods which look at spectral bands stand in contrast with those of wave-turbulence theory (WTT), which approaches energy exchange on a mode-by-mode basis by looking at spectral interaction triads \citep{lvov10}.  When WTT methods are applied to realistic 3D numerical simulation output, they have so far been computationally limited to select specific test modes in the triads, as in \cite{pan20}.  

This paper will utilize a decomposed version of the spectral energy fluxes and transfers that has some of the benefits of each of the methods discussed above: the isolation of different spectral bands, a decomposition into interaction triads, and spatial resolution of the spectral exchanges in the direction orthogonal to the spectra computed (see section \ref{sec:methods}\ref{sec:methods:budgets}).  Also, while some other methods such as WTT have previously exclusively considered energy exchange among sets of three modes (see appendix \ref{app:triads}), the spectral-flux decomposition presented in this paper will use an unsymmetrized version of scattering triads that isolates the exchange of energy between two modes mediated by a third mode, allowing for the direction and magnitude of energy exchange between any two scales to be exactly discerned.  Through this spectral flux decomposition, this paper will attempt to bring theoretical understanding of IW interactions into alignment with the internal wave dynamics as revealed in the previously mentioned high-resolution regional models.  It will also attempt to use such methods and numerical simulations to interpret observations of spectral transfers that are dimensionally limited, specifically in \cite{sun12}.  A distinguishing feature of this work from the aforementioned utilizing spectral transfer methods is that it will focus on isolating the supertidal band of internal gravity waves rather than separating mesoscale from submesoscale \citep[as in ][]{barkan21} or near-inertial from semi-diurnal waves \citep[as in ][]{wagner16}.  Additionally, this work will identify nonlinear energy transfers involving both IWs and eddy fields through mechanisms that are not described by in previous work of WTT.

\subsection{Resonant nonlinear interactions in theoretical IW models}
\label{sec:IW_theory}

The GM spectrum \citep{garrett75} does not attempt to describe the means of forcing, dissipation, spectral inhomogeneities due to tidal harmonics, or spatial inhomogeneities.  It also is completely uninvolved with details of the nonlinear interactions that give rise to such a spectrum.  Numerous theoretical attempts have been made to describe the rise of the GM spectrum through nonlinear interactions of IWs.   Many of these attempts involve assumptions of weak nonlinearity as a pretext for identifying specific resonant triadic interactions \citep{muller86}, although there is evidence that this is not always a valid assumption \citep{mccomas81a}. 

\cite{mccomas77a} identified different types of nonlinear-interaction mechanisms likely to be important to the development of the IW continuum spectrum. For instance, elastic scattering ($\mathcal{M}_{\textrm{ES}}$) involves high-frequency, high-vertical-wavenumber waves scattering off of low-frequency, high-vertical-wavenumber waves in a way that acts to relax vertically anisotropic IWs towards the vertical isotropy of the GM spectrum.   Parametric Subharmonic Instability ($\mathcal{M}_{\textrm{PSI}}$) is a frequency-halving mechanism moving energy from high-frequency and low-vertical-wavenumber to low-frequency and high-vertical-wavenumber IWs.  Because of the jump in frequency, $\mathcal{M}_{\textrm{PSI}}$ is strongly latitudinally dependent based on whether semidiurnal tides can decay above or below the inertial frequency; however, $\mathcal{M}_{\textrm{PSI}}$ may play an important role at higher frequencies at any latitude.  \cite{pan20} found $\mathcal{M}_{\textrm{ES}}$ and $\mathcal{M}_{\textrm{PSI}}$ to play a subdominant role in IW KE transfers at high wavenumbers and frequencies.  The focus of this paper will be on the other mechanisms, introduced below.  This is also in part because it is difficult to clearly identify $\mathcal{M}_{\textrm{ES}}$ and $\mathcal{M}_{\textrm{PSI}}$ with a frequency-based spectral flux decomposition. 

Induced diffusion ($\mathcal{M}_{\textrm{ID}}$) involves high-wavenumber, high-frequency waves scattering off of near-inertial, low-wavenumber waves and inducing diffusion of wave action, $\mathcal{A}=\frac{KE}{\omega}$, forward in vertical wavenumber space.  Here, $\omega$ denotes the flow frequency.  \cite{mccomas77a} found using two different theoretical methods that wave action should be conserved in the supertidal band under $\mathcal{M}_{\textrm{ID}}$.  The associated energy diffusion depends, then, on the direction of the frequency cascade.  The direction of the frequency cascade, in turn, depends (through the dispersion relation) on the aspect ratio of the change in horizontal-to-vertical-wavenumber following the energy flow within the supertidal IW continuum through the interaction (see figure 5 in \cite{mccomas81a} and the pertinent discussion).  \cite{mccomas77b} argues that the horizontal wavenumber is approximately constant and therefore the associated frequency diffusion is backward and that some energy necessarily has to be moved into the near-inertial waves (NIWs) from the high-frequency IWs to conserve $\mathcal{A}$.  On the other hand, \cite{dematteis22} argue that the horizontal wavenumber ``keeps pace'' with the vertical wavenumber, and therefore the associated frequency cascade is neutral-to-forward depending on the specific IW spectral slope.  Therefore, the compensating energy will either be absent or will be exchanged from the near-inertial frequencies into the high frequencies\footnote{\cite{dematteis22} use a specific set of parameters to define the IW spectrum that is consistent with the stationary solution of the WTT collision integral.  For a frequency spectrum of $E\left(\omega\right) \sim \omega^{-2}$, their parameter choices imply a vertical wavenumber energy spectrum that is $E\left(m\right) \sim m^{-2}$, which is not necessarily consistent with all parts of the ocean or the simulations in the paper (compare with Fig. \ref{fig:fig1b}).  However, \cite{dematteis22} argue that it is necessary for this set of parameters to reflect dominant processes in the ocean in order to resolve the "oceanic ultraviolet catastrophe," that is, for the spectral frequency fluxes associated with induced diffusion to be forward and not inverse given the lack of high-frequency energy sources to feed the IW continuum \citep{polzin17}.  For this reason, we compare the present numerical simulations with predictions made for $E\left(m\right) \sim m^{-2}$ in \cite{dematteis22}.} 

This paper differentiates between such conflicting accounts of $\mathcal{M}_{\textrm{ID}}$ by decomposing the downscale vertical\footnote{A vertical spectral energy flux indicates energy transfer from low-to-high vertical wavenumber, not a spectra of vertical spatial energy flux.  The term ``vertical spectral flux" is used throughout this paper to denote a flux from low-to-high vertical wavenumber.}  spectral kinetic-energy flux into components based on the frequency band of the energy source.  Vertical spectral fluxes are used because the various theoretical models of IW cascades are consistent in their predictions that within the IW ``inertial range,'' energy flow is to smaller vertical wavenumbers \citep{muller86, dematteis22}.  Our use of frequency-decomposed spectral fluxes also allows for the traditional definition of $\mathcal{M}_{\textrm{ID}}$ (which conserves and diffuses wave action in the supertidal band) to be decomposed into two components: (1) downscale kinetic energy diffusion within the supertidal band ($\mathcal{M}_{\textrm{ID}_{\textrm{diff}}}$) and (2) the compensating energy exchanged with near-inertial and tidal frequencies ($\mathcal{M}_{\textrm{ID}_{\textrm{comp}}}$).  A separate but conceptually related mechanism, downscale kinetic energy diffusion that is induced by eddy fields instead of wave fields, is referred to as $\mathcal{M}_{\textrm{ID}_{\textrm{eddy}}}$.  While such labels are included here for reference, they are not rigorously defined until section \ref{sec:methods}\ref{sec:methods:budgets}\ref{sec:flux}. 

Finally, \cite{dematteis21} point out that local interactions ($\mathcal{M}_{\textrm{LI}}$) (that is, high-wavenumber, high-frequency waves interacting among themselves independent of other parts of the flow) may also be a dominant mechanism for moving energy down vertical scales within the IW continuum.  The fluxes due to $\mathcal{M}_{\textrm{LI}}$ and $\mathcal{M}_{\textrm{ID}}$ in particular are computed in the present work, thereby testing the prediction that the former is of greater magnitude than the latter (albeit, within the limits of the resolution of this model.)

\subsection{Observations of resonant nonlinear interactions}

\cite{sun12} used sonar and CTD observations near Hawaii to directly quantify nonlinear energy transfers among supertidal internal waves.  They found coherent energy transfers that were inconsistent with $\mathcal{M}_{\textrm{ID}}$.  They were also unable to find a signal where they expected to see $\mathcal{M}_{\textrm{ID}}$ on vertical-wavenumber bispectra.  This paper produces a similar bispectrum in section \ref{sec:results}\ref{sec:results_bispectra} in an attempt to reproduce the most prominent feature from their observations while also explaining why \cite{sun12} do not observe $\mathcal{M}_{\textrm{ID}}$.  In computing bispectra from model output, the assumption of \cite{sun12} that the vertical-gradient component of the spectral KE transfer represents the entire transfer is tested.

\subsection{Resonant Nonlinear Interactions in IW-Permitting Simulations}
\label{sec:intro:plan}

Regional models that permit a partially resolved IW continuum must account for significant spatial energy flux of IWs across the regional boundaries \citep{mazloff20}.  Numerous regional models have a vigorous mesoscale eddy field but lack robust internal-wave and internal-tide forcing at the boundary conditions, (e.g. \cite{nugroho18,renault21,wang21}).   A straightforward means of accounting for this is to use IW-permitting global-model output to set the regional boundary conditions.  Global models require simultaneous atmospheric and tidal forcing as well as sufficient horizontal and vertical resolution in order to energize a robust IW field, conditions that have only begun to be implemented recently \citep{arbic10, muller15, rocha16, rocha16b, arbic18, arbic22}.  \cite{nelson20}, \cite{pan20}, and \cite{thakur22} used a regional model with tidal and wind forcing, and, importantly, imposed IW forcing at the boundaries from an IW-permitting global model.  These are the first and, at the time of submission, the only published regional models that attempt to resolve the IW boundary flux issue noted by \cite{mazloff20}. The present paper will continue using this approach.  

As IW-permitting models adopt higher resolutions, they begin to resolve more of the IW continuum, both in global models \citep{muller15, savage17a} and regional models \citep{nelson20}.  \cite{pan20} finds that IW dispersion curves in a high-resolution regional model are clearly defined over two orders of magnitude of frequency and horizontal wavenumbers, much greater than in available global models; they further find evidence for a dynamically significant $\mathcal{M}_{\textrm{ID}}$.    

High-resolution IW-permitting regional models now have the potential to validate recent advances in theoretical understanding and to study challenges faced in observations of internal wave scattering.  Some open questions that can now be addressed are: 
\begin{itemize}
\item At what model resolutions do $\mathcal{M}_{\textrm{ID}}$ and $\mathcal{M}_{\textrm{LI}}$ become significant?
\item Are $\mathcal{M}_{\textrm{ID}}$ and $\mathcal{M}_{\textrm{LI}}$ associated with forward or inverse frequency cascades?
\item How does $\mathcal{M}_{\textrm{ID}_{\textrm{diff}}}$ compare with  $\mathcal{M}_{\textrm{ID}_{\textrm{comp}}}$?
\item Why do \cite{sun12} not see $\mathcal{M}_{\textrm{ID}}$ in their observations?  What do they see?
\item How important is $\mathcal{M}_{\textrm{ID}_{\textrm{eddy}}}$ and other mechanisms that don't fit into existing categories of nonlinear wave-wave interactions?
\end{itemize}
This paper will attempt to shed light on these questions using a vertical spectral kinetic energy analysis. A follow-up paper will also include a horizontal wavenumber spectral energy analysis and will differentiate between various dissipation mechanisms used by the model and existing IW closures.


\section{Methods}
\label{sec:methods}

\subsection{The model}
\label{sec:model}

\begin{figure}[]
\includegraphics[width=0.5\textwidth]{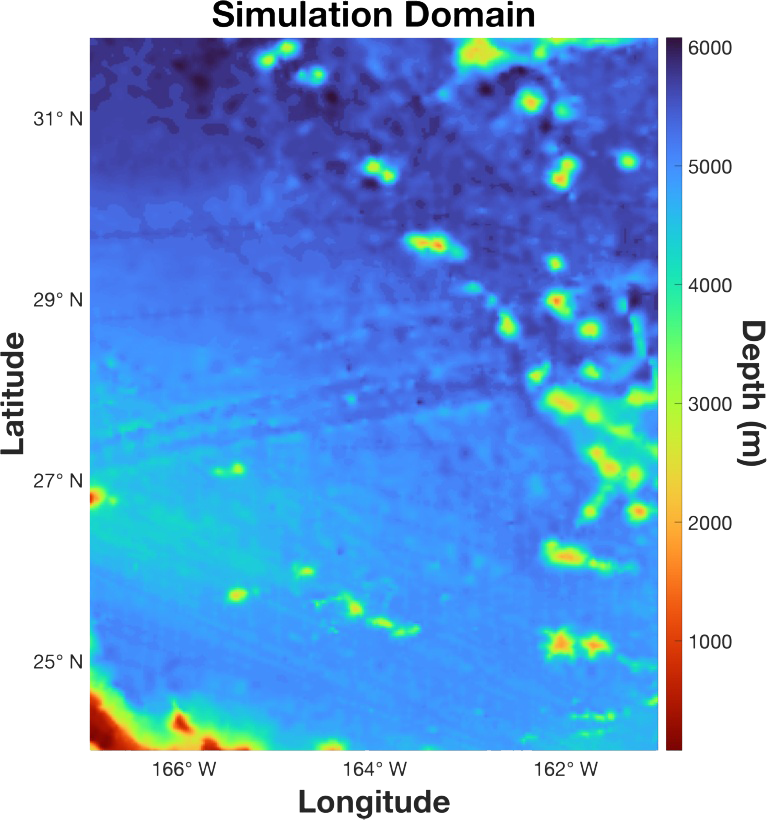}
\centering \caption{Bathymetry of the Region of Study.}
\label{fig:bathymetry}
\end{figure}

The model simulates a region north of the Hawaiian archipelago (Fig. \ref{fig:bathymetry}). It includes wind and astronomical forcing and, as a result, a partially resolved spectrum of near-inertial oscillations, internal tides and supertidal internal waves.  The study area features a northward-propagating internal-tidal beam generated in the French Frigate Shoals as well as interactions with the Musician Seamounts. The region also features areas north and south of the critical latitude for $\mathcal{M}_{\textrm{PSI}}$ at 28.8$^\circ$N \citep{alford07}. We use the Massachusetts Institute of Technology general circulation model \citep[MITgcm;][]{mitgcm}. Rigidly imposed nested boundary conditions are provided from a similarly forced global ocean simulation of MITgcm, often denoted LLC4320 \citep{rocha16,arbic18}.  Previous work based on wavenumber and frequency spectra has demonstrated that this regional Hawaii simulation permits a partial representation of the internal-wave cascade \citep{nelson20, pan20, thakur22}.  

The simulation employs a hydrostatic finite volume solver. The horizontal eddy-viscosity scheme is a modified form of the Leith scheme \citep{fox-kemper08}, while interior and mixed-layer dissipation and diffusivity is handled by the K-profile parameterization \citep[KPP;][]{large94}.  The Leith scheme and KPP are the dominant mechanisms acting in the ocean interior to dissipate KE in the modeled IW continuum.  Wind forcing is updated every 6 hours, while boundary conditions in the regional model are updated hourly.  The model grid is stretched substantially in the vertical direction (vertical grid spacing is $0.333$m at the surface and 133m at the domain bottom in the highest-resolution run).  The regional model was run for 106 days, from March 1, 2012 until June 15, 2012, and output was collected every 20 minutes for the last 7 days of this period and used for the analysis presented in this work.  The regional model was run with four different resolutions (as in \cite{nelson20}):  
\begin{enumerate}
\item 88 vertical levels and 2km horizontal grid spacing,
\item 88 vertical levels and 256m horizontal grid spacing,
\item 264 vertical levels and 2km horizontal grid spacing,
\item 264 vertical levels and 256m horizontal grid spacing.
\end{enumerate}

\subsection{Spectral KE budgets}
\label{sec:methods:budgets}

The present work is focused on attaining a clear picture of nonlinear IW triadic scattering mechanisms as far from spectrally inhomogeneous forcing as possible, that is, deep within the ``inertial range'' of the IW continuum.  The flow field is decomposed into spectral bands: the low-pass band (LP) with $\omega < 0.8 f_0$, a band-pass of near-inertial and tidal frequencies (BP) with $0.8 f_0 < \omega < 2.5 f_0$, and a supertidal high-pass band (HP) with $2.5 f_0 < \omega$.  The HP band at high vertical wavenumber is of primary interest for the IW-continuum energetics.  Here, $f_0$ denotes the Coriolis frequency at the central latitude of the regional domain.  To the extent that an inertial range exists in the IW continuum, forcing and dissipation should exhibit scale separation and energy move through the intermediate scales via nonlinear interactions among waves.  The dissipation mechanisms and the vertical spectral flux should be in approximate balance at small scales.  In order to compute these vertical spectral kinetic energy budgets, the velocity evolution due to each term in the governing equations must first be interpolated to a uniform vertical grid that has the same number of grid points.  This sampling rate is based on numerical experiments with synthetic spectra.

\subsubsection{Terms and boundary conditions}
\label{sec:terms}

Consider the horizontal velocity evolution equation
\begin{align}
\partial_t \mathbf{u} &= \underbrace{-\left(\mathbf{v}\cdot\boldsymbol{\nabla}\right) \mathbf{u}}_{\mathsf{A}}  \;\; \underbrace{-\;\;  \boldsymbol{\nabla}_h \frac{p_{hyd}}{\rho_0}}_{\mathsf{P}} \;\;  \underbrace{-\;\;\left(\textrm{$\boldsymbol{f}$}\times\mathbf{u}\right)_h}_{\mathsf{C}} + \mathsf{F} + \mathsf{D} + \textrm{BCs} \label{eq:horz_eq} \\
\mathsf{D} &= \mathsf{D}_{\textrm{Leith}} + \mathsf{D}_{\textrm{KPP ML}} + \mathsf{D}_{\textrm{KPP Shear}} \nonumber \\
& \; \; \;  + \mathsf{D}_{\textrm{KPP Conv.}} + \mathsf{D}_{\textrm{KPP BG}} + \mathsf{D}_{\textrm{QBD}} \quad \textrm{,} \label{eq:diss}
\end{align}
\noindent where $\mathbf{u}$ is the 2D horizontal velocity vector, $\mathbf{v}$ is the 3D velocity vector $\boldsymbol{f} = 2 \Omega \textrm{Sin}\left(\theta\right) \hat{\mathbf{z}}$ is the Coriolis frequency vector, $\Omega$ is the frequency of Earth's rotation, $\theta$ is latitude, $p_{hyd}$ is the hydrostatic pressure, $\rho_0$ is the constant reference density, $\boldsymbol{\nabla}_h$ is the 2D horizontal gradient operator, $\left(\right)_h$ is the horizontal 2D component of the term within the parentheses, $\mathsf{A}$ is the advective term, $\mathsf{P}$ is the pressure term, $\mathsf{F}$ is forcing, $\mathsf{D}$ is  dissipation, $\mathsf{C}$ is the Coriolis force, and BCs are spatial boundary forcing terms.  Dissipation is decomposed into individual terms in equation \ref{eq:diss}.  We are interested in terms and boundary conditions that may be active at small vertical scales.  Under the assumption that the IW continuum acts as a wave-turbulence cascade with scale separation between forcing and dissipation, the forcing terms will be ignored.  Notes on the implementation of spectral diagnosis of the terms and boundary conditions of interest, which are computed for a partial vertical integrated spectral KE budget in section \ref{sec:methods}\ref{sec:methods:budgets}\ref{sec:flux}, are listed in appendix \ref{app:terms}.

\subsubsection{Spectral flux and transfers}
\label{sec:flux}

In order to compute partial vertical spectral budgets, it is useful to write the horizontal velocity evolution (equation \ref{eq:horz_eq}) in terms of a generic sum of terms and boundary conditions
\begin{align}
\partial_t u_i &= \sum_{\textrm{terms}} \partial_t (u_i)_{\textrm{terms}} \label{eq:horz_ev} \\
\partial_t (u_i)_{\mathsf{A}} &= - \varv_j \partial x_j u_i \quad \textrm{,} \label{eq:adv}
\end{align}
where $u_i$ are the 2D horizontal velocity field components, $\varv_j$ are the 3D velocity field components, subscripts sum over all applicable dimensions, and the summation runs over all  terms and spatial boundary conditions (for example, $\mathsf{D}_{\textrm{KPP Shear}}$).  The advective term ($\mathsf{A}$) is explicitly written out as an important contribution to the horizontal velocity evolution in equation \ref{eq:adv}.  

Local spectral budgets, $\mathcal{T}_{\textrm{term}}$, are the change in KE of a specific spectral mode resulting from one of the terms in equation \ref{eq:horz_ev}.  The terminology is adopted from \cite{scott07}, but others refer to this as a ``spectral transfer function'' \citep{arbic14, muller15}, a label that will be reserved strictly for the advective local spectral budget term, as discussed below.  This can be written (see appendix \ref{app:lsb}):
\begin{equation}
\mathcal{T}_{term}\left(m,\omega\right) = \left<u_i{\left(m,\omega\right)} \; \partial_t (u_i)_{\textrm{term}}\right> \quad \textrm{,} \label{eq:T}
\end{equation}
where\footnote{In equation \ref{eq:u_def}, $u_i\left(m,\omega\right)$ is a shorthand for  $u_i\left(x,y,z,m,t,\omega\right)$, not a distinct vector field.}
\begin{align}
u_i\left(m,\omega\right) =& \; u_i \left(x,y,z,m,t,\omega\right)  \\
=& \int \hat{u}_i\left(x,y,m',\omega'\right) \delta\left(|m'|-|m|\right) \delta\left(|\omega'|-|\omega|\right) \nonumber \\
& \qquad \qquad e^{2 \pi \imag (m' z-\omega' t)} dm' d\omega' \label{eq:u_def}  \\
\hat{u}_i \left(x,y,m',\omega'\right) &= \int u_i\left(x,y,z,t\right) e^{-2 \pi \imag \left(m' z-\omega' t\right)} dz \; dt \quad \textrm{.}
\end{align}
Here, $m$ is the vertical wavenumber,  $\delta$ is the Dirac delta function, $\imag = \sqrt{-1}$, $u_i{\left(m,\omega\right)}$ is the component of the velocity field with spectral modes ($\pm m, \pm \omega$) so that it is real valued in position space, as indicated in equation \ref{eq:u_def}. The angle brackets denote an average over time and spatial domain over which the spectral components are defined.   Note that $u_i$ is a function of 3D spatial coordinates and time as well as the vertical wavenumber and frequency, \{$x$,$y$,$z$,$m$,$t$,$\omega$\}, while the spectral amplitudes, $\hat{u}_i$, are a function of \{$x$,$y$,$m'$,$\omega'$\}.  One can integrate equation \ref{eq:T} to obtain contributions to the kinetic energy evolution at all vertical wavenumbers above a threshold.  These ``integrated spectral budget terms'' \citep{scott07} can be written as
\begin{equation}
\Pi_{term}^>\left(m,\omega\right) = \left<u_i^{>}{\left(m,\omega\right)} \; \partial_t (u_i)_{\textrm{term}}\right> \quad \textrm{,} \label{eq:Pi}
\end{equation}
where $u_i^{>}$ refers to a hard cutoff high-pass Fourier filter on $u_i$ above vertical wavenumber $m$:
\begin{align}
u_i^{>}\left(m,\omega\right) =& \; u_i^{>} \left(x,y,z,m,t,\omega\right)  \\
 =& \int \hat{u}_i\left(x,y,m',\omega'\right) \delta\left(|\omega'|-\omega\right) \Theta\left(|m'|-|m|\right) \nonumber \\
 & \qquad \qquad e^{2 \pi \imag (m' z-\omega' t)} dm' d\omega' \quad \textrm{,} \nonumber
\end{align}
\noindent where $\Theta$ is the Heaviside step function.  As in equation \ref{eq:T}, $u_i^{>}$ includes both positive and negative valued wavenumbers and frequencies so that the $u_i^>$ is real-valued in position space.  The integrated spectral budgets, (equation \ref{eq:Pi}) will be used for evaluating the spectral flux integrated over bands of frequencies rather than individual frequencies.  

The advective term conserves energy away from boundary effects on a divergence-less flow. Because of this, the local spectral budget contribution of advection is to simply transfer energy from one wavenumber to another. The contribution of advection to the integrated spectral budget is a flux of KE between low wavenumbers to high wavenumbers.  The local and integrated contributions from the advective term in particular are therefore referred to as spectral transfers and fluxes, respectively.  These can be written out explicitly using equations \ref{eq:T} and \ref{eq:Pi}\footnote{If the frequency dependence is ignored by integrating over all $\omega$, then for an incompressible flow, for which $\partial x_i \varv_i = 0$, the latter is equivalent to the form given by \cite{frisch95} and \cite{scott05}:
\begin{equation}
\Pi_{\mathsf{A}}^>\left(m\right) = - \left<u_i^{>}{\left(m\right)} \; \left(\varv_j \partial x_j u_i^{<}\left(m\right) \right)\right> \quad \textrm{.} \label{eq:adv_flx}
\end{equation} }:
\begin{align}
\mathcal{T}_{\mathsf{A}}\left(m,\omega\right) &= - \left<u_i{\left(m,\omega\right)} \; \left(\varv_j \partial x_j u_i \right)\right> \label{eq:adv_transf} \\
\Pi_{\mathsf{A}}^>\left(m,\omega\right) &= - \left<u_i^{>}{\left(m,\omega\right)} \; \left(\varv_j \partial x_j u_i \right)\right>  \quad \textrm{.} \label{eq:adv_flx_one_side}
\end{align} 

For incompressible flows, the three velocity fields written in equation \ref{eq:adv_flx_one_side} can be organized in spectral space into triads with each filling a distinct role.  In equations \ref{eq:adv_transf} and \ref{eq:adv_flx_one_side}, the advecting 3D field, $\varv_j$ acts as a ``catalyst'' and contributes no energy to the spectral flux.  All KE transfer comes from the rightmost ``source'' $u_i$ field, while the energy goes into the leftmost ``destination'' mode of $u_i^>\left(m,\omega\right)$, as shown in appendix \ref{app:triads}. These terms of ``catalyst'', ``source'', and ``destination'' modes will be used throughout the paper.  The terminology is useful in that it allows for the separation of different sources of KE flux within different frequency bands.  For example, from equation \ref{eq:adv_flx_one_side}:
\begin{equation}
\Pi_{\textrm{LP} \xrightarrow{\textrm{BP}} \textrm{HP, } \mathsf{A}}^> \left(m\right) = - \left<u_i^{>, \textrm{HP}} \; \left(\varv_j^{\textrm{BP}} \partial x_j u_i^{\textrm{LP}} \right)\right> \label{eq:adv_flx_sources}
\end{equation}
is the vertical spectral energy flux from the LP (eddy) frequencies scattering off of the catalyst BP (tidal and near-inertial) frequencies into the HP (internal-wave-continuum) frequencies above a cutoff vertical wavenumber $m$.  Here, $u_i^{>, \textrm{HP}}$ refers to a hard high-pass Fourier filter on $u_i$ above vertical wavenumber $m$ and a high-pass frequency filter as defined in section \ref{sec:methods}\ref{sec:methods:budgets}. The average must be computed over the same time period used to define the frequency filter. Error introduced from overlap of the bands, but it is determined to be small through numerical experiments.  A 22\%-tapered Tukey window is applied to the time series. The Tukey window also introduces error in the band isolation and reduces the magnitude of the transfers being computed. Through experiments with different window lengths, we determine that the findings of this paper are insensitive to errors arising from the Tukey window.  Hereafter, the $\mathsf{A}$ indicating the advective (or flux) terms, will be dropped but implied for all $\Pi$ and $\mathcal{T}$ that indicate directional KE exchange (e.g. $\textrm{BP}\xrightarrow{\textrm{all}}\textrm{HP}$).

\begin{figure*}[]
\centering
\includegraphics[width=\textwidth]{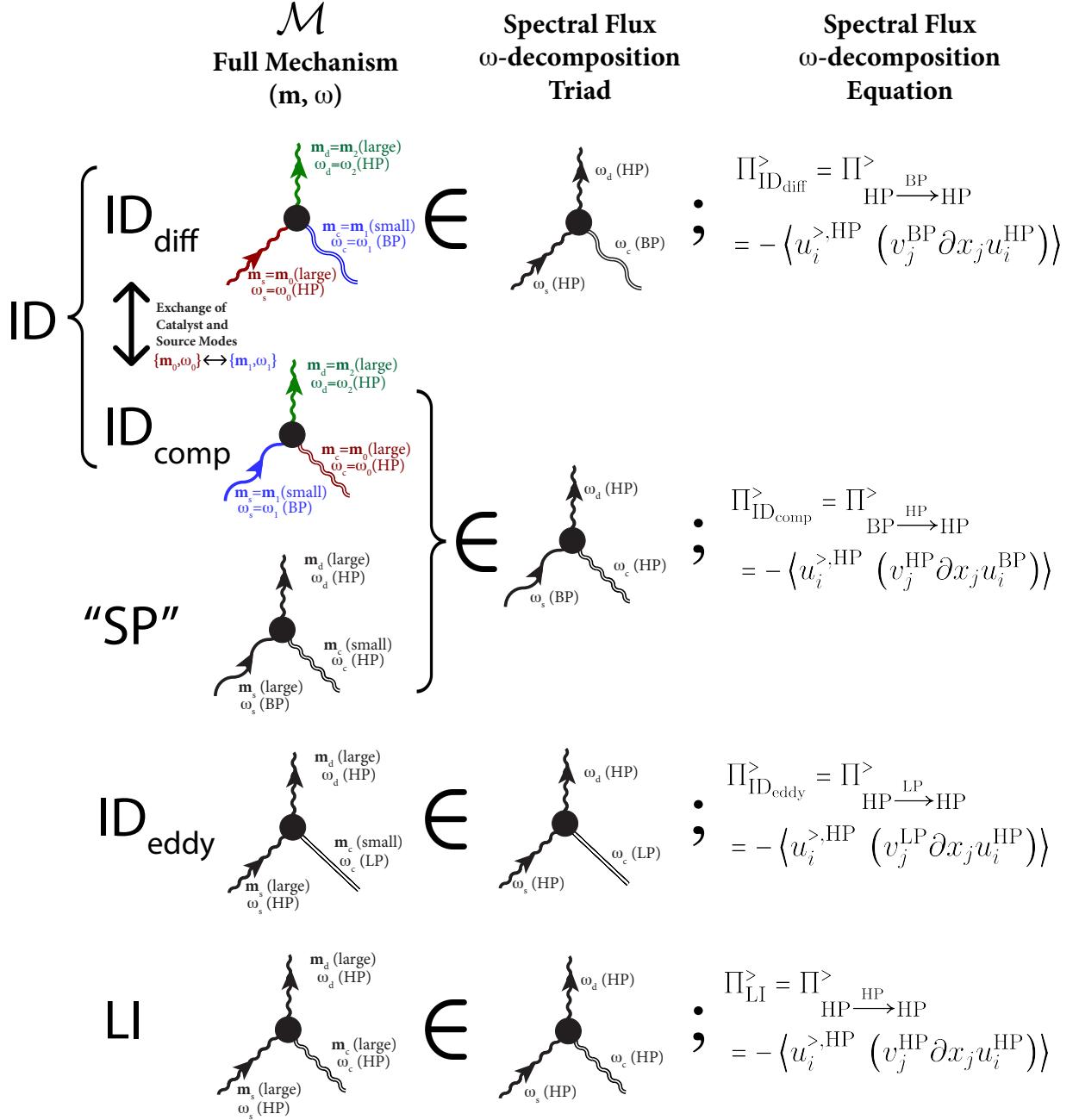}
\centering \caption{Scattering triads, introduced in appendix \ref{app:triads}, for various mechanisms discussed in this paper.  The left column shows the triads for the full mechanisms, $\mathcal{M}_i$, which constrain both the vertical wavenumber and frequencies of the interacting modes.  The center column shows the scattering diagrams that only constrain frequency, which, when integrated, comprise the the decomposition of spectral flux (equation \ref{eq:adv_flx_sources}), displayed in the right column.  "SP" refers to a mechanism ($\mathcal{M}_{\textrm{SP}}$), that was observed by \cite{sun12} (see section \ref{sec:results}\ref{sec:sp12}) and will be introduced in the context of the present results in section \ref{sec:results}\ref{sec:bispectra}.  As described in appendix \ref{app:triads}, the bottom-left mode of each scattering diagram, $m_s$, is the source of KE in the triad, the bottom-right mode, $m_c$, is the catalyst mode in the triad and contributes no KE, and the top mode, $m_d$, is the destination of KE in the triad. Thus, each diagram represents the evolution of mode $m_d$ and not the others.  The line colors in the $\mathcal{M}_{\textrm{ID}}$ scattering diagrams each correspond to a specific mode, e.g. $m_1, \omega_1$, in order to emphasize the exchange of catalyst and source modes.}
\label{fig:mechanisms}
\end{figure*}

In this framework, energy transfer into the supertidal (HP) band can be decomposed:
\begin{alignat}{2}
&\Pi_{\textrm{all} \xrightarrow{\textrm{all}} \textrm{HP}}^> \left(m\right) &&= - \left<u_i^{>, \textrm{HP}} \; \left(\varv_j^{\textrm{all}} \partial x_j u_i^{\textrm{all}} \right)\right> \nonumber \\
&&&= - \left<u_i^{>, \textrm{HP}} \; \left(\varv_j^{\textrm{LP+BP+HP}} \partial x_j u_i^{\textrm{LP+BP+HP}} \right)\right> \nonumber \\
&\color{gray} (\Pi_{\textrm{ID}_{\textrm{diff}}}): \color{black}&&= - \left<u_i^{>, \textrm{HP}} \; \left(\varv_j^{\textrm{BP}} \partial x_j u_i^{\textrm{HP}} \right)\right> \nonumber \\
&\color{gray} (\Pi_{\textrm{ID}_{\textrm{comp}}}): \color{black}&& \quad - \left<u_i^{>, \textrm{HP}} \; \left(\varv_j^{\textrm{HP}} \partial x_j u_i^{\textrm{BP}} \right)\right> \nonumber \\
&\color{gray} (\Pi_{\textrm{ID}_{\textrm{eddy}}}): \color{black}&& \quad - \left<u_i^{>, \textrm{HP}} \; \left(\varv_j^{\textrm{LP}} \partial x_j u_i^{\textrm{HP}} \right)\right> \nonumber \\
&\color{gray} (\Pi_{\textrm{LI}}): \color{black}&& \quad - \left<u_i^{>, \textrm{HP}} \; \left(\varv_j^{\textrm{HP}} \partial x_j u_i^{\textrm{HP}} \right)\right> \nonumber \\
&\color{gray} (\Pi_{\textrm{BP}_{\textrm{other}}}): \color{black}&& \quad - \left<u_i^{>, \textrm{HP}} \; \left(\varv_j^{\textrm{BP+LP}} \partial x_j u_i^{\textrm{BP}} \right)\right> \nonumber \\
&\color{gray} (\Pi_{\textrm{LP}}): \color{black}&& \quad - \left<u_i^{>, \textrm{HP}} \; \left(\varv_j^{\textrm{HP+BP+LP}} \partial x_j u_i^{\textrm{LP}} \right)\right>.
\label{eq:hp_flx_decomp}
\end{alignat}
It should be stressed that the notation $\mathcal{M}_i$ describes the abstract nonlinear interaction mechanisms $i$, which are defined in terms of wavenumber as well as frequency.  The spectral flux components, written $\Pi_{\textrm{i}}$ where $i$ refers to different scattering mechanisms, only constrain the frequency of the interaction modes.  Yet $\Pi_{\textrm{i}}$ are used to approximately measure $\mathcal{M}_i$ in this paper.  Bispectra, introduced in section \ref{sec:methods}\ref{sec:methods:budgets}\ref{sec:bispectra} constrain the vertical wavenumbers and are better able to isolate a given mechanism.  The distinction between the mechanisms, $\mathcal{M}_i$ and the spectral flux components, $\Pi_i$, is depicted diagrammatically for several important processes in Fig. \ref{fig:mechanisms}.    The induced-diffusion mechanism, as it is defined \cite{mccomas77a}, can be broken up into two subprocesses,  $\mathcal{M}_{\textrm{ID}} = \mathcal{M}_{\textrm{ID}_{\textrm{diff}}} + \mathcal{M}_{\textrm{ID}_{\textrm{comp}}}$.

$\mathcal{M}_{\textrm{ID}_{\textrm{diff}}}$ is the kinetic-energy transfer within the supertidal band (HP) caused by scattering off the catalyst field of near-inertial and tidal waves (BP).  $\mathcal{M}_{\textrm{ID}_{\textrm{diff}}}$ is contained in the flux component $\Pi_{\textrm{ID}_{\textrm{diff}}} = \Pi_{\textrm{HP} \xrightarrow{\textrm{BP}} \textrm{HP}}^> \left(m\right)$. $\mathcal{M}_{\textrm{ID}_{\textrm{comp}}}$ is the compensating energy that must be exchanged between the near-inertial and tidal waves (BP) and the supertidal continuum (HP) in order to conserve supertidal wave action, as discussed in section \ref{sec:intro}\ref{sec:IW_theory}.  It is contained in $\Pi_{\textrm{ID}_{\textrm{comp}}} = \Pi_{\textrm{BP} \xrightarrow{\textrm{HP}} \textrm{HP}}^> \left(m\right)$.  The compensating energy triad ($\mathcal{M}_{\textrm{ID}_{\textrm{comp}}}$) is related to the kinetic-energy diffusion triad ($\mathcal{M}_{\textrm{ID}_{\textrm{diff}}}$) by an exchange of the catalyst and source modes, as depicted in Fig. \ref{fig:mechanisms}.  Both of these triads are treated together in wave turbulence theory because that theory (as it has been previously implemented) imposes symmetry between the source and catalyst modes, as discussed in appendix \ref{app:triads}.  Taken together, the two components of $\mathcal{M}_{\textrm{ID}}$ diffuse wave action in the supertidal band (but neither do this individually).  Also note that $\Pi_{\textrm{ID}}$ may not actually conserve wave action in the model evolution as it may not perfectly reflect the idealized $\mathcal{M}_{\textrm{ID}}$. 

Separately, one can consider wave-action diffusion that is induced by catalyst modes in the eddy field (LP) rather than the wave field (BP), which we label $\mathcal{M}_{\textrm{ID}_{\textrm{eddy}}}$.  This would be contained in $\Pi_{\textrm{ID}_{\textrm{eddy}}} = \Pi_{\textrm{HP} \xrightarrow{\textrm{LP}} \textrm{HP}}^> \left(m\right)$.  On the other hand, $\mathcal{M}_{\textrm{LI}}$ is the energy transfer within the supertidal band that scatters off of other supertidal modes, such that it would be contained in $\Pi_{\textrm{LI}} = \Pi_{\textrm{HP} \xrightarrow{\textrm{HP}} \textrm{HP}}^> \left(m\right)$.  Together, $\Pi_{\textrm{LI}}$, $\Pi_{\textrm{ID}_{\textrm{diff}}}$, and $\Pi_{\textrm{ID}_{\textrm{eddy}}}$ constitute all of the energy transfer within (meaning coming from and remaining in) the supertidal band and represent a fully nonlinear inertial range.  Other energy transfer into HP (coming from different frequency bands) is represented by $\Pi_{\textrm{ID}_{\textrm{comp}}}$ and $\Pi_{\textrm{Other}} = \Pi_{\textrm{BP} \xrightarrow{\textrm{LP+BP}} \textrm{HP}}^> \left(m\right) + \Pi_{\textrm{LP} \xrightarrow{\textrm{all}} \textrm{HP}}^> \left(m\right)$, which can be further subdivided if need be.  $\mathcal{M}_{\textrm{ES}}$ primarily involves an exchange of energy between different upward and downward propagating waves rather than from scale to scale, so we do not expect to see an $\mathcal{M}_{\textrm{ES}}$ signal in this analysis.  $\mathcal{M}_{\textrm{PSI}}$ would be contained in the spectral flux components $\Pi_{\textrm{PSI}} = \Pi_{\textrm{BP} \xrightarrow{\textrm{BP}} \textrm{BP}}^> \left(m\right) + \Pi_{\textrm{HP} \xrightarrow{\textrm{BP}} \textrm{BP}}^> \left(m\right) + \Pi_{\textrm{HP} \xrightarrow{\textrm{HP}} \textrm{HP}}^> \left(m\right)$, although we only see evidence of $\mathcal{M}_{\textrm{PSI}}$ in the first of these in the present results (see section \ref{sec:results}\ref{sec:m-omega}).

\subsubsection{Bispectra}
\label{sec:bispectra}

The advective spectral transfer (equation \ref{eq:adv_transf}) can also be written as a function of the vertical wavenumbers of two of the interacting fields instead of one.  Such bispectra allow for the discernment of the spectral transfer as a function of specific catalyst and source modes, as well as the destination modes accessible from the single-wavenumber version in equation \ref{eq:adv_transf}.  Determining the vertical wavenumbers in the energy exchanges along with the frequency bands allows for the identification of specific types of wave-wave interactions that cannot be isolated in the spectral flux frequency-band decomposition (e.g. equation \ref{eq:adv_flx_sources}).  As mentioned in section \ref{sec:intro}\ref{sec:IW_theory}, \cite{sun12} computed observational bispectra and did not find an $\mathcal{M}_{\textrm{ID}}$ signal where they expected it.  The bispectrum that they use can be written as:
\begin{align}
&\tilde{\mathcal{B}}_{\textrm{BP}\xrightarrow{\textrm{\textbf{HP}}} \textrm{\textbf{HP}}}  \left(m_{cat},m_{dest}\right)_z = \nonumber \\
& \quad 2 \left< \textrm{Re} \left( m_{src} \hat{u}_i^\textrm{HP}\left(m_{dest}\right) \hat{\varv}_3^\textrm{HP}\left(m_{cat}\right) \hat{u}_i^{*\textrm{BP}}\left(m_{src}\right)  \right)\right> \; \; \; \textrm{.} \label{eq:spectral_bispectra}
\end{align}
Here, the bold frequency-band labels indicate the two fields, in this case the catalyst and destination fields, that the independent wavenumbers reference.  This form of the bispectrum neglects the horizontal gradients in the advective term as an approximation based on horizontal homogeneity in the flow, an assumption that is neither applicable nor theoretically correct (see section \ref{sec:results}\ref{sec:results_bispectra}).  Also note that these interactions are constrained such that $m_{src}=m_{cat}+m_{dest}$, and therefore the bispectra computed in terms of any two of the field wavenumbers contain identical information. Computing such a bispectrum from 3-dimensional data would be prohibitively computationally expensive.  Therefore, we use a related form that is computed as an integral over position space:
\begin{align}
  &\mathcal{B}_{\textrm{BP}\xrightarrow{\textrm{\textbf{HP}}} \textrm{\textbf{HP}}}  \left(m_{cat},m_{dest}\right) = \nonumber \\
  & \qquad -  \left<u_i^{\textrm{HP}}\left(m_{dest}\right) \; \left(\varv_j^{\textrm{HP}}\left(m_{cat}\right) \partial x_j u_i^{\textrm{BP}} \right)\right> \; \; \; \textrm{.} \label{eq:bispectra_cat}
\end{align}
One difference between equation \ref{eq:spectral_bispectra} used by \cite{sun12} and equation \ref{eq:bispectra_cat}, used here, is that, because the latter is computed in position space, the positive and negative signs of the wavenumbers cannot be separated.  Thus the bispectra computed in this paper are equivalent to those in which the positive- and negative- $m_{cat}$ parts of equation \ref{eq:spectral_bispectra} are averaged together.  An implication of this averaging is that our bispectra will not pick up a signal from mechanisms that change sign with the sign of the wavenumber.  For example, $\mathcal{M}_\textrm{ES}$ requires complex-valued modes to discern upward and downward propagating waves exchanging energy.  The other key difference is that \cite{sun12} are only able to use the vertical-gradient component in equation \ref{eq:spectral_bispectra}, whereas we have written our bispectra in equation \ref{eq:bispectra_cat} as containing both gradient directions unless stated otherwise. 

Now that these bispectra have been defined, they can be interpreted along the same lines as the spectral flux decomposition in section \ref{sec:methods}\ref{sec:methods:budgets}\ref{sec:flux}.  Induced diffusion, as defined in \cite{mccomas77a}, would be contained in $\mathcal{B}_{\textrm{ID}_{\textrm{diff}}} = \mathcal{B}_{\textrm{HP}\xrightarrow{\textrm{BP}} \textrm{HP}}$ and $\mathcal{B}_{\textrm{ID}_{\textrm{comp}}} = \mathcal{B}_{\textrm{BP}\xrightarrow{\textrm{HP}} \textrm{HP}}$, which correspond to the diffusion of wave energy in the supertidal band and the compensating energy delivered from BP, respectively.  So, \cite{sun12}'s bispectrum (equation \ref{eq:spectral_bispectra}) only captures $\mathcal{M}_{\textrm{ID}_{\textrm{comp}}}$ and not the energy diffusion within the supertidal band.  As previously mentioned, \cite{dematteis22} reason that in $\mathcal{M}_{\textrm{ID}}$, the horizontal wavenumber should keep pace with the vertical in order to resolve the "oceanic ultraviolet catastrophe", implying a constant or forward frequency cascade, as opposed to an inverse cascade.  Importantly, this means that compensating energy, and therefore induced diffusion may not even be visible in the bispectra computed by \cite{sun12} if the associated frequency cascade is neutral.  In section \ref{sec:methods}\ref{sec:methods:budgets}\ref{sec:methods:m-omega} and section \ref{sec:results}\ref{sec:m-omega} spectra transfers as a function of both frequency and vertical wavenumber are introduced and presented, respectively, that aim to identify the direction of the frequency cascade associated with $\mathcal{M}_{\textrm{ID}}$ and other processes.  This will aid in interpreting $\mathcal{B}_{\textrm{ID}_{\textrm{comp}}}$. 

The reason that the bispectra of \cite{sun12} was not identified by the authors as measuring  $\mathcal{M}_{\textrm{ID}_{\textrm{comp}}}$ and not $\mathcal{M}_{\textrm{ID}_{\textrm{diff}}}$ is likely that wave turbulence theory (WTT) and many other theoretical approaches to turbulence utilize scattering coefficients that are symmetrized between the source and catalyst wavenumbers, as discussed in appendix \ref{app:triads} and depicted in Fig. \ref{fig:wtt}.  Put another way, in the WTT framework, $\left<u_i^{\textrm{HP}} \left(\varv_j^{\textrm{BP}} \partial x_j u_i^{\textrm{HP}} \right)\right>$ and $\left<u_i^{\textrm{HP}} \left(\varv_j^{\textrm{HP}} \partial x_j u_i^{\textrm{BP}} \right)\right>$ are treated interchangeably. However, the observations of \cite{sun12} do not symmetrize between the catalyst and source modes.  They simply measured $\left<u_i^{\textrm{HP}} \left(\varv_3^{\textrm{HP}} \partial x_3 u_i^{\textrm{BP}} \right)\right>$.  

To address the question of whether the bispectra of \cite{sun12} were consistent with the model output used in this paper, the catalyst bispectra, equation \ref{eq:bispectra_cat}, is computed and presented in section \ref{sec:results}\ref{sec:sp12}. However, in order to observe $\mathcal{M}_{\textrm{ID}_{\textrm{diff}}}$, bispectra using the energy source rather than the catalyst mode are also computed. The latter formulation allows for the discernment of energy transfer between modes of similar wavenumber\footnote{If $\mathcal{B}_{\textrm{ID}_{\textrm{diff}}}$ were instead plotted as a function of the catalyst mode (in the manner of equation \ref{eq:bispectra_cat}), the positive and negative catalyst modes would be expected to have oppositely signed bispectra.  When the positive- and negative- catalyst-mode bispectra are averaged to generate a real-valued field to be integrated in position space (again, refer to equation \ref{eq:bispectra_cat}), the oppositely signed contributions would cancel.}. Such a source-destination bispectrum (containing $\mathcal{M}_{\textrm{ID}_{\textrm{diff}}}$) can be written as:
\begin{align}
\mathcal{B}_{\textrm{\textbf{HP}} \xrightarrow{\textrm{BP}} \textrm{\textbf{HP}}}^> \left(m_{src},m_{dest}\right) &= - \left<u_i^{\textrm{HP}}\left(m_{dest}\right) \; \left(\varv_j^{\textrm{BP}} \partial x_j u_i^{<, \textrm{HP}}\left(m_{src}\right) \right)\right> \label{eq:src_bispectra} 
\end{align}

\subsubsection{Spatial distributions of integrated spectral budgets}
\label{sec:spatial_dist}
The horizontal distributions of the vertical spectral flux (in equation \ref{eq:adv_flx_sources}) can also be computed by omitting a horizontal average.  The analogous component with horizontal spatial dependence looks like: 
\begin{align}
\Pi_{\textrm{ID}_{\textrm{diff}}}^>\left(m,x,y,\omega\right) &= - \left<u_i^{>,\textrm{HP}}{\left(\omega\right)} \; \left(\varv_j^{\textrm{BP}} \partial x_j u_i^{\textrm{HP}} \right)\right>_{z,t} \label{eq:adv_flx_xy} 
\end{align}
where now the brackets indicate a vertical and time average.  This will be used in Fig. \ref{fig:regions_of_study} to select subregions of interest.  Detailed comparison of spatial distributions of spectral flux as well as various dissipation mechanisms will be reserved for a future paper.

\subsubsection{Spectral transfers vs wavenumber and frequency}
\label{sec:methods:m-omega}

As has been mentioned, compensating energy flux associated with induced diffusion ($\Pi_{\textrm{ID}_{\textrm{comp}}}$) may be forward, backward, or approximately zero depending on how the aspect ratio of the IW wavenumbers changes between the source and destination modes, which in turn corresponds to the direction of energy flow in frequency space through the IW dispersion relation.  A forward frequency cascade associated with induced diffusion within the supertidal band (i.e. specifically within $\Pi_{\textrm{ID}_{\textrm{diff}}}$) would require energy exchange from BP to HP ($\Pi_{\textrm{ID}_{\textrm{comp}}}>0$) while an inverse frequency cascade would require energy exchange from HP to BP ($\Pi_{\textrm{ID}_{\textrm{comp}}}<0.$).  The direction of frequency exchange would be computationally prohibitive to determine directly.  Therefore, we instead compute the spectral transfer as a function of destination vertical wavenumber and frequency for ID$_{\textrm{diff}}$:  
\begin{align}
\mathcal{T}_{\textrm{ID}_{\textrm{diff}}}\left(m,\omega\right) &= \mathcal{T}_{\textrm{HP}\xrightarrow{\textrm{BP}}\textrm{all}}\left(m,\omega\right) \\
&= \left<u_i{\left(m,\omega\right)} \; \varv_j^\textrm{BP} \partial x_j(u_i)^{\textrm{HP}} \right> \label{eq:T_decomp}
\end{align}
The transfer in equation \ref{eq:T_decomp} conserves energy in the supertidal band.  Thus, it should be possible to infer the direction of the cascade by looking at where the sources and sinks of KE are in the HP band.   Spectral transfers are also computed for the other terms in order to see, for example, if there is evidence of $\mathcal{M}_{\textrm{PSI}}$, how energy is exchanged between potential and kinetic energy, and at what frequencies energy is entering and leaving the subdomains of interest.


\section{Results and discussion}
\label{sec:results}

\begin{figure}[]
\centering
\begin{subfigure}{0.4\textwidth}
\includegraphics[width=\textwidth]{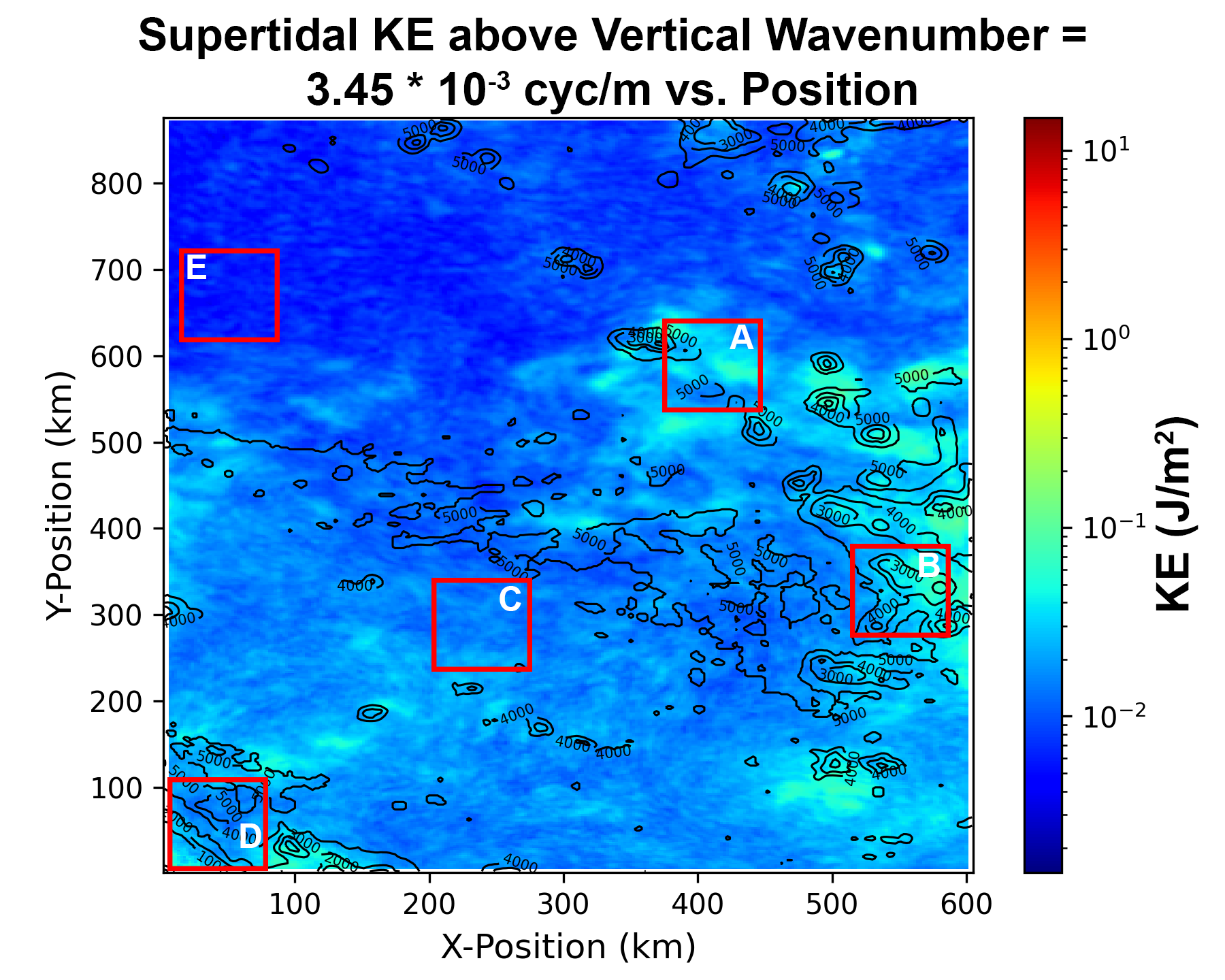}
\centering \caption{Supertidal (HP) kinetic energy above vertical wavenumber $3.45 \times 10^{-3} \textrm{m}^{-1}$.}
\label{fig:HP_KE}
\end{subfigure}

\begin{subfigure}{0.4\textwidth}
\centering
\includegraphics[width=\textwidth]{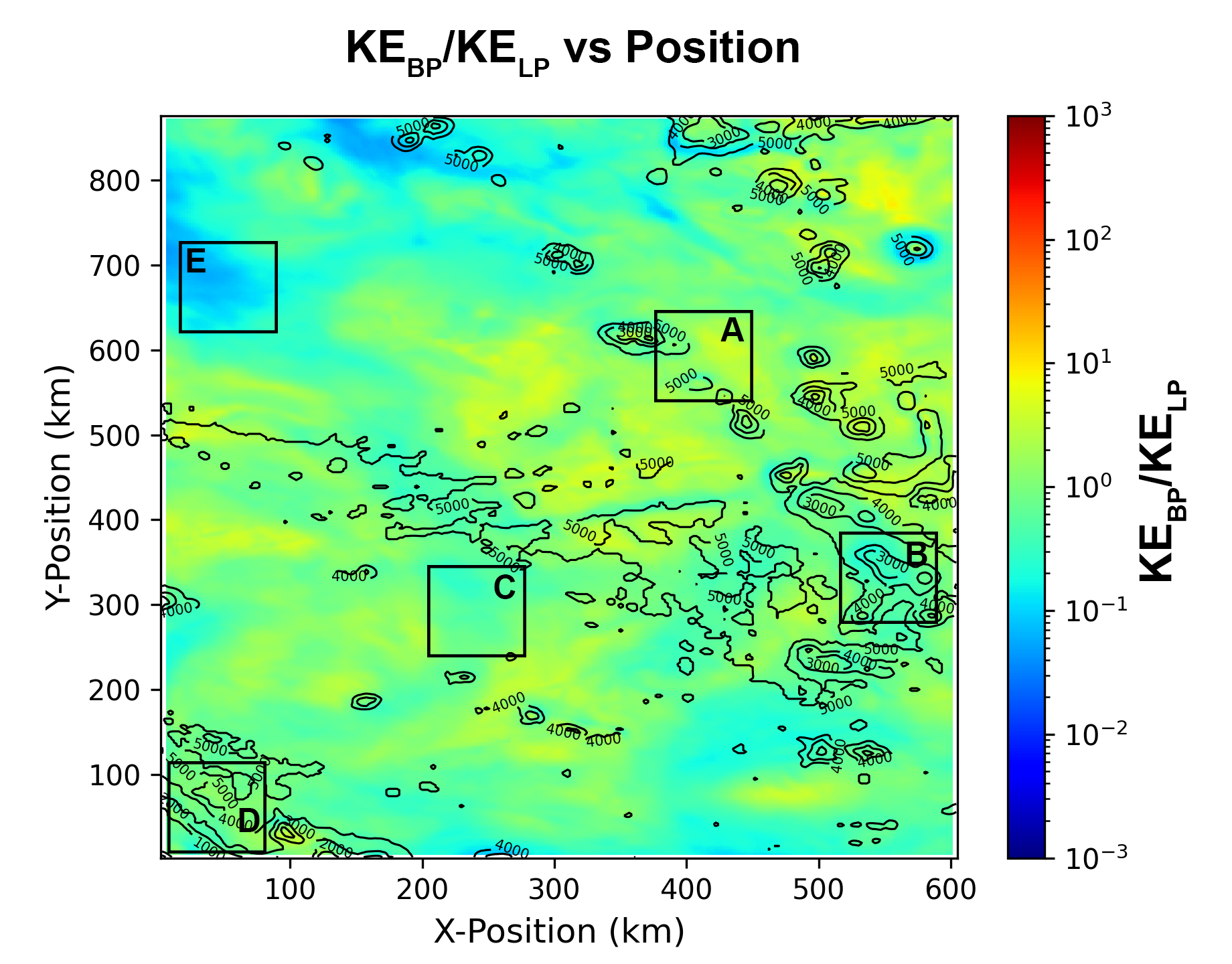}
\centering \caption{The ratio of BP (sum of near-inertial and tidal) kinetic energy to LP (eddy) kinetic energy.}
\label{fig:BP_to_LP}
\end{subfigure}

\begin{subfigure}{0.4\textwidth}
\centering
\includegraphics[width=\textwidth]{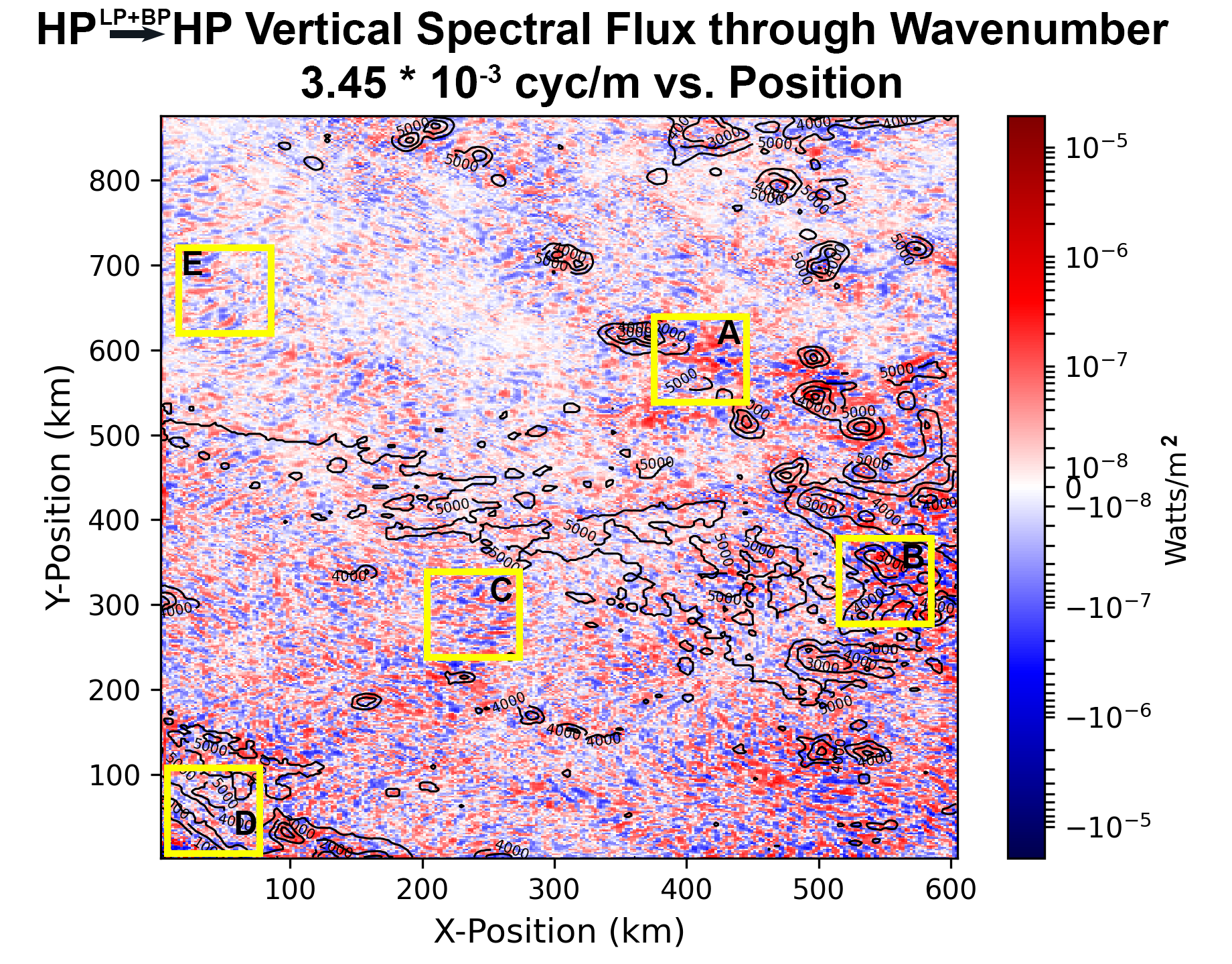}
\centering \caption{Downscale vertical induced-diffusion-type spectral kinetic-energy flux through wavenumber $3.45 \times 10^{-3} \textrm{m}^{-1}$, computed as in equation \ref{eq:adv_flx_xy}.}
\label{fig:hp_flux}
\end{subfigure}
\centering \caption{Maps of various diagnostics (see subfigure captions) in the regions of study. Contours indicate 1000-meter levels of the bathymetry.}
\label{fig:regions_of_study}
\end{figure}

All results in this paper are computed from an average of five subregions of interest, shown in Fig. \ref{fig:regions_of_study}.  Subregions, rather than the entire model domain, are used to compute results because of the prohibitive computational cost and data management requirements associated with the high-resolution case, which has 2.3 billion points of 3D data over the entire model domain. These subregions were chosen to capture a sample of rough and flat topography (visible in the contours) and a range of supertidal-frequency (HP) kinetic energies (Fig. \ref{fig:HP_KE}), supertidal (HP) vertical spectral kinetic energy fluxes (Fig. \ref{fig:hp_flux}), and ratios of near-inertial and tidal to eddy (BP/LP) kinetic energy (Fig. \ref{fig:BP_to_LP}).  All results in this paper are computed with units of inverse volume as opposed to mass.  We use a background density of $\rho_0 = 1027.5$ kg m$^{-3}$.  When applicable, the domain-averaged Coriolis frequency, $f_0 = 6.85 \times 10^{-5}$ s$^{-1}$ is also used.

\begin{figure}[]
\centering
\begin{subfigure}{0.5\textwidth}
\includegraphics[width=\textwidth]{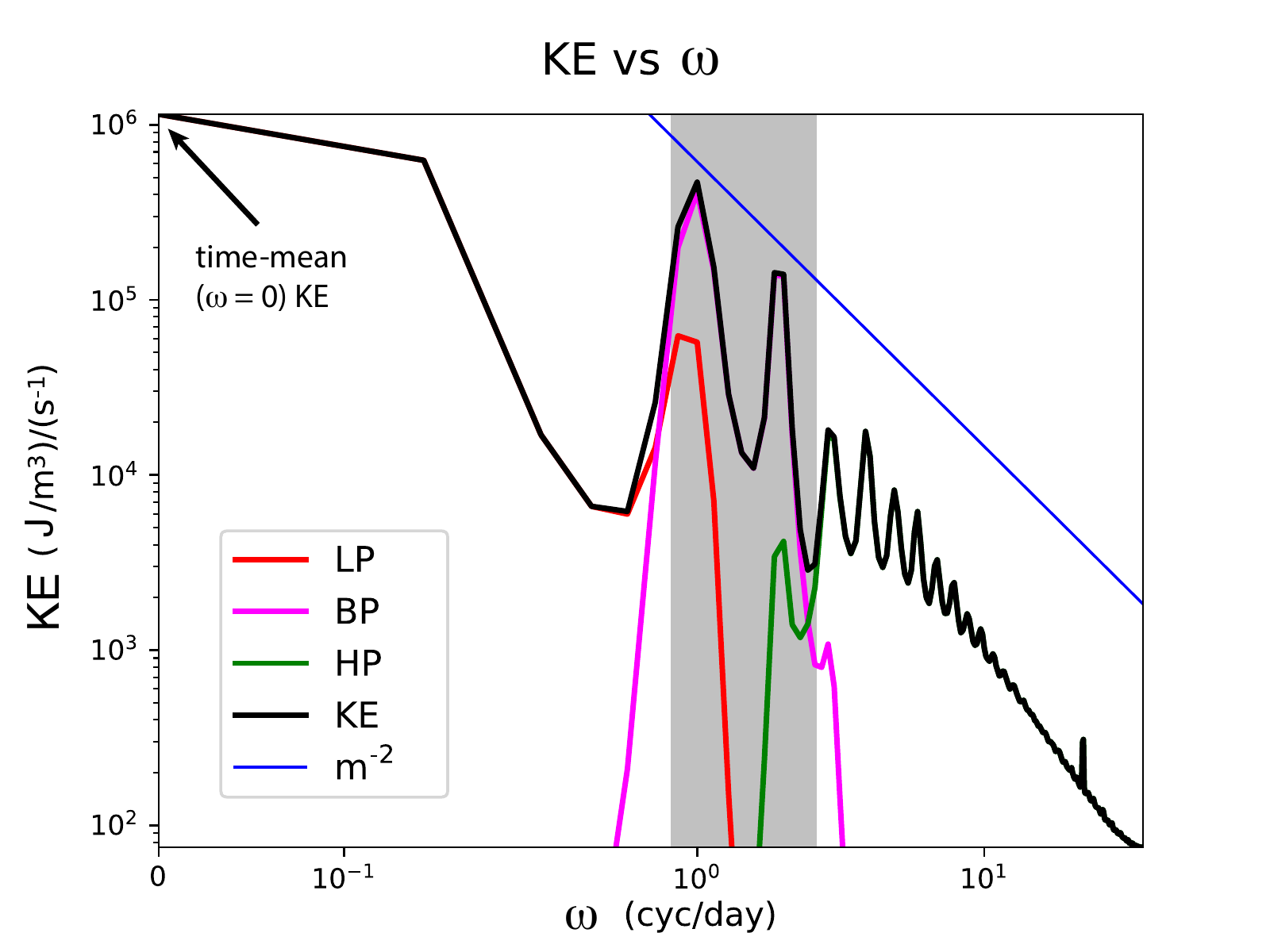}
\centering \caption{Frequency spectrum of the high-resolution case averaged across the subregions of interest, displayed in Fig. \ref{fig:regions_of_study}.  The band of $0.8 f_0 < \omega < 2.5 f_0$ is blocked out in gray. The low-pass/band-pass/high-pass decomposition (see text) of the overall frequency spectrum (black) is shown using the red, magenta, and green curves, respectively. The decomposition was performed using a 12th-order bandpass.  The blue line indicates the asymptotic GM slope value of $\omega^{-2}$.}
\label{fig:fig1a}
\end{subfigure}

\begin{subfigure}{0.5\textwidth}
\centering
\includegraphics[width=\textwidth]{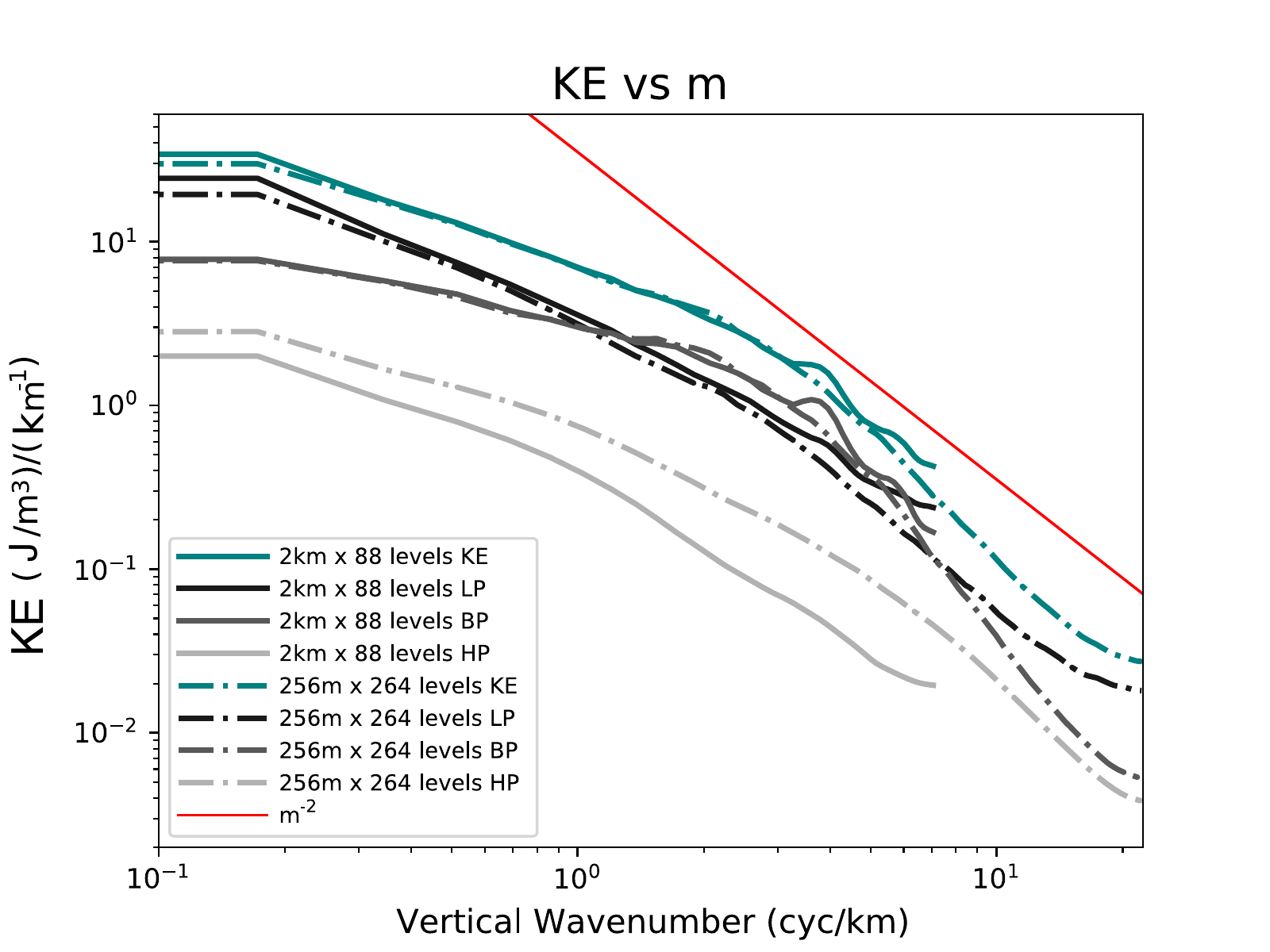}
\centering \caption{Vertical-wavenumber spectrum of the different frequency bands in the high- and low-resolution cases averaged across the subregions of interest, shown in Fig. \ref{fig:regions_of_study}.}
\label{fig:fig1b}
\end{subfigure}
\centering \caption{Kinetic-energy spectra}
\label{fig:fig1}
\end{figure}

The frequency spectrum of the high-resolution case (256m x 264 vertical levels) is shown in Fig. \ref{fig:fig1a} with shading and separate curves indicating the three frequency bands discussed in section \ref{sec:methods}\ref{sec:methods:budgets}.  The band lines are shown at all frequencies to convey the overlap left after a 12th-order bandpass filter is applied.  \cite{nelson20} have previously compared frequency spectra of these runs to observations and the GM spectrum \citep{garrett75}. The supertidal frequency spectrum in the high resolution simulation is slightly steeper than the asymptotic $\omega^{-2}$ prediction of GM while the spectrum from the low-resolution model (not shown) falls off much more steeply.  In vertical wavenumber space (Fig. \ref{fig:fig1b}), the eddy (LP) frequencies are energetically dominant at low and high wavenumbers while the BP frequencies are energetically dominant at intermediate wavenumbers.  HP contains the least kinetic energy - an order of magnitude less than BP.  There is some energy buildup at the vertical grid-scale in both simulations in Fig. \ref{fig:fig1b}.  This buildup is likely a reflection of the sharp gradients that exist at the top and bottom of the domain and of the (intentional) choice to analyze the entire water column rather than tapering it, as is done in \cite{thakur22, nelson20, pan20}.  Finally, note that KE of the HP band is much greater at higher resolution, while the energy levels of BP and LP do not change very much between the different resolution cases.

\subsection{Integrated vertical spectral KE budgets}
\label{sec:partial_budget_results}
\label{sec:frequency-decomposition}

As described in section \ref{sec:methods}\ref{sec:methods:budgets}, the downscale vertical spectral KE flux and integrated vertical spectral dissipation transfers should be in balance at high vertical wavenumbers, at least to the extent that scale-separation exists in the internal wave continuum.  Assessing details of the budget at these wavenumbers is of primary importance. (In contrast, forcing terms are active at the lowest vertical wavenumbers such that the spectral flux and dissipation should not be in balance at those wavenumbers).  In order to emphasize details at high wavenumbers, budgets are presented on linear axes as opposed to logarithmic axes, the latter of which would emphasize the lowest wavenumbers.  

\begin{figure*}[]
\centering
\begin{subfigure}{0.45\textwidth}
\includegraphics[width=\textwidth]{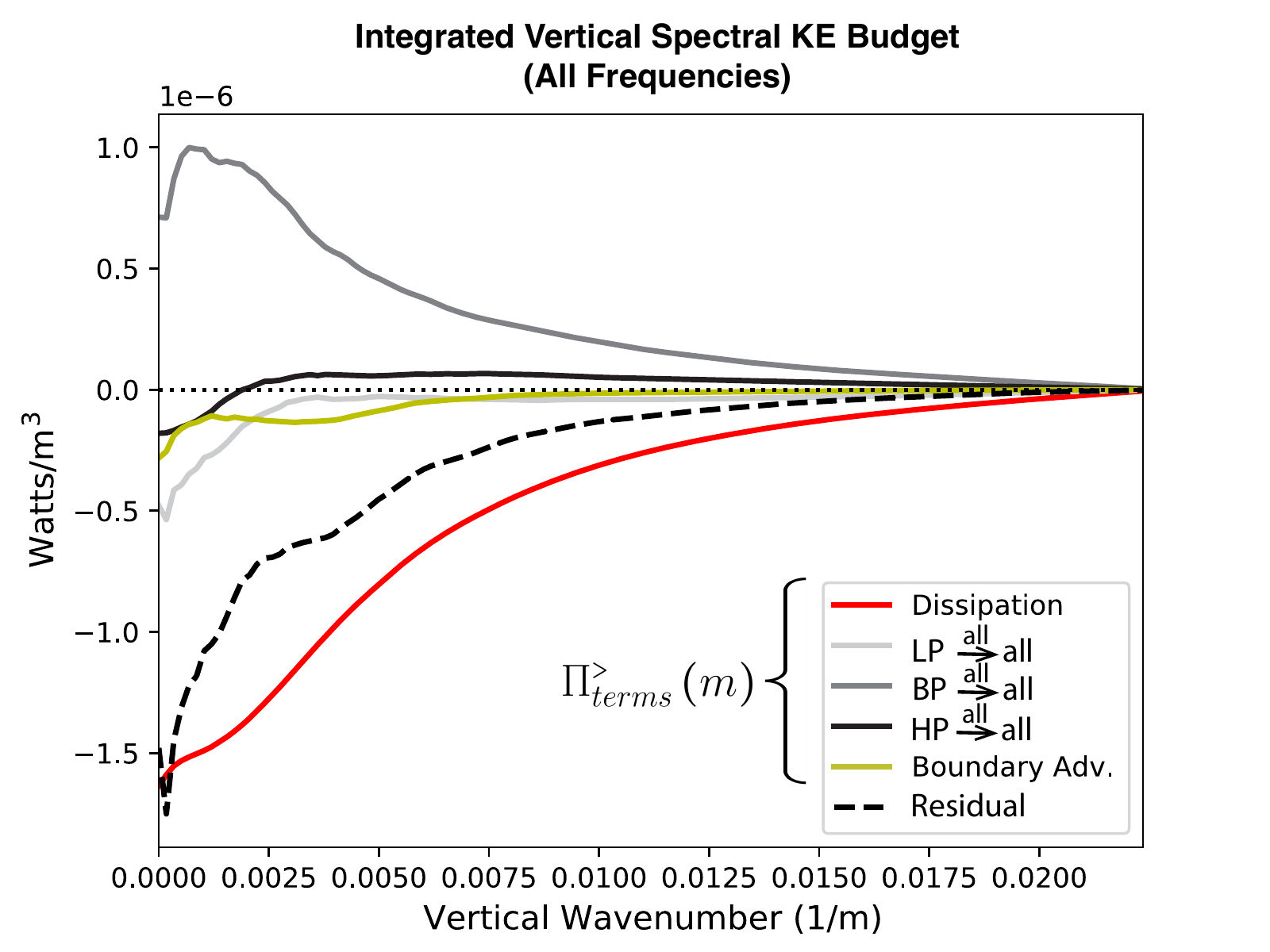}
\centering \caption{All Frequencies}
\label{fig:high_res_budget_all}
\end{subfigure}
\begin{subfigure}{0.45\textwidth}
\centering
\includegraphics[width=\textwidth]{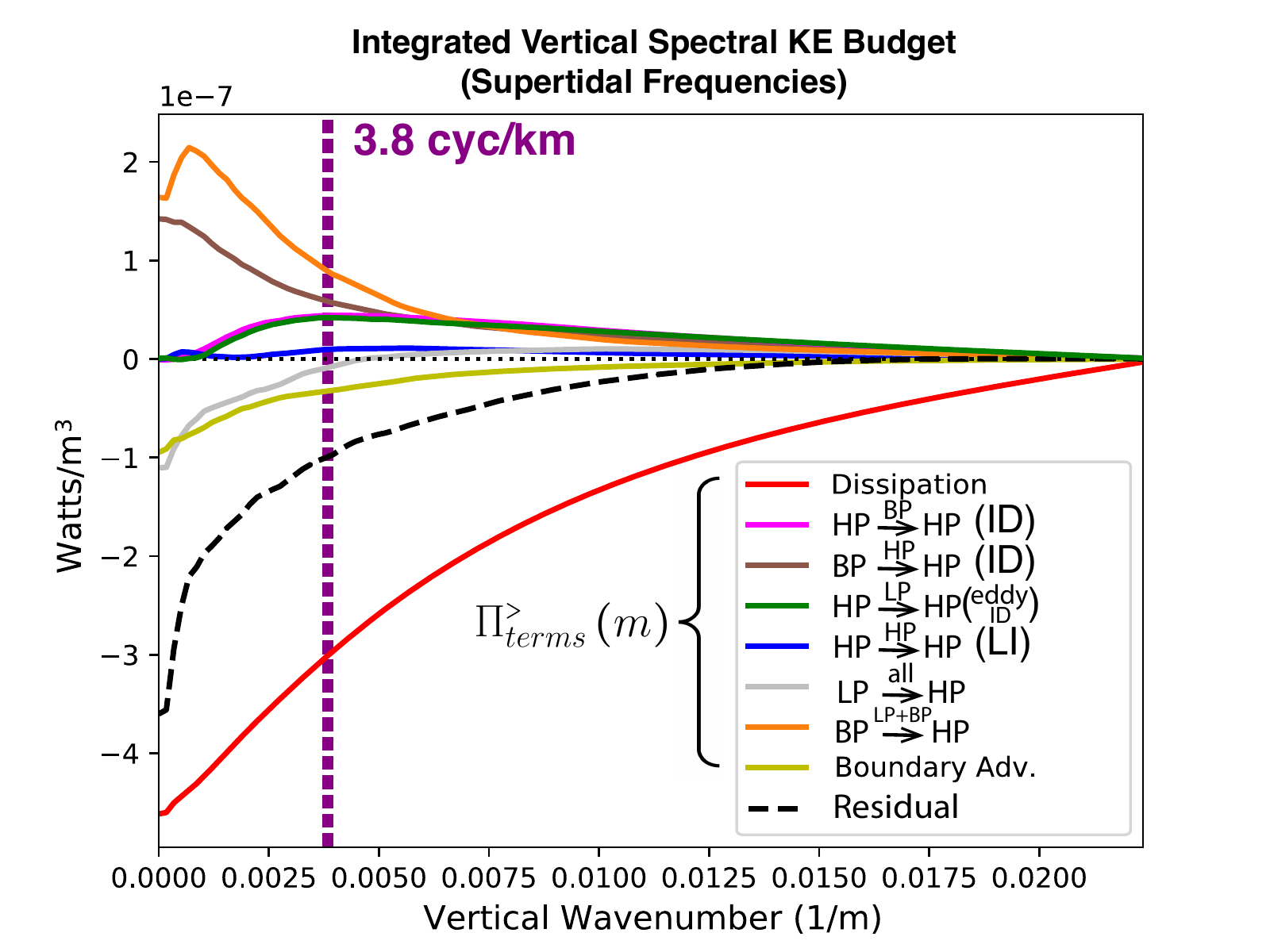}
\centering \caption{HP Band ($\omega > 2.5 f_0$)}
\label{fig:high_res_budget_hp}
\end{subfigure}
\centering \caption{Partial integrated spectral budgets of KE for the high-resolution run averaged across the subregions of interest, shown in Fig. \ref{fig:regions_of_study}.  Note that the local spectral budget (the change in energy of a given wavenumber) is proportional to the slope of the curve at that wavenumber.  The purple line indicates the vertical wavenumber at which the terms are sampled for the resolution and subregional comparison in Fig. \ref{fig:decomp}.}
\label{fig:high_res_budgets}
\end{figure*}

The partial integrated vertical spectral kinetic-energy budget of the high-resolution case is shown in Fig. \ref{fig:high_res_budgets}, with the left and right panels corresponding to all and HP frequencies.  The overall amount of dissipation can be assessed by the value of the dissipation curve at $m=0$, while the amount of energy that is dissipated at a given vertical wavenumber is proportional to the slope of the dissipation curve.  The dissipation, spectral flux, and advection into the domain are in balance at the highest wavenumbers for both all frequencies (Fig. \ref{fig:high_res_budget_all}) and the supertidal (Fig. \ref{fig:high_res_budget_hp}) frequencies.  The latter indicates that the internal waves are energized at small and intermediate but not high vertical wavenumbers; energy in the IW field at high vertical wavenumbers gets there through nonlinear interactions among waves and eddies in the flow.  This does not indicate that scale separation exists between forcing and dissipation; there is no classic inertial range at the resolutions used in this paper.  However, the absence of forcing at the highest wavenumbers indicates that the IW continuum has some cascade-like properties and that an LES-type IW closure may be appropriate.   

\begin{figure}[]
\includegraphics[width=0.5\textwidth]{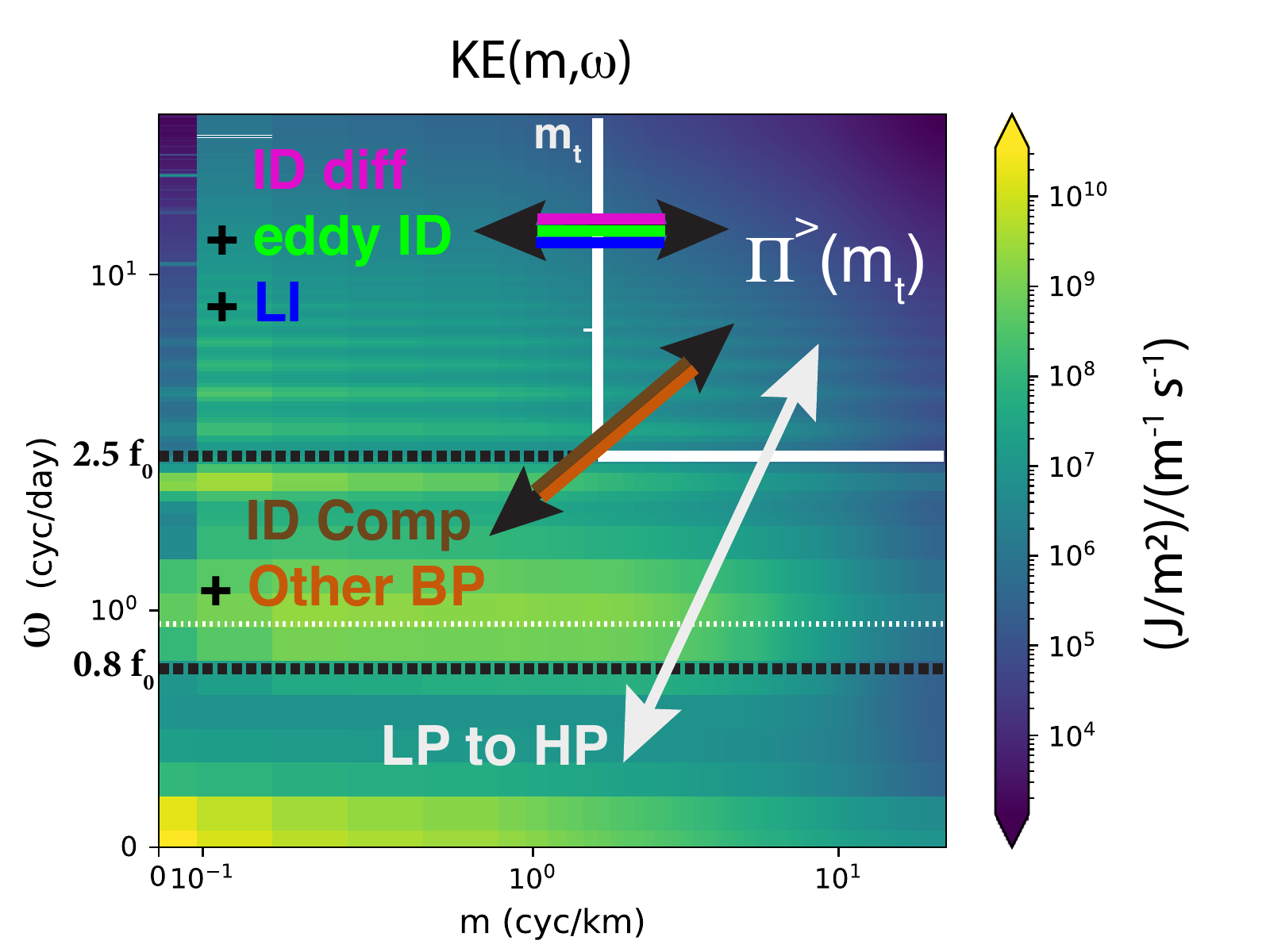}
\centering \caption{KE m-$\omega$ spectra and guide for different mechanisms averaged across the subregions of interest, shown in Fig. \ref{fig:regions_of_study}.  The white horizontal dotted line corresponds to the inertial frequency, $f_0$, averaged over the domain.}
\label{fig:KE_m_omega}
\end{figure}

The frequency-decomposed vertical spectral KE-flux reflects specific types of nonlinear interactions that underlie the ocean's IW continuum.  Notably, within the HP band, $\mathcal{M}_{\textrm{ID}}$ and $\mathcal{M}_{\textrm{LI}}$ are expected to play a dominant role \citep{dematteis21,dematteis22}, at least among wave interactions.  As described in section \ref{sec:methods}\ref{sec:methods:budgets}\ref{sec:flux}, spectral flux into the HP band is decomposed into six components (see equation \ref{eq:hp_flx_decomp} and Fig. \ref{fig:mechanisms}).  These six components are presented in Fig. \ref{fig:high_res_budget_hp} along with dissipation and boundary advection.  The HP-to-HP-flux components are exchanges between large and small vertical scales on either side of vertical wavenumber $m$.  On the other hand, the BP-to-HP and LP-to-HP flux components can come from any vertical wavenumber; these directional exchanges are depicted in wavenumber-frequency space in Fig. \ref{fig:KE_m_omega}.  Taken together, the components of the spectral flux shown in figures \ref{fig:high_res_budget_hp} and \ref{fig:KE_m_omega} constitute a complete decomposition of all advective energy exchange with supertidal (HP) modes above a given vertical wavenumber, $m$.  

\begin{figure}[]
\centering
\begin{subfigure}{0.5\textwidth}
\includegraphics[width=\textwidth]{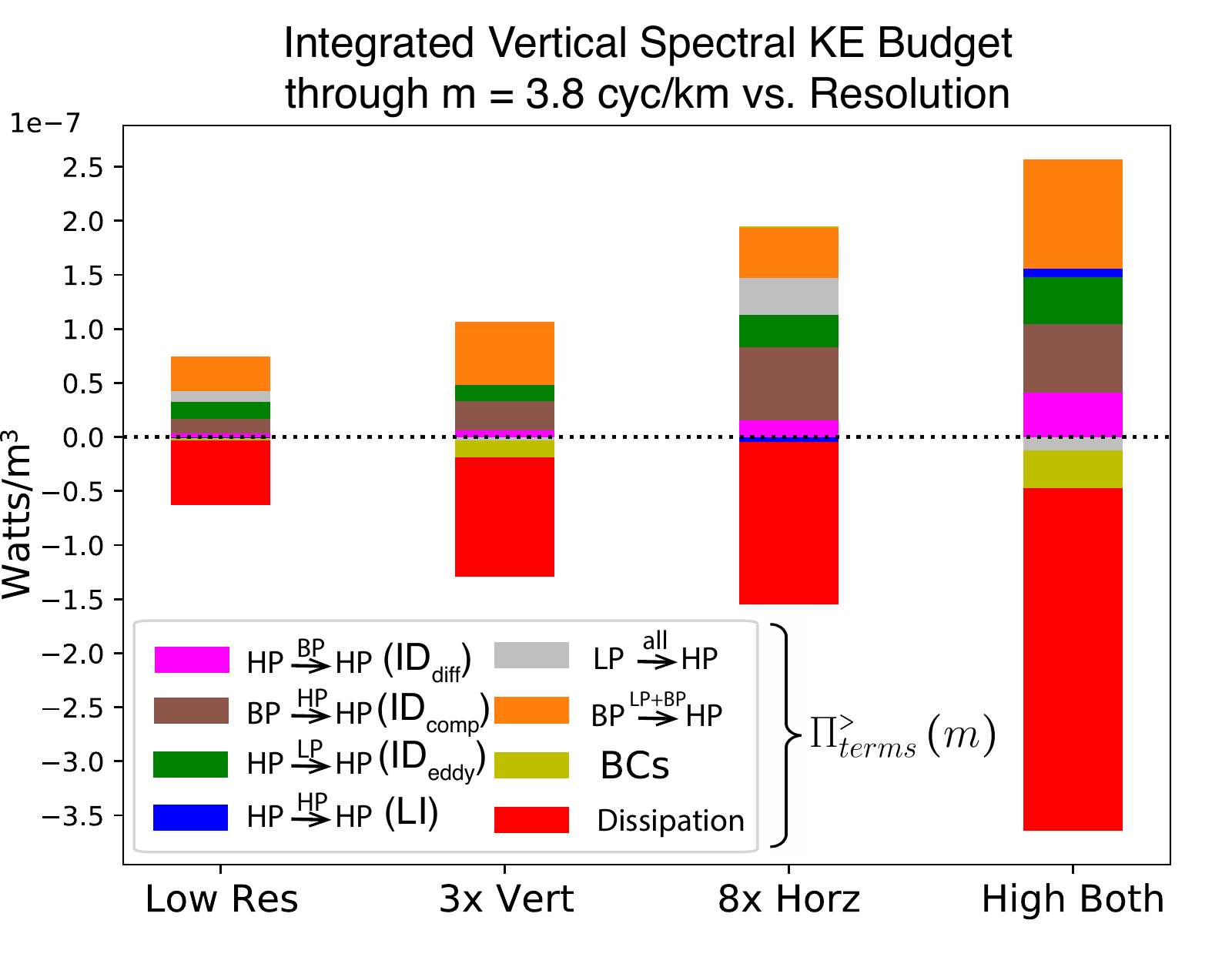}
\centering \caption{vs. resolution.}
\label{fig:decomp_res}
\end{subfigure}

\begin{subfigure}{0.5\textwidth}
\centering
\includegraphics[width=\textwidth]{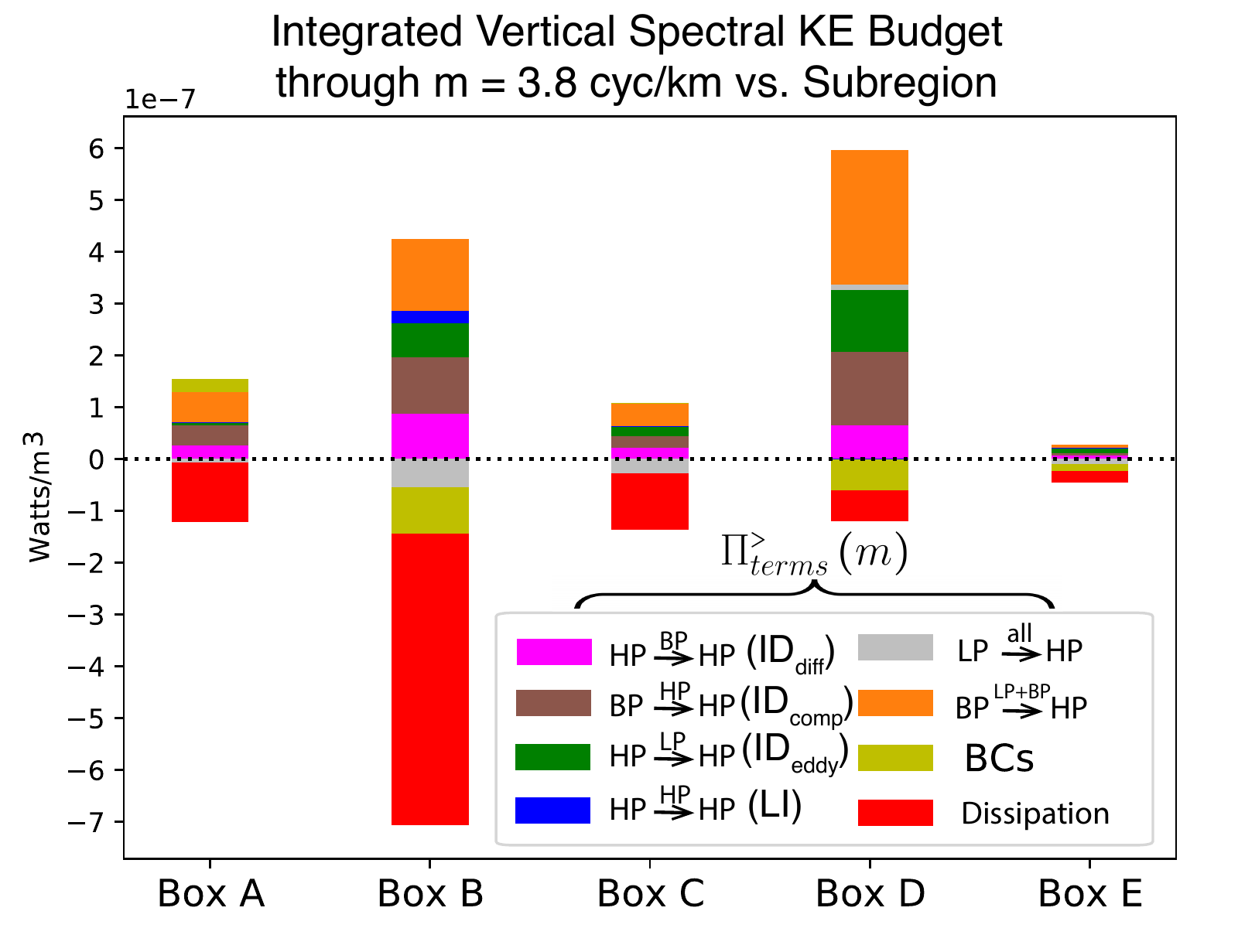}
\centering \caption{vs. location, for the high-resolution case.  Box location is shown in Fig. \ref{fig:regions_of_study}}
\label{fig:decomp_loc}
\end{subfigure}
\centering \caption{A partial integrated vertical spectral KE budget (equation \ref{eq:adv_flx_sources}) at supertidal (HP) frequencies through 3.8 cyc/m (indicated by the purple lines in Fig. \ref{fig:high_res_budgets}) versus location and resolution.  Spectral flux decomposition terms are included along with dissipation and boundary flux.  Source terms and KPP's mixed-layer dissipation are omitted.  The PE-to-KE conversion is excluded but is confirmed to be small in these breakdowns.}
\label{fig:decomp}
\end{figure}

The individual components of spectral KE flux can be more easily compared across simulations and regions by looking at just a single vertical wavenumber.  Integrated spectral budgets through a specific wavenumber of $m = 3.8$ cyc/km are compared across resolutions and subregions in Fig. \ref{fig:decomp}.  This wavenumber was chosen to capture a strong signal from the nonlinear KE scattering mechanisms in both the high-resolution run (in which the peak $\Pi_{\textrm{ID}}$ is at a slightly higher vertical wavenumber) and the low-resolution run (in which the peak $\Pi_{\textrm{ID}}$ is at a slightly lower vertical wavenumber).  In all cases, BP-to-HP flux, which is decomposed into $\Pi_{\textrm{ID}_{\textrm{comp}}}$ and $\Pi_{\textrm{BP}_{\textrm{other}}}$, is the dominant advective energy transfer to the HP band relative to energy flux from the eddy (LP) and supertidal (HP) bands.  Note that at higher vertical wavenumbers in the high-resolution case, the HP-to-HP flux becomes comparable to the BP-to-HP flux (seen in Fig. \ref{fig:high_res_budget_hp}), but this cannot be meaningfully compared to the lower resolutions which do not resolve such wavenumbers.  

In all cases in Fig. \ref{fig:decomp}, ($\Pi_{\textrm{LI}}$) is an insignificant portion of the vertical spectral KE flux, and intraband energy diffusion associated with $\Pi_{\textrm{ID}_{\textrm{diff}}}$ is only significant in the highest resolution case.  The spectral flux component that includes $\mathcal{M}_{\textrm{ID}_{\textrm{comp}}}$ from near-inertial and tidal frequencies ($\Pi_{\textrm{ID}_{\textrm{comp}}}$) is much larger than and not in consistent proportion to the one containing supertidal energy diffusion, suggesting that this component is capturing mechanisms other than $\mathcal{M}_{\textrm{ID}}$ that should be apparent in the bispectral analysis.  These results stand in contrast to the prediction by \cite{dematteis22} that $\mathcal{M}_{\textrm{LI}}$ would be the largest mechanism moving energy downscale in the IW continuum.  For the parameters in that study, $\mathcal{M}_{\textrm{LI}}$ is responsible for moving about five times as much supertidal energy as $\mathcal{M}_{\textrm{ID}}$.  It should be noted that the vertical grid spacing in the steepest part of the pycnocline is very fine at approximately 10 meters.  It is unclear if significant further improvement in the representation of the IW scattering mechanisms will result from increased resolution alone.  It is possible that a nonhydrostatic model is required to permit greater spectral KE flux due to $\mathcal{M}_{\textrm{LI}}$ that is not captured with a hydrostatic model.   

The decomposition in Fig. \ref{fig:decomp_res} indicates that the spectral flux due to eddy-induced diffusion ($\Pi_{\textrm{ID}_{\textrm{eddy}}} = \Pi_{\textrm{HP} \xrightarrow{\textrm{LP}} \textrm{HP}}^>$) is larger than regular wave-induced energy diffusion ($\Pi_{\textrm{ID}_{\textrm{diff}}}$) in all cases except for the highest resolution case.  Perhaps a mechanistic understanding of the interactions that give rise to the GM spectrum cannot be gained without accounting for these eddy interactions, even though they are frequently omitted in wave-only frameworks of WTT. 

Combined, the HP-to-HP spectral flux ($\Pi_{\textrm{ID}_{\textrm{diff}}}$, $\Pi_{\textrm{ID}_{\textrm{eddy}}}$ and $\Pi_{\textrm{LI}}$) contribute significantly (> 20\% of the total downscale contributions to the integrated budget) for the two cases with high horizontal resolution, and only modestly to the low horizontal resolution cases (< 20\%).  Increasing the vertical resolution does not noticeably increase the HP-to-HP spectral flux as a portion of the whole.  This relatively small sensitivity to vertical resolution suggests that horizontal resolution is more important for representing these processes (particularly the first two) and, by extension, a fully nonlinear IW continuum with direct energy exchange between supertidal modes. This finding is consistent with \cite{nelson20}, and may be due to the requirement for high horizontal resolution to activate higher vertical modes in the continuum \citep{thakur22}.  

An important implication of the resolution dependence of HP-to-HP spectral flux is that at the lower resolutions (2km $\times$ 88 levels, a resolution that is computationally feasible in an IW-permitting global model such as MITgcm LLC4320), the IW continuum will be effectively (generalized) quasilinear\footnote{"quasilinear" is a technical term that does not mean approximately linear.} \citep{marston16} around the lower-frequency tides, near-inertial waves, and eddy fields in which interactions of the type $\Pi_{\textrm{HP} \xrightarrow{\textrm{all}} \textrm{HP}}^>$ are not moving a significant portion of the energy in the supertidal band.  The quasilinear nature of the continuum in global models does not necessarily mean that their mixing and transport properties are inaccurate. In various turbulent flows, generalized quasilinear solutions can be useful approximations of the full nonlinear solutions, e.g. \cite{marston16}.  Nonetheless, the relative weakness of nonlinearity in the IW continuum in the new generation of global IW-permitting models should be accounted for when developing and applying closures and mixing schemes appropriate for such models.  The use of higher resolutions in regional models permit a different, fully nonlinear regime of the IW continuum that (1) will be useful as a model of ``truth'' for understanding and improving IW handling in coarser resolution global models and (2) may also be optimally handled with a different dissipation mechanism that accounts for the different nature of the IW flows in regional vs. global models.  

The vertical spectral KE flux decomposition in the high-resolution case is compared across the different subregions in Fig. \ref{fig:decomp_loc}.  Box B, which is over rugged topography (like box D) and has high supertidal (HP) KE (like box A), clearly contributes the most to vertical spectral KE flux and dissipation of the 5 regions. In particular, both $\Pi_{\textrm{ID}_{\textrm{diff}}}$ and $\Pi_{\textrm{ID}_{\textrm{diff}}}$ are each larger in magnitude in Box B than in the other boxes, while $\Pi_{\textrm{ID}_{\textrm{eddy}}}$ is largest in both boxes B and D, the two with the roughest bathymetry.  

One possible reason for the markedly different breakdown of energy transfers in box D is its proximity to the domain boundary.  This proximity would limit the amount of energy advecting into that region at the highest vertical wavenumbers and frequencies. Another possible reason is that the rough topography in box D is rather shallow, coming to within 150 meters of the surface, whereas it is at great depth in box B (see Fig. \ref{fig:bathymetry}).  The latter reason would indicate enhanced flux (and wave-induced dissipation and mixing) occurring at depth through bottom interactions. Vertical distributions of horizontal spectral KE flux and dissipation mechanisms will be examined in a subsequent paper and will shed light on fluxes at depth.  

The boxes (B and D) overlying rough topography also have negative contributions of HP KE to their budgets from the boundary advection (BC) term.  Supertidal (HP) IW energy is generated in these regions and moves outward into the rest of the domain (such as boxes A and C).

\subsection{Bispectra of nonlinear scattering mechanisms}
\label{sec:results_bispectra}

\begin{figure*}[]
\centering
\begin{subfigure}{0.45\textwidth}
\includegraphics[width=\textwidth]{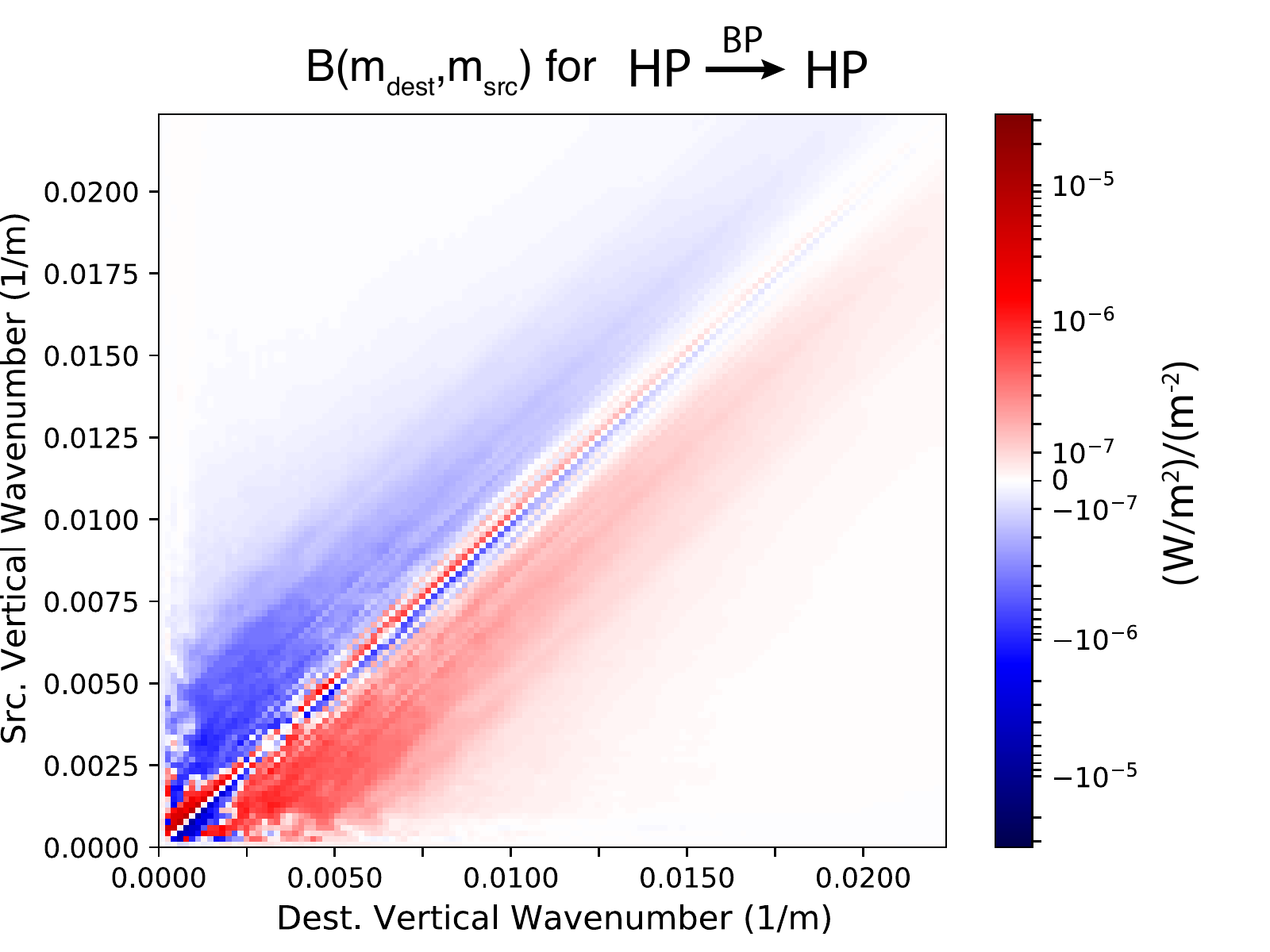}
\centering \caption{Induced Diffusion: HP $\xrightarrow{BP}$ HP}
\label{fig:Bid}
\end{subfigure}
\begin{subfigure}{0.45\textwidth}
\includegraphics[width=\textwidth]{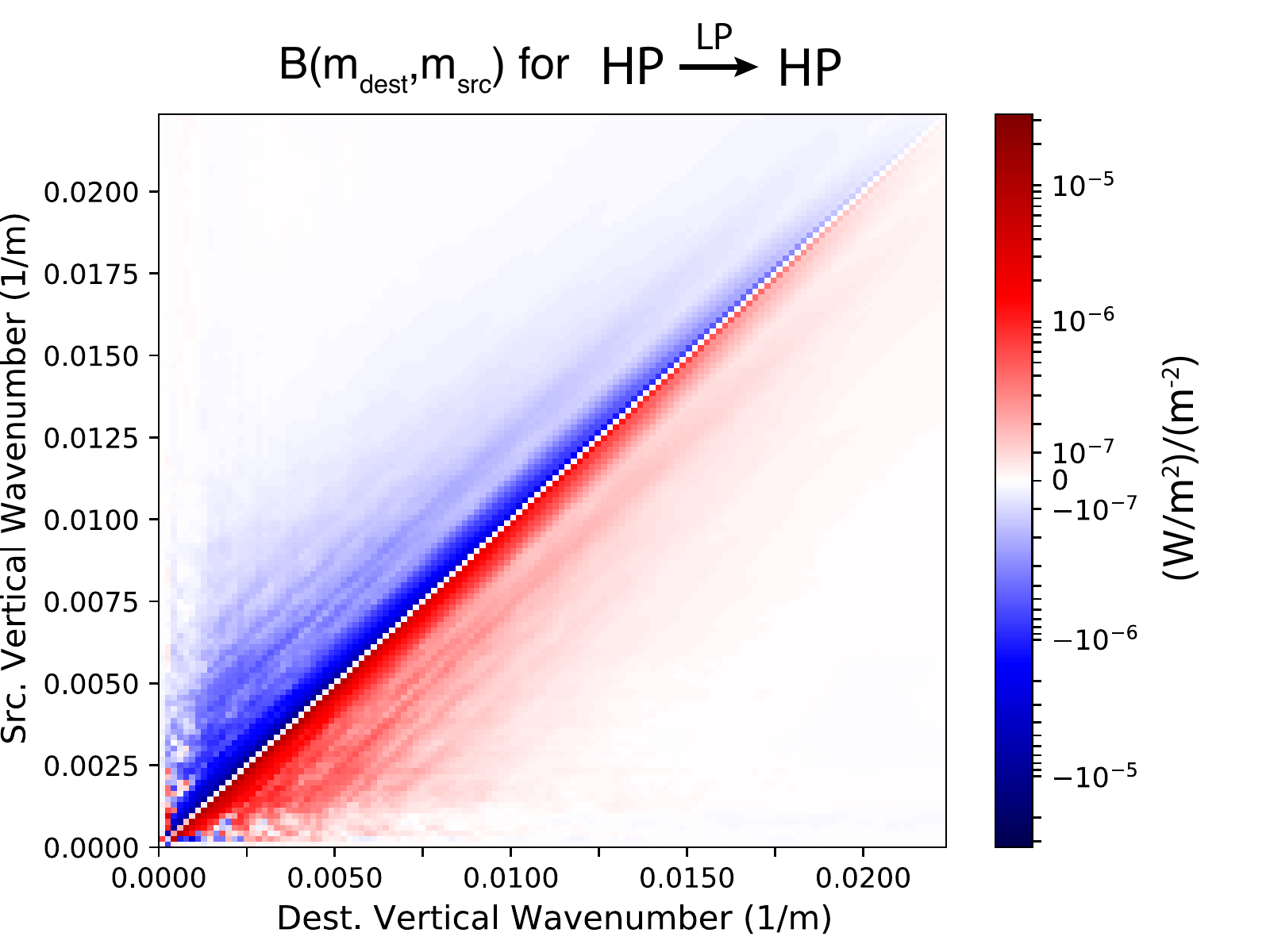}
\centering \caption{Induced Diffusion: HP $\xrightarrow{LP}$ HP}
\label{fig:Bide}
\end{subfigure}

\begin{subfigure}{0.45\textwidth}
\centering
\includegraphics[width=\textwidth]{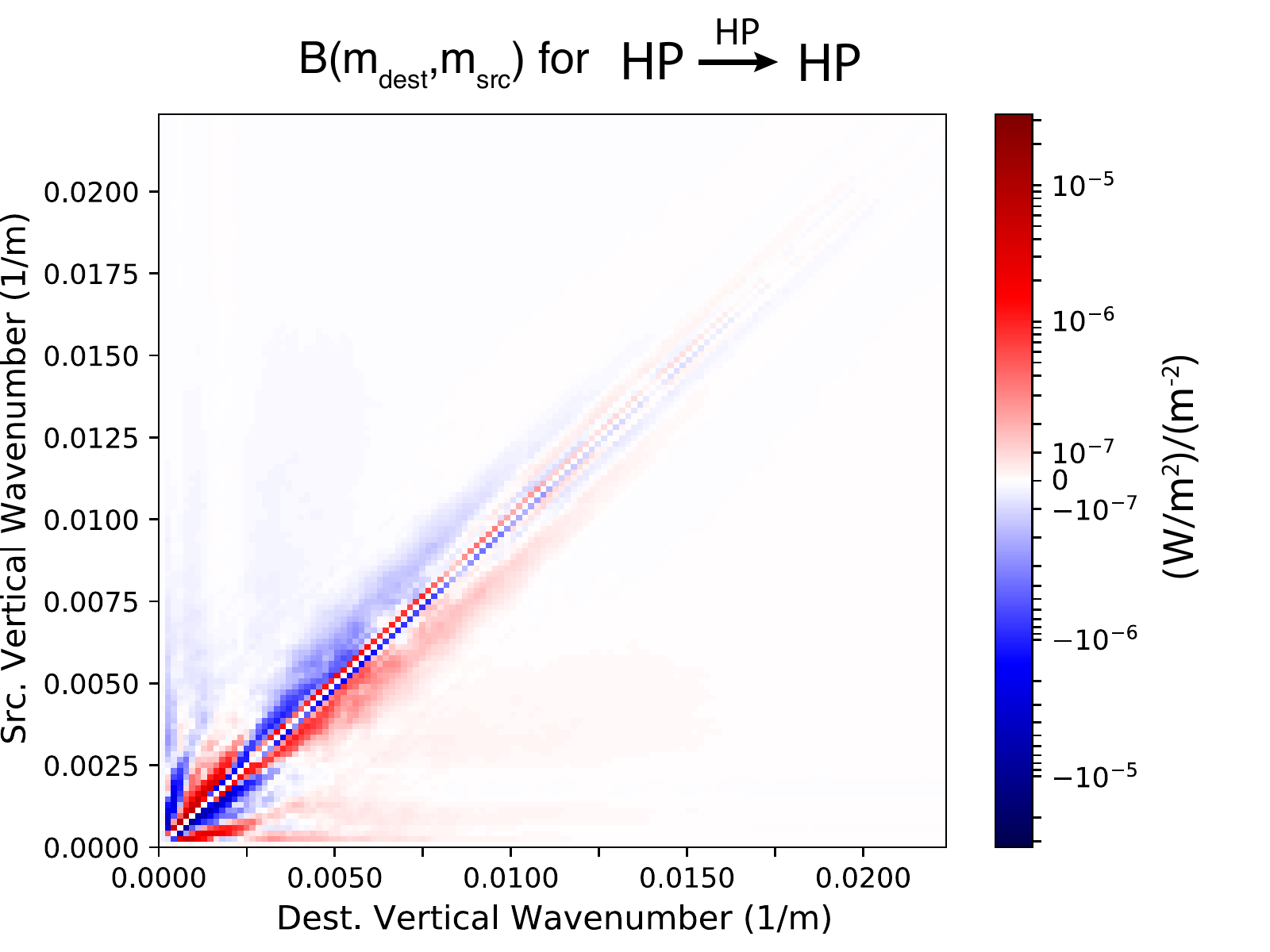}
\centering \caption{Local Interaction: HP $\xrightarrow{HP}$ HP}
\label{fig:Bli}
\end{subfigure}
\begin{subfigure}{0.45\textwidth}
\includegraphics[width=\textwidth]{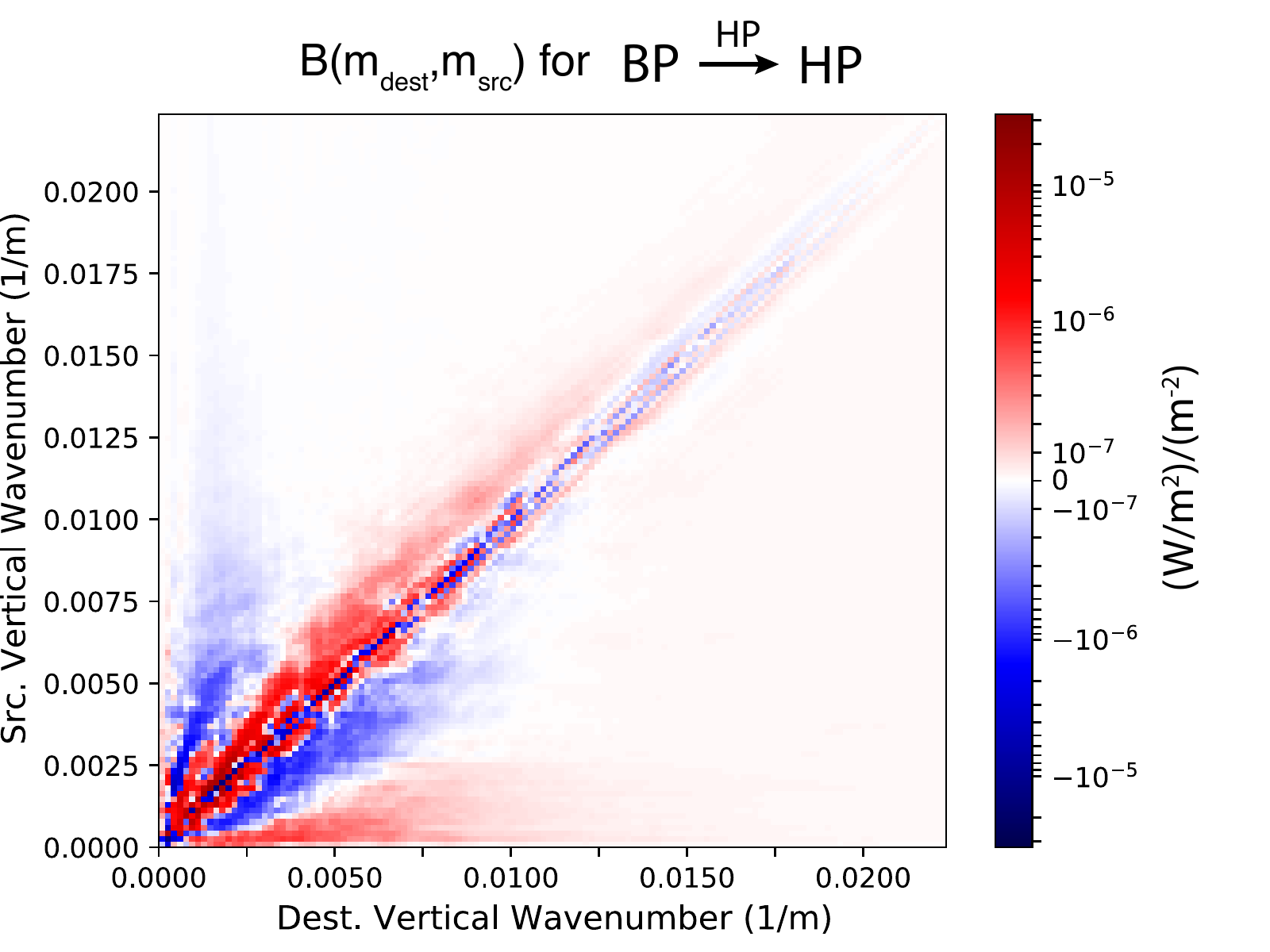}
\centering \caption{BP $\xrightarrow{HP}$ HP.}
\label{fig:Bcomp}
\end{subfigure}
\centering \caption{Bispectra into the supertidal (HP) frequencies for the high-resolution case (case 4 in section \ref{sec:methods}\ref{sec:model}) averaged over the 5 regions of interest in Fig. \ref{fig:regions_of_study}. Bispectra are defined as the kinetic-energy transfer from one vertical wavenumber to another, as defined in equation \ref{eq:src_bispectra}.  The four contributions shown above constitute a complete decomposition.  Fig. \ref{fig:Bcomp} is related to the bispectrum of figure 17 in \cite{sun12} except that we use the KE source vertical wavenumber instead of the catalyst mode and we use both vertical and horizontal components of the gradient.}
\label{fig:bispectra}
\end{figure*}

Bispectra, introduced in equation \ref{eq:src_bispectra} in section \ref{sec:methods}\ref{sec:methods:budgets}\ref{sec:flux}, are displayed in Fig. \ref{fig:bispectra}.  $\mathcal{B}_{\textrm{ID}_{\textrm{diff}}} = \mathcal{B}_{\textrm{HP} \xrightarrow{\textrm{BP}} \textrm{HP}}$ (Fig. \ref{fig:Bid}) exhibits very clear downscale flux, with energy always leaving modes at smaller source vertical wavenumber than their destination wavenumber (indicated by the negative top-left half of the figure and positive in the bottom-right).  Additionally, the strongest interactions occur near the diagonal, at which the source and destination wavenumbers are comparable.  This is consistent with the two being separated by a small-wavenumber catalyst mode, as in the theoretical definition of $\mathcal{M}_{\textrm{ID}}$ \citep{mccomas77a, pan20, dematteis22}. 

$\mathcal{B}_{\textrm{ID}_{\textrm{eddy}}} = \mathcal{B}_{\textrm{HP} \xrightarrow{\textrm{LP}} \textrm{HP}}$ (Fig. \ref{fig:Bid}) also exhibits very clear downscale flux, but in this case the catalyst modes are even smaller than in $\mathcal{B}_{\textrm{ID}_{\textrm{diff}}}$ as indicated by the tighter proximity to the positive and negative bands to the diagonal in Fig. \ref{fig:Bide}.   Also apparent is a thin line of upscale energy transfer hugging the diagonal in $\mathcal{B}_{\textrm{\textrm{ID}}_{\textrm{diff}}}$ in Fig. \ref{fig:Bid}.  When $\mathcal{B}_{\textrm{ID}_{\textrm{diff}}}$ (Fig. \ref{fig:Bid}) and $\mathcal{B}_{\textrm{ID}_{\textrm{eddy}}}$ (Fig. \ref{fig:Bide}) are combined, they form continuous bands of downscale flux, meaning that the upscale flux occurring with small catalyst modes in $\mathcal{B}_{\textrm{ID}_{\textrm{diff}}}$ is working against a larger downscale flux in $\mathcal{B}_{\textrm{ID}_{\textrm{eddy}}}$, effectively erasing it. 

The bispectrum of $\mathcal{B}_{\textrm{LI}} = \mathcal{B}_{\textrm{HP} \xrightarrow{\textrm{HP}} \textrm{HP}}$ is generally weaker than the induced-diffusion bispectra, consistent with the finding that $\mathcal{B}_{\textrm{LI}}$ is smaller than $\mathcal{B}_{\textrm{ID}_{\textrm{diff}}}$ at all resolutions and locations (see Fig. \ref{fig:decomp} and discussions in section \ref{sec:results}\ref{sec:frequency-decomposition}). As with $\mathcal{B}_{\textrm{ID}_{\textrm{diff}}}$, $\mathcal{B}_{\textrm{LI}}$ is strongest near the diagonal axis.  Besides this, it exhibits polarity reversal as the wavenumbers increase along the diagonal axis, with several alternating anti-symmetric bands of forward and backward spectral transfer.  The strong signal along the diagonal reflects high-frequency, high-wavenumber IWs scattered by high-frequency and low-wavenumber IWs.  Thus the strongest interaction is only local in frequency and not vertical wavenumber, implying that much of $\Pi^>_{\textrm{LI}}$ is not really $\mathcal{M}_{\textrm{LI}}$ (see Fig. \ref{fig:mechanisms}).  However, weaker signals off of the diagonal are present and reflect locality in both frequency and wavenumber.  

The compensating energy bispectra, $\mathcal{B}_{\textrm{ID}_{\textrm{comp}}} = \mathcal{B}_{\textrm{BP} \xrightarrow{\textrm{HP}} \textrm{HP}}$, is shown in Fig. \ref{fig:Bcomp}.  For a forward frequency cascade\footnote{the direction of the frequency cascade will be determined in section \ref{sec:results}\ref{sec:m-omega}}, a positive signal is expected at small source wavenumbers and larger destination wavenumbers, as seen at the bottom of the plot.  However, a positive signal is also seen along the diagonal.  The positive signal corresponds to energy coming from high vertical wavenumbers in near-inertial and tidal frequencies (BP) scattering off of low-vertical-wavenumber modes in the supertidal band (HP), a mechanism that doesn't fit cleanly into existing wave interaction categories described in wave-turbulence approaches such as \cite{mccomas77a}.  However, \cite{wagner16} found a process through which near-inertial modes of high vertical wavenumber are energized through $\mathcal{M}_{\textrm{PSI}}$, and perhaps these modes are acting in $\Pi^>_{\textrm{ID}_{\textrm{comp}}}$ triads in the present model.  This mechanism, which we will label $\mathcal{M}_{\textrm{SP}}$, will be discussed in greater detail in the context of the results of \cite{sun12} in section \ref{sec:results}\ref{sec:sp12}. 

\begin{figure*}[]
\centering
\begin{subfigure}{0.45\textwidth}
\includegraphics[width=\textwidth]{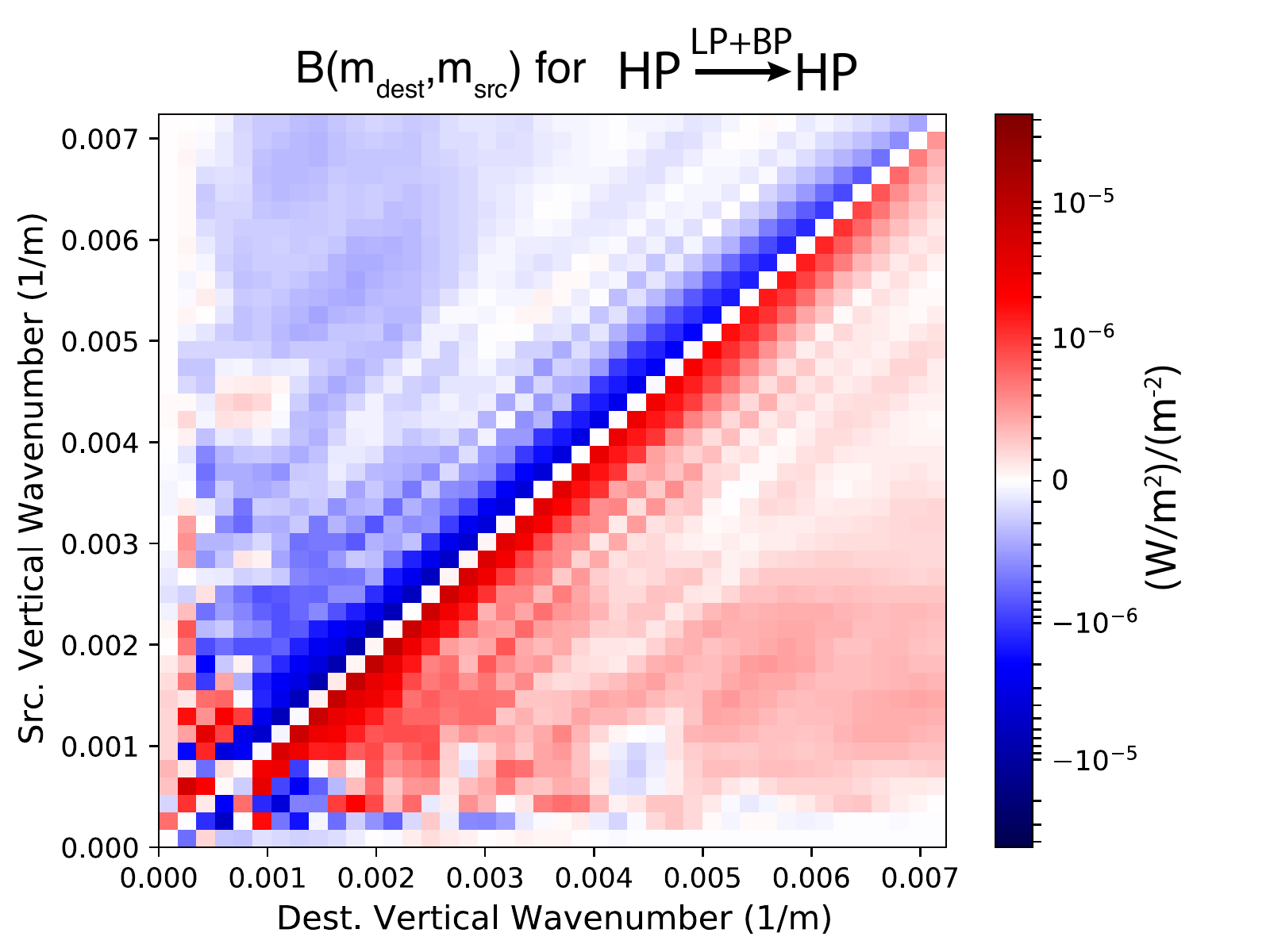}
\centering \caption{low resolution}
\label{fig:bispectra_low}
\end{subfigure}
\begin{subfigure}{0.45\textwidth}
\centering
\includegraphics[width=\textwidth]{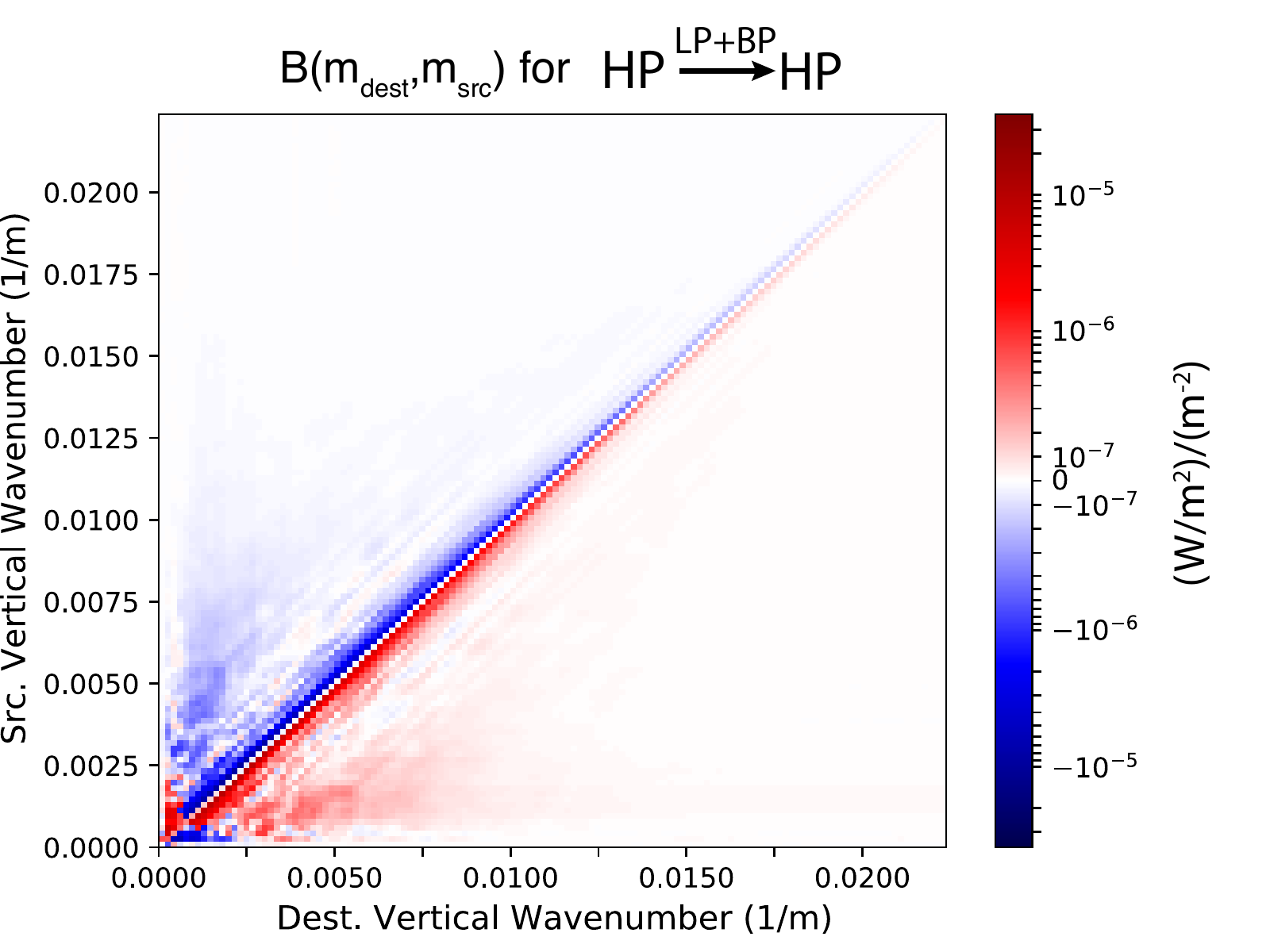}
\centering \caption{3x vertical}
\label{fig:bispectra_vert}
\end{subfigure}

\begin{subfigure}{0.45\textwidth}
\includegraphics[width=\textwidth]{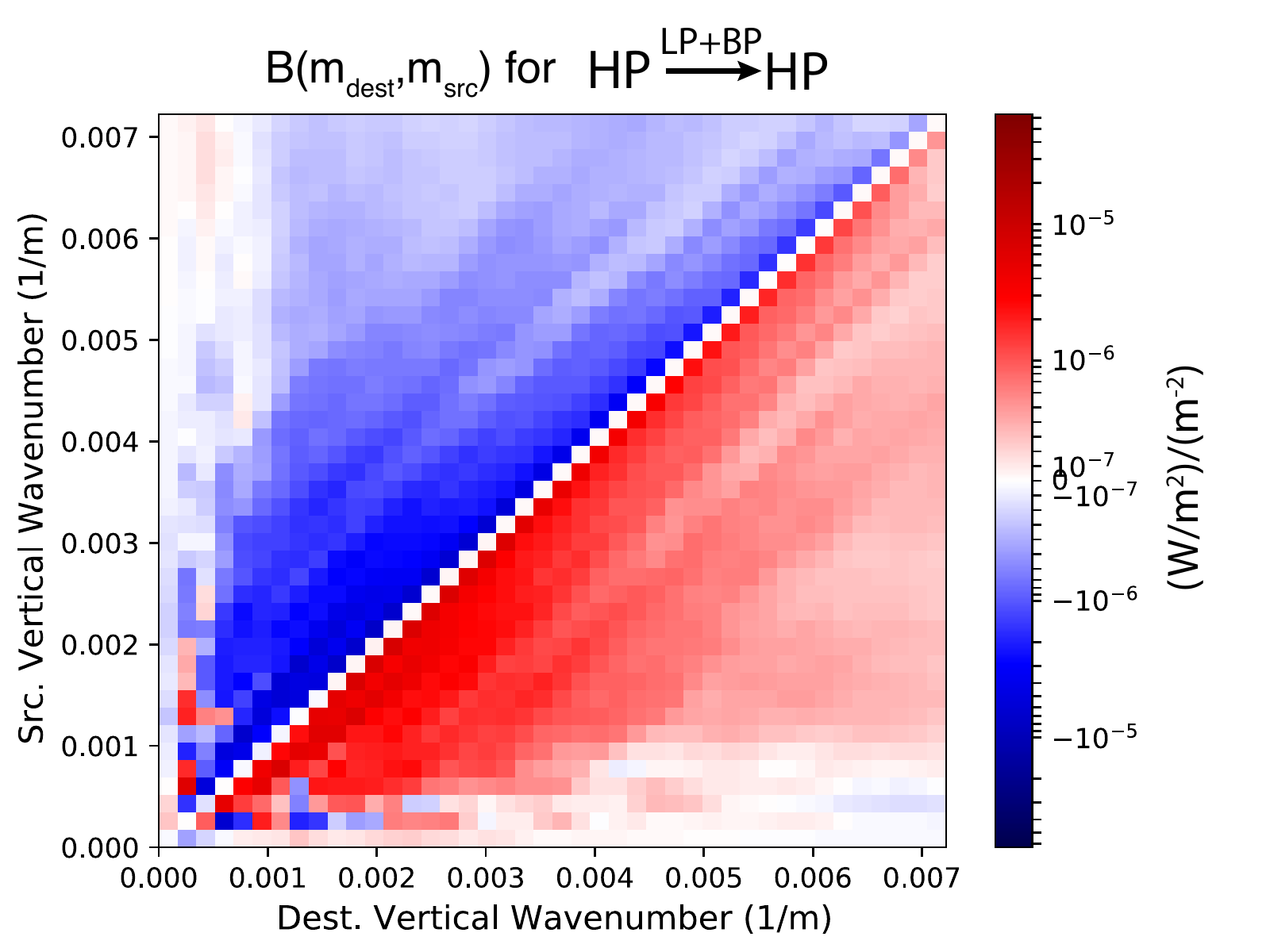}
\centering \caption{8x horizontal}
\label{fig:bispectra_horz}
\end{subfigure}
\begin{subfigure}{0.45\textwidth}
\centering
\includegraphics[width=\textwidth]{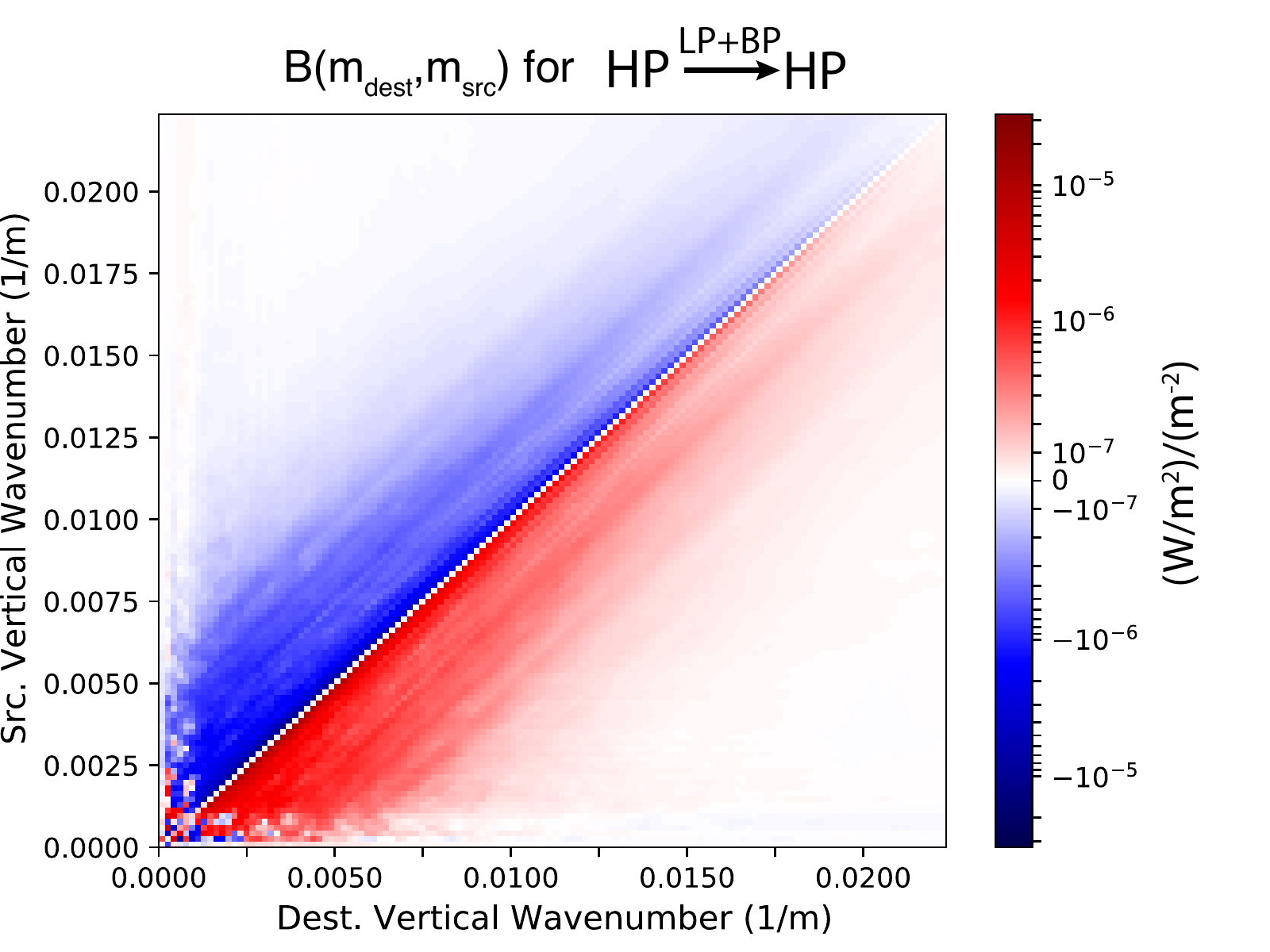}
\centering \caption{3x vertical and 8x horizontal}
\label{fig:bispectra_high}
\end{subfigure}
\centering \caption{Resolution comparison of $\mathcal{B}_{\textrm{ID}_{\textrm{diff}}}$ (equation \ref{eq:src_bispectra}) computed over the subregions of interest, shown in Fig. \ref{fig:regions_of_study}. The axes of the left plots span a range of wavenumbers that is one third the range in the right plots.  Increasing horizontal resolution results in more off-diagonal induced diffusion, thus implying a larger-wavenumber catalyst mode (in the lower-frequency bands).}
\label{fig:bispectra_res}
\end{figure*}

Bispectra of $\mathcal{B}_{\textrm{ID}_{\textrm{diff}}} + \mathcal{B}_{\textrm{ID}_{\textrm{eddy}}}$ at the four resolutions studied are shown in Fig. \ref{fig:bispectra_res}.  These two mechanisms are combined because they exhibit more uniform spectral bands of downscale flux that way.  The most striking thing about these plots is that at low horizontal resolution, the positive and negative bands along the diagonal are very narrow and close together.  This means that the catalyst mode at low horizontal resolution has relatively low vertical wavenumber.  In contrast, the band is much broader at higher horizontal resolutions, which therefore permit scattering off of higher vertical-wavenumber catalyst modes that are predominantly in the near-inertial and tidal (BP) frequencies (not shown). $\mathcal{M}_{\textrm{ID}}$ scattering off of the lowest wavenumber catalyst modes (predominantly in LP, not shown) is not stronger at higher horizontal resolution.  The increase in $\mathcal{M}_{\textrm{ID}}$ is strictly due to interaction with higher-wavenumber BP catalyst modes.

\subsection{m-$\omega$ spectra}
\label{sec:m-omega}

\begin{figure*}[]
\centering
\begin{subfigure}{0.45\textwidth}
\includegraphics[width=\textwidth]{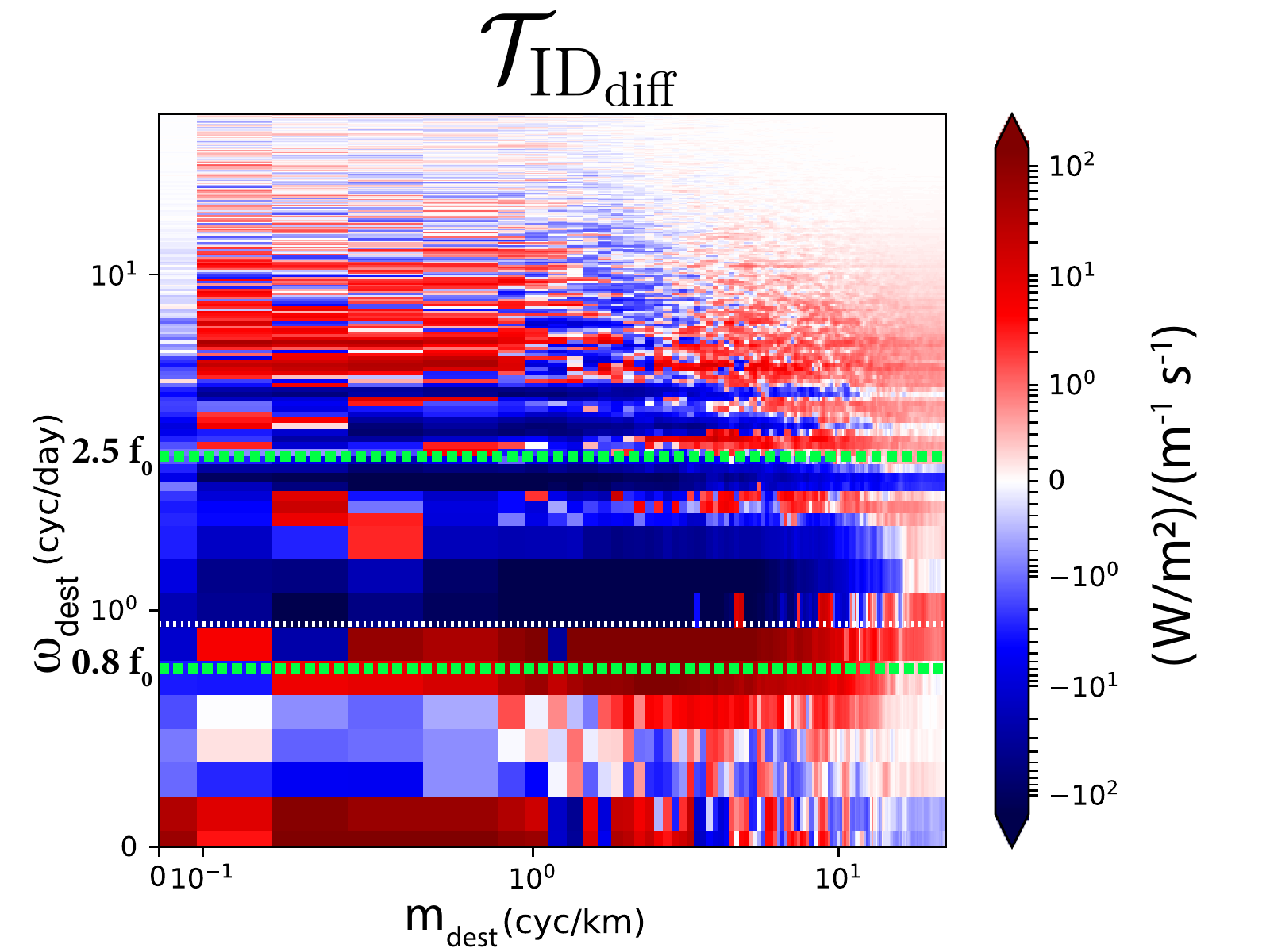}
\centering \caption{$\mathcal{T}_{\textrm{ID}_{\textrm{diff}}}$: HP $\xrightarrow{\textrm{BP}}$ all}
\label{fig:m_omega_ID}
\end{subfigure}
\begin{subfigure}{0.45\textwidth}
\includegraphics[width=\textwidth]{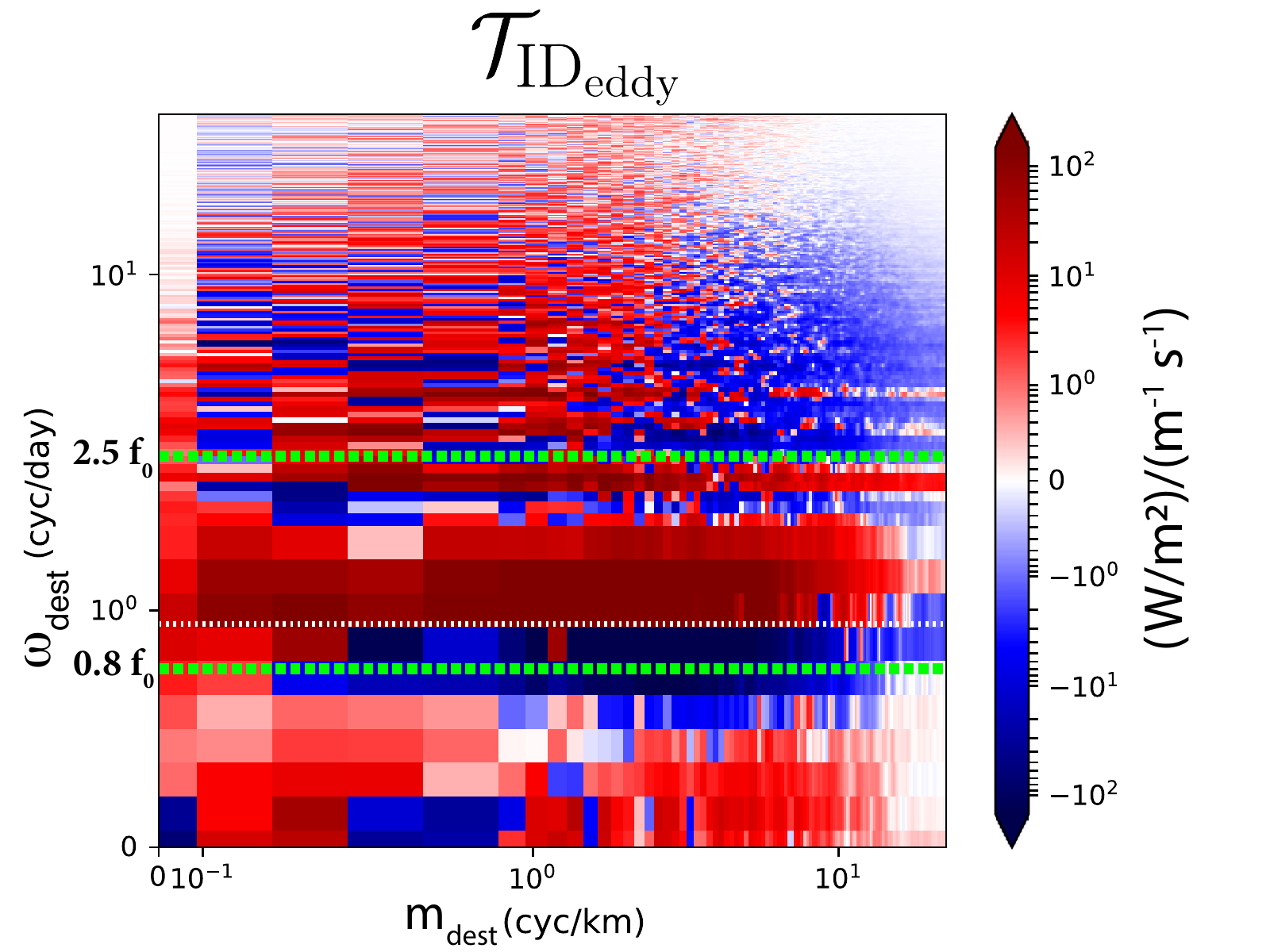}
\centering \caption{$\mathcal{T}_{\textrm{ID}_{\textrm{eddy}}}$: HP $\xrightarrow{\textrm{LP}}$ all}
\label{fig:m_omega_ID_lp}
\end{subfigure}

\begin{subfigure}{0.45\textwidth}
\includegraphics[width=\textwidth]{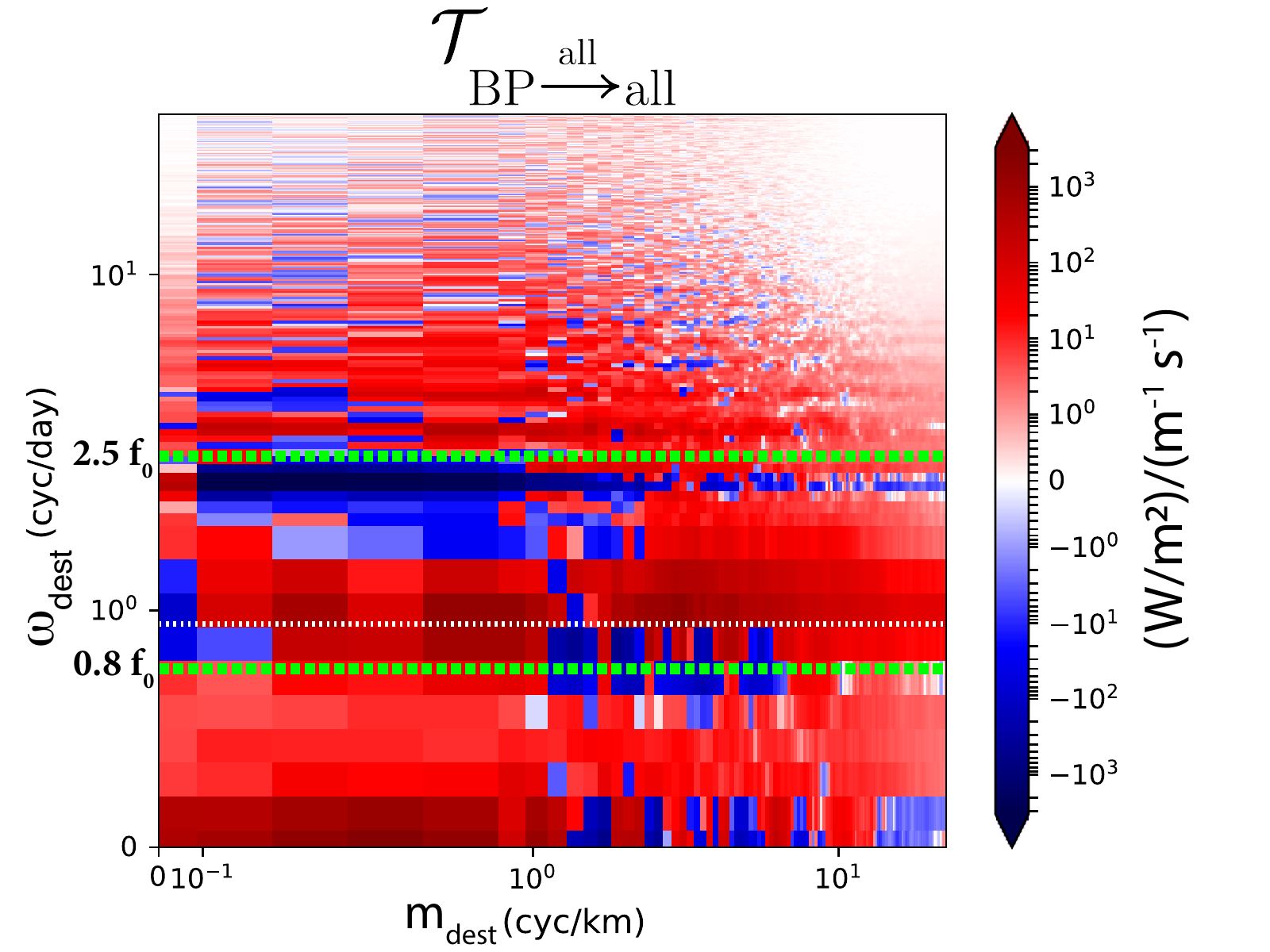}
\centering \caption{$\mathcal{T}_{\textrm{BP} \xrightarrow{\textrm{all}} \textrm{all}}$}
\label{fig:m_omega_BP_to_HP}
\end{subfigure}
\begin{subfigure}{0.45\textwidth}
\centering
\includegraphics[width=\textwidth]{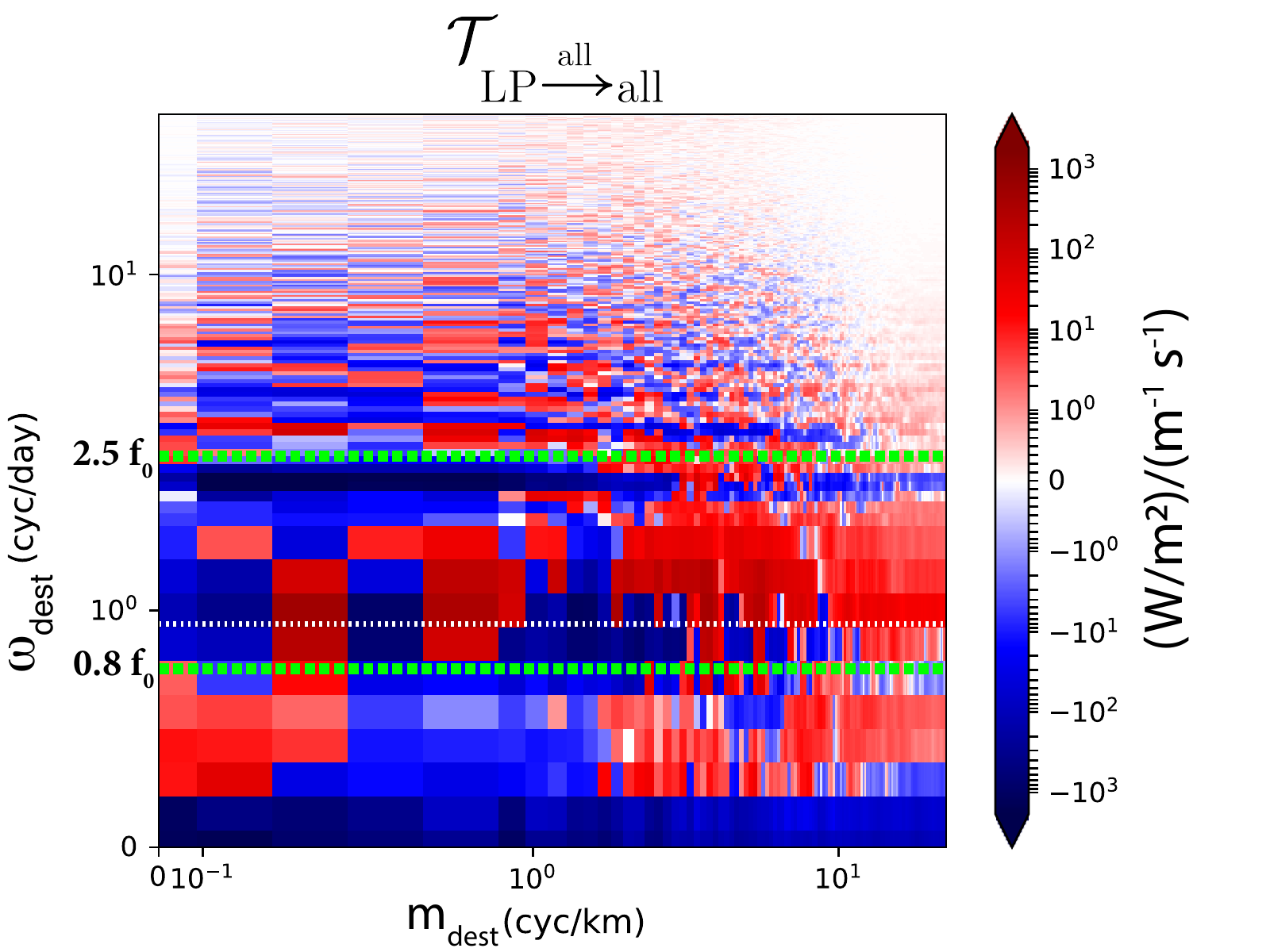}
\centering \caption{$\mathcal{T}_{\textrm{LP} \xrightarrow{\textrm{all}} \textrm{all}}$}\label{fig:m_omega_LP_to_HP}
\end{subfigure}

\begin{subfigure}{0.3\textwidth}
\centering
\includegraphics[width=\textwidth]{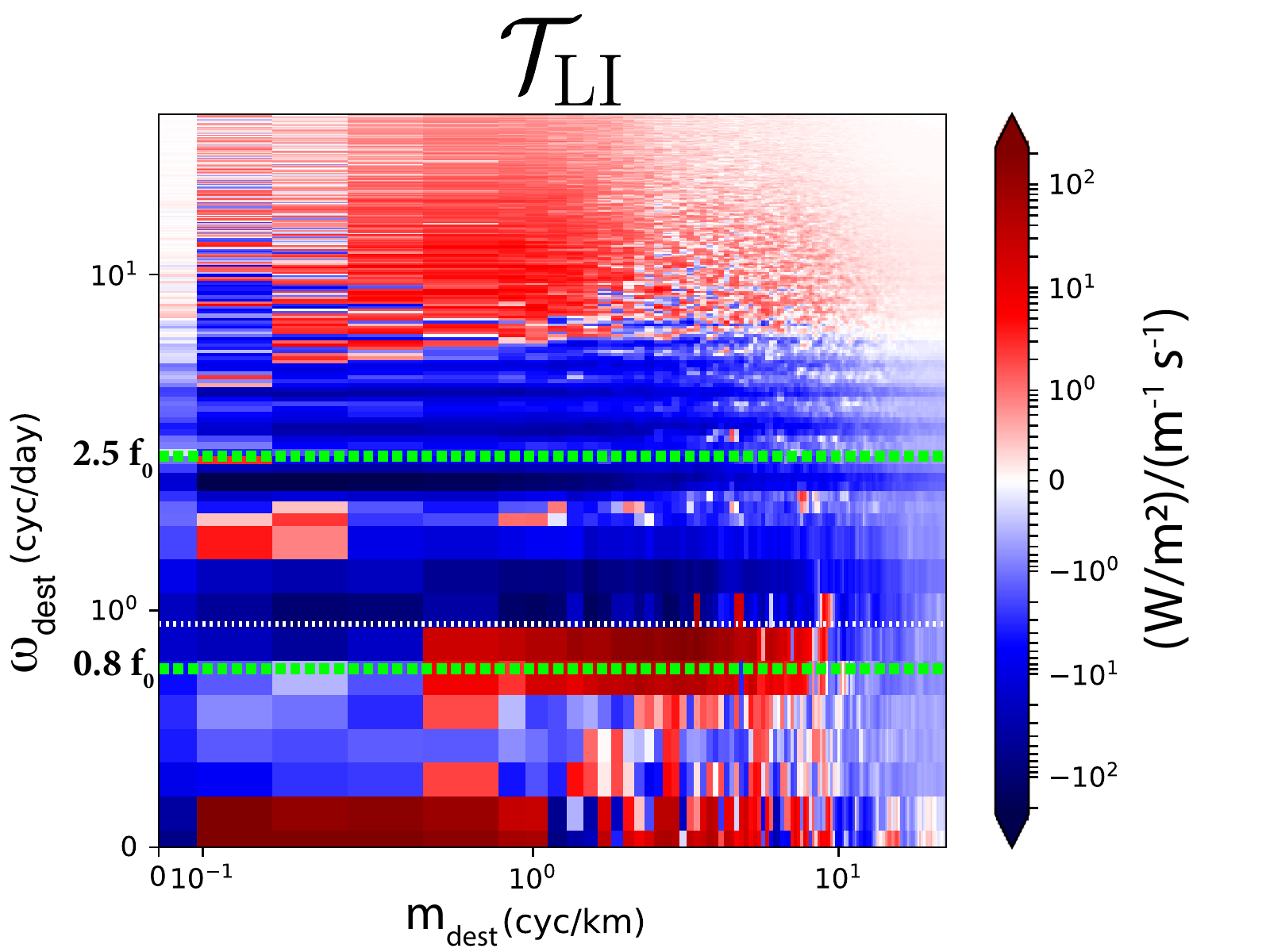}
\centering \caption{$\mathcal{T}_{\textrm{LI}}$: HP $\xrightarrow{\textrm{HP}}$ all}
\label{fig:m_omega_LI}
\end{subfigure}
\begin{subfigure}{0.3\textwidth}
\includegraphics[width=\textwidth]{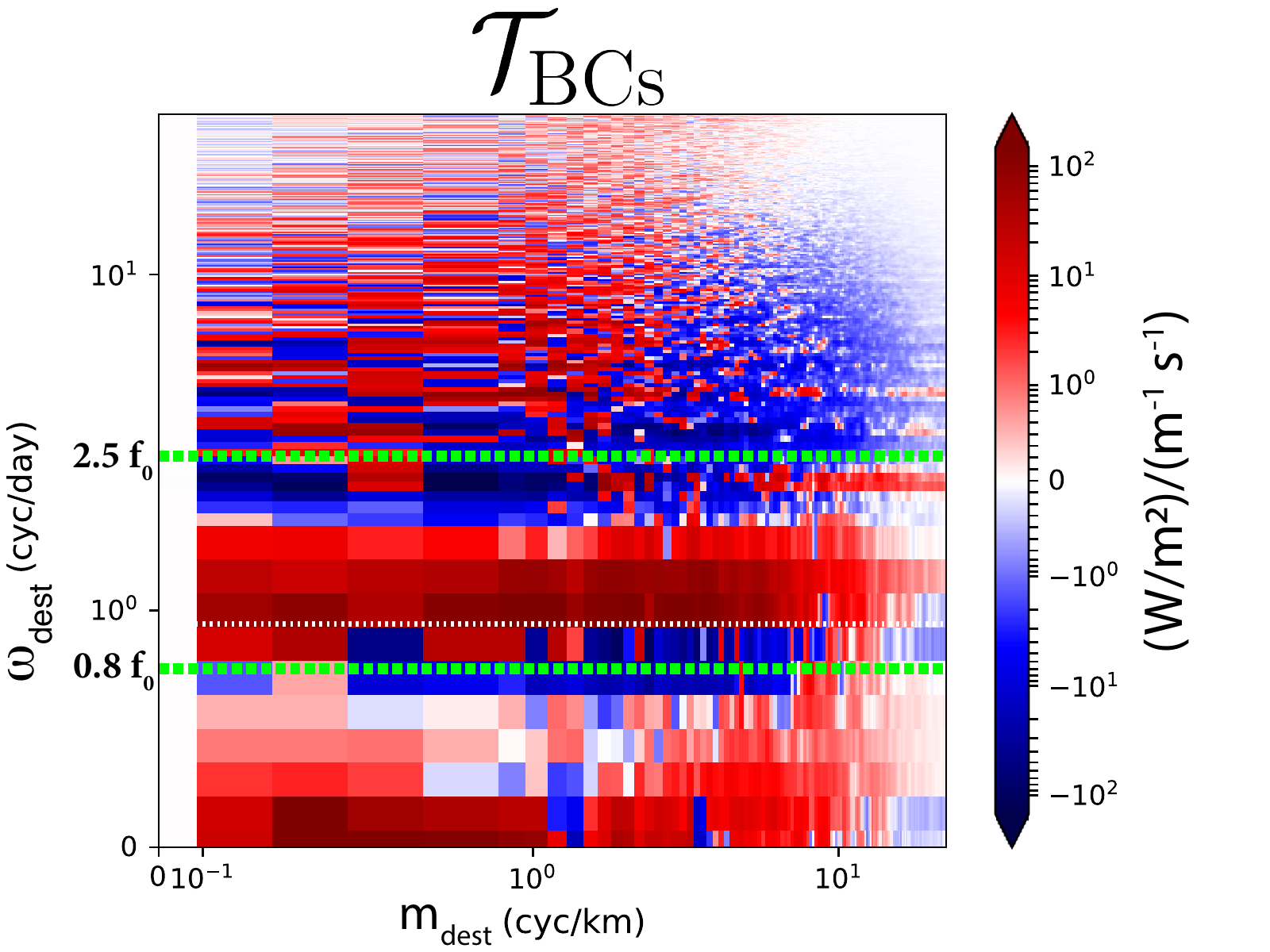}
\centering \caption{$\mathcal{T}_{\textrm{BCs}}$}
\label{fig:m_omega_BCs}
\end{subfigure}
\begin{subfigure}{0.3\textwidth}
\includegraphics[width=\textwidth]{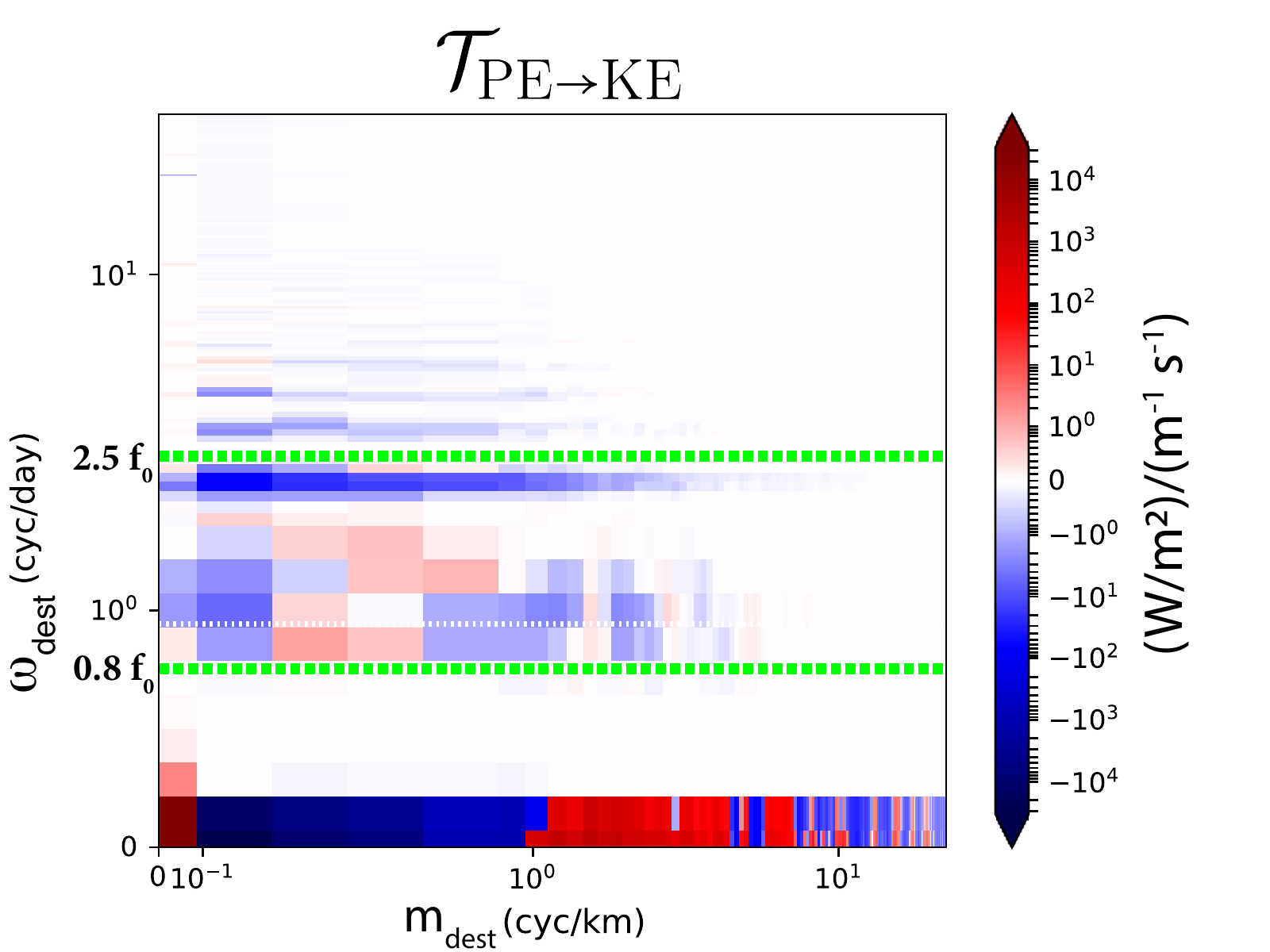}
\centering \caption{$\mathcal{T}_{\textrm{PE}\rightarrow\textrm{KE}}$}
\label{fig:m_omega_PE}
\end{subfigure}
\centering \caption{$m_{\textrm{dest}}$-$\omega_{\textrm{dest}}$ spectra of various local spectral budget terms (equation \ref{eq:T_decomp}) averaged across the subregions of interest, shown in Fig. \ref{fig:regions_of_study}.  Labels for spectral transfers of specific mechanisms (e.g. $\mathcal{T}_{\textrm{LI}}$) refer to the high-pass part of the destination frequency, but the plots actually include all destination frequencies.  The first 5 figures constitute a complete decomposition of nonlinear advective scattering within any given frequency band.  Note that the advective term is a spectral transfer as opposed to a spectral flux. Computing the flux would involve integrating the transfer in the wavenumber domain.  The direction of frequency flux associated with a given transfer mechanism can be approximately discerned within a subspace that conserves energy under the evolution of that mechanism (for example, in the HP band for $\mathcal{M}_{\textrm{ID}_{\textrm{diff}}}$ in panel a); if a sink (blue) is identified in the energy-conserving spectral band, energy must be transferred from those modes to the source (red) modes in that band.  In this way, for example, $\mathcal{M}_{\textrm{ID}_{\textrm{diff}}}$ in panel a can be seen to move KE from low to high $\omega$.}
\label{fig:m_omega}
\end{figure*}

Two-dimensional wavenumber-frequency local spectral budgets, introduced in section \ref{sec:methods}\ref{sec:methods:budgets}\ref{sec:methods:m-omega} are shown in Fig. \ref{fig:m_omega}.  The primary motivation for computing these is to discern the direction of the frequency cascade associated with different nonlinear wave interactions.  The advective spectra are decomposed in figures \ref{fig:m_omega_ID}-\ref{fig:m_omega_LI}.  The energy transfer decomposition of the supertidal (HP) frequencies exists in the top portion of each figure, with $\omega > 2.5 f_0$.  Note that these spectra constitute a local (or transfer), as opposed to an integrated (or flux), budget.   Also note that the plots for $\mathcal{T}_{\textrm{ID}_{\textrm{diff}}}$, $\mathcal{T}_{\textrm{ID}_{\textrm{eddy}}}$ and $\mathcal{T}_{\textrm{LI}}$ in figures \ref{fig:m_omega_ID}, \ref{fig:m_omega_ID_lp}, and \ref{fig:m_omega_LI}, respectively, must conserve energy within the supertidal band, such that the positive and negative (blue and red) transfers above the $\omega = 2.5 f_0$ line must be balanced.  

For $\mathcal{T}_{\textrm{ID}_{\textrm{diff}}}$, transfers in the HP band increase along $m$ and are approximately homogeneous in $\omega$ with a small sink at the lowest frequencies; this can be interpreted as moving energy predominantly from low to high vertical wavenumber, and slightly forward in frequency, all of which is consistent with the argument of \cite{dematteis22} that the horizontal wavenumber ``keeps pace'' with the vertical wavenumber under induced diffusion.  The computed spectral transfers are inconsistent with the theory of \cite{mccomas81a} in which ID is associated with an inverse-frequency cascade for supertidal IWs. It also means that compensating energy, $\Pi_{\textrm{ID}_{\textrm{comp}}}$, is expected to be weakly positive\footnote{To reprise some relevant points previously made in section \ref{sec:intro:plan}, wave action is expected to be conserved in the supertidal band under $\mathcal{M}_{\textrm{ID}}$.  The definition of wave action, $\mathcal{A} = \frac{E}{\omega}$, implies the direction of compensating energy that must be supplied from the catalyst band in order to conserve wave action is the same as that of the frequency cascade associated with $\mathcal{M}_{\textrm{ID}}$, a point that is made clear in the discussion of figure 5 in \cite{mccomas81a}).  \cite{dematteis22} study spectra that they expect share properties with the real ocean that has both a weak forward frequency cascade (top row in figure 5 of that paper) and neutral frequency cascades (middle row in figure 5 of that paper) under $\mathcal{M}_{\textrm{ID}}$, indicating compensating energy from the catalyst band to the supertidal band that is absent or weak.} in the present model results in order to conserve wave action among supertidal IWs. 

$\mathcal{T}_{\textrm{LI}}$ (Fig. \ref{fig:m_omega_LI}), on the other hand, exhibits a clear forward frequency cascade at mid-to-high vertical wavenumber (indicated by KE leaving the lower frequencies in the HP band and entering the higher frequencies.)  The forward direction of the $\mathcal{T}_{\textrm{LI}}$ frequency cascade is inconsistent with the predictions of \cite{dematteis22} in which $\mathcal{M}_{\textrm{LI}}$ is expected to transfer energy more in the horizontal spectral direction relative to $\mathcal{M}_{\textrm{ID}}$.  $\mathcal{M}_{\textrm{LI}}$ should therefore have a less pronounced forward frequency transfer (see figure 6 in that paper).  This inconsistency, along with the fact that $\mathcal{B}_{\textrm{LI}}$ energy transfers are mostly nonlocal in wavenumber, indicates that $\mathcal{T}_{\textrm{LI}}$, $\Pi_{\textrm{LI}}$, and $\mathcal{B}_{\textrm{LI}}$ are mostly capturing mechanisms that are distinct from those identified in \cite{dematteis22}.  

Energy transfer from BP (near-inertial and tidal frequencies) into HP (the supertidal band), depicted in Fig. \ref{fig:m_omega_BP_to_HP} is positive throughout the HP band.  Energy transfer from LP (eddy frequencies) into HP, depicted in Fig. \ref{fig:m_omega_LP_to_HP}, shows some KE transfer from HP to LP at the lower vertical wavenumbers and only forward-frequency (LP-to-HP) transfer at the highest vertical wavenumbers.  Note that in these two plots, the transfer need not sum to zero in the supertidal (HP) band. 

KE advected into the regions of study is also decomposed into two-dimensional wavenumber-frequency spectra in Fig. \ref{fig:m_omega_BCs}.  This implies that, at least for the five subregions of interest, there is advection of supertidal KE into these regions at low-to-mid vertical wavenumber and advection out of these regions at high vertical wavenumbers.  

Some theories of the IW continuum (e.g. \cite{mccomas77b}) suggest that wind forcing injects energy at high-$\omega$, low-$m$, at which point KE is moved through nonlinear interactions among waves to smaller $\omega$.  Signatures of this are not visible as an inverse frequency cascade at low vertical wavenumber in the advective spectra of transfer within the HP band, figures \ref{fig:m_omega_ID}-\ref{fig:m_omega_LI}, where KE might be leaving the highest frequencies.  They are also not visible in the top-left of the PE-to-KE conversion spectral, Fig. \ref{fig:m_omega_PE}, where some energy might be injected from SSH perturbations.  These characteristics of the $m-\omega$ spectral transfers may be because wind-forcing is limited to being updated only every 6 hours.  Higher-frequency forcing, or perhaps coupling to an atmospheric model, are likely necessary to force the flow in a manner consistent with the picture developed by \cite{mccomas77b} and will be explored in a future study.  Such high frequency forcing could also impact the overall direction of the $\mathcal{M}_{\textrm{ID}}$ or $\mathcal{M}_{\textrm{LI}}$ cascades that are observed. On the other hand, there is a $\mathcal{M}_{\textrm{PSI}}$ signal at tidal frequencies.  The plot of KE transfer from BP (Fig. \ref{fig:m_omega_BP_to_HP}) conserves energy within the BP band and the strong negative signal at low vertical wavenumber and the semidiurnal frequency is consistent with $\mathcal{M}_{\textrm{PSI}}$.  This likely $\mathcal{M}_{\textrm{PSI}}$ signal is about an order of magnitude lower in box E, which is north of the critical latitude for $\mathcal{M}_{\textrm{PSI}}$, than in boxes B, C, and D, which are south of the critical latitude (not shown).  The signal in Box A, which straddles the critical latitude, is also much larger than in box E.

\subsection{Interpretation of and comparison with observations of \cite{sun12}}
\label{sec:sp12}

\begin{figure}[]
\centering
\begin{subfigure}{0.35\textwidth}
\includegraphics[width=\textwidth]{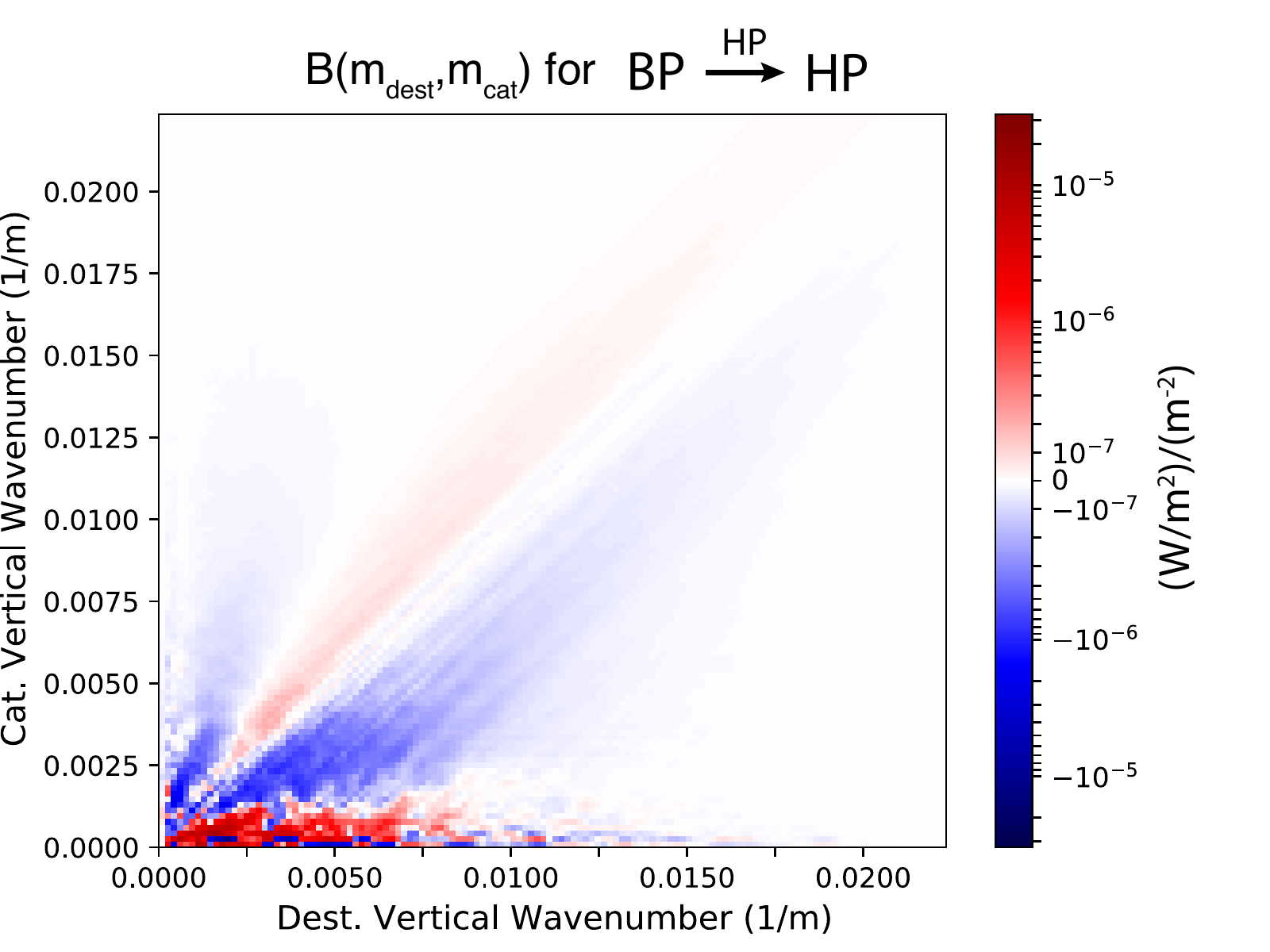}
\centering \caption{The bispectrum from equation \ref{eq:bispectra_cat} similar to that computed in \cite{sun12} figure 17.  One key difference from \cite{sun12} is that the contributions at the positive and negative vertical catalyst wavenumber are added together, making the bispectrum much easier to compute using 3D model output.  This is computed as an average over the subregions of interest, displayed in Fig. \ref{fig:regions_of_study}.}
\label{fig:sp12}
\end{subfigure}

\begin{subfigure}{0.35\textwidth}
\centering
\includegraphics[width=\textwidth]{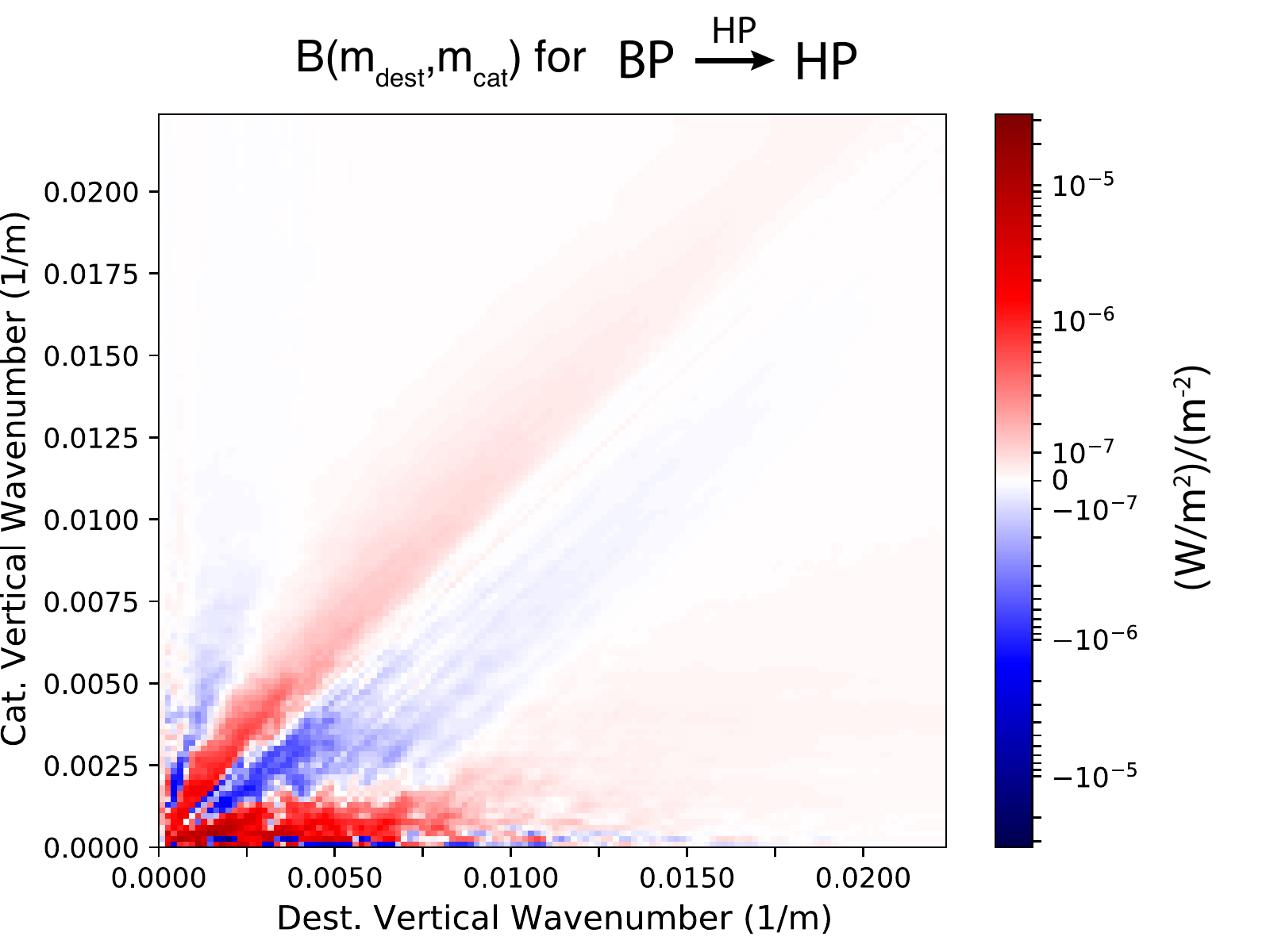}
\centering \caption{As with Fig. \ref{fig:sp12} except with the horizontal gradient included.}
\label{fig:sp12_both}
\end{subfigure}

\begin{subfigure}{0.35\textwidth}
\centering
\includegraphics[width=\textwidth]{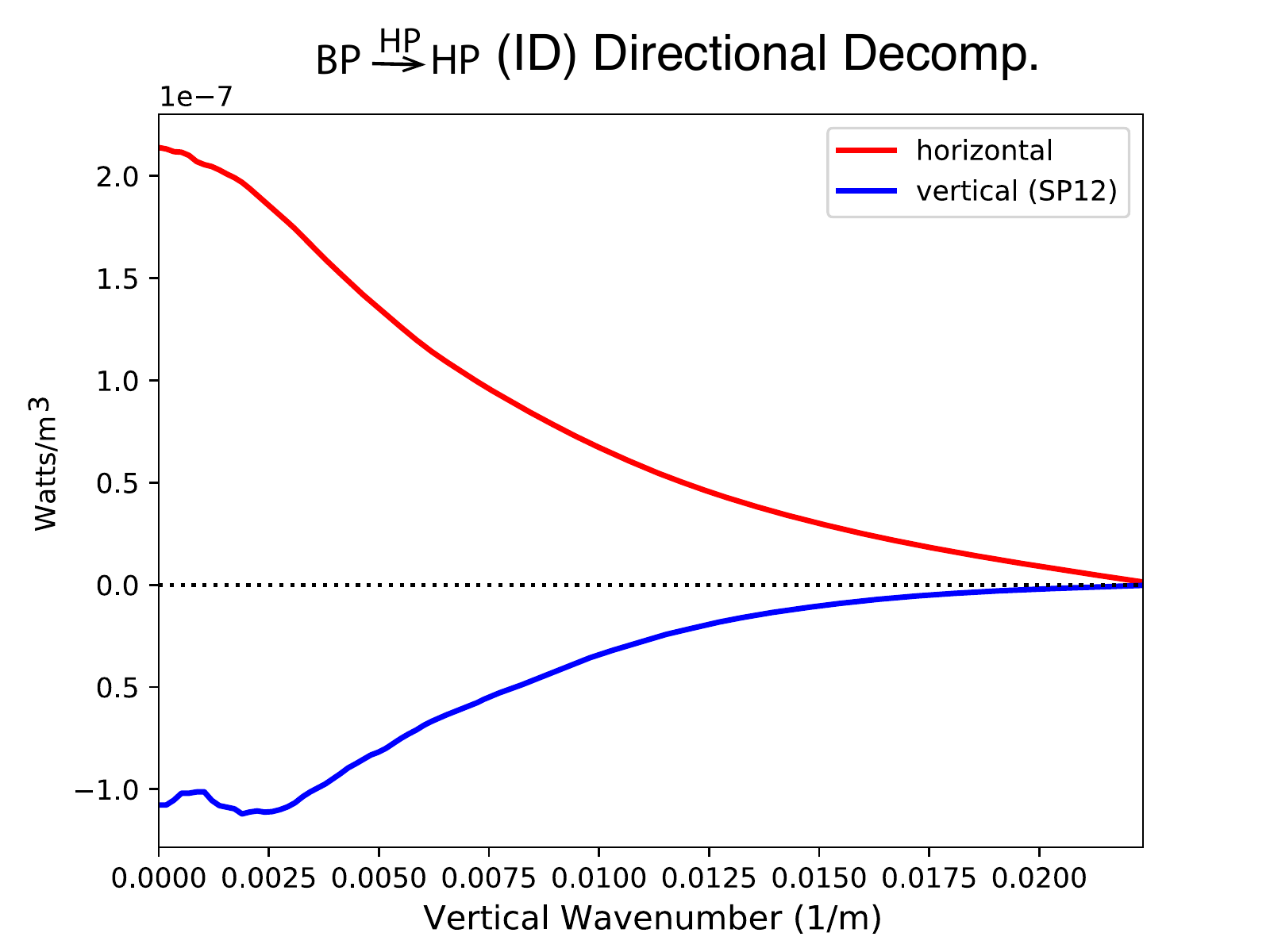}
\centering \caption{Spectral fluxes associated with the bispectra of figures \ref{fig:sp12} and \ref{fig:sp12_both}.}
\label{fig:vert_horz}
\end{subfigure}
\centering \caption{The bispectra of the catalyst mode and destination wavenumber, as presented in figure 17 of \cite{sun12} (see equation \ref{eq:bispectra_cat} in this paper). Subplot c displays an integrated flux of these two contributions, thus highlighting their differences. }
\label{fig:sp12plots}
\end{figure}

\cite{sun12} use observational data to compute bispectra in an attempt to identify $\mathcal{M}_{\textrm{ID}}$.  Specifically, they compute bispectra using the catalyst mode and using only the vertical component of the gradient in the advective scattering (see equation \ref{eq:spectral_bispectra}).  Related bispectra (equation \ref{eq:bispectra_cat}) are reproduced from the present model data (averaged over the 5 regions of interest in the domain, depicted here in Fig. \ref{fig:regions_of_study}) in Fig. \ref{fig:sp12plots}.  To reiterate important points made in section \ref{sec:methods}\ref{sec:methods:budgets}\ref{sec:bispectra}, an advantage of the method used in this paper is that unlike in \cite{sun12}, the bispectra can include the vertical and horizontal gradients (as in Fig. \ref{fig:sp12_both}).  At the same time, the method used in this paper requires averaging over positive and negative values of the source and destination modes, implying that it cannot identify vertical anisotropies in scattering mechanisms such as through $\mathcal{M}_{\textrm{ES}}$.  Additionally, \cite{sun12} separate their supertidal (HP) band from their low-pass background field (LP) by removing intermediate frequencies whereas the bispectra in figures \ref{fig:sp12plots} do not. 

It is worth discussing whether it is appropriate to use only the vertical-gradient component of a bispectrum such as in equation \ref{eq:spectral_bispectra}.  \cite{sun12} point out that such an approximation is only valid in horizontally homogeneous flows and noted that would not be applicable in their region of study.  We further point out that the horizontal-gradient part of these bispectra and advective tendencies in general should only be zero in horizontally homogeneous flows if the source and destination fields are the same (as it is in \cite{gargett84}, which \cite{sun12} cite to support their assumption).  To put this another way, if $\mathbf{f}$, $\mathbf{g}$, and $\mathbf{h}$ are distinct general incompressible 3-dimensional velocity fields, then under horizontally homogeneous statistics, $\overline{\mathbf{f} \cdot \left(\left(\mathbf{g} \cdot \boldsymbol{\nabla}_h\right) \mathbf{f}\right)} = 0$ while $\overline{\mathbf{f} \cdot \left(\left(\mathbf{g} \cdot \boldsymbol{\nabla}_h\right) \mathbf{h}\right)} \ne 0$, where the overline indicates a time average and $\boldsymbol{\nabla}_h$ is the horizontal gradient operator.  In the compensating-energy bispectra used by \cite{sun12}, the source field is of near-inertial frequencies while the destination field is of supertidal frequencies.  Thus, even in horizontally homogeneous flows, there will still be a nonzero value of these terms that represent the transfer of energy from low to high frequencies.  

\begin{figure*}[]
\centering
\begin{subfigure}{0.45\textwidth}
\includegraphics[width=\textwidth]{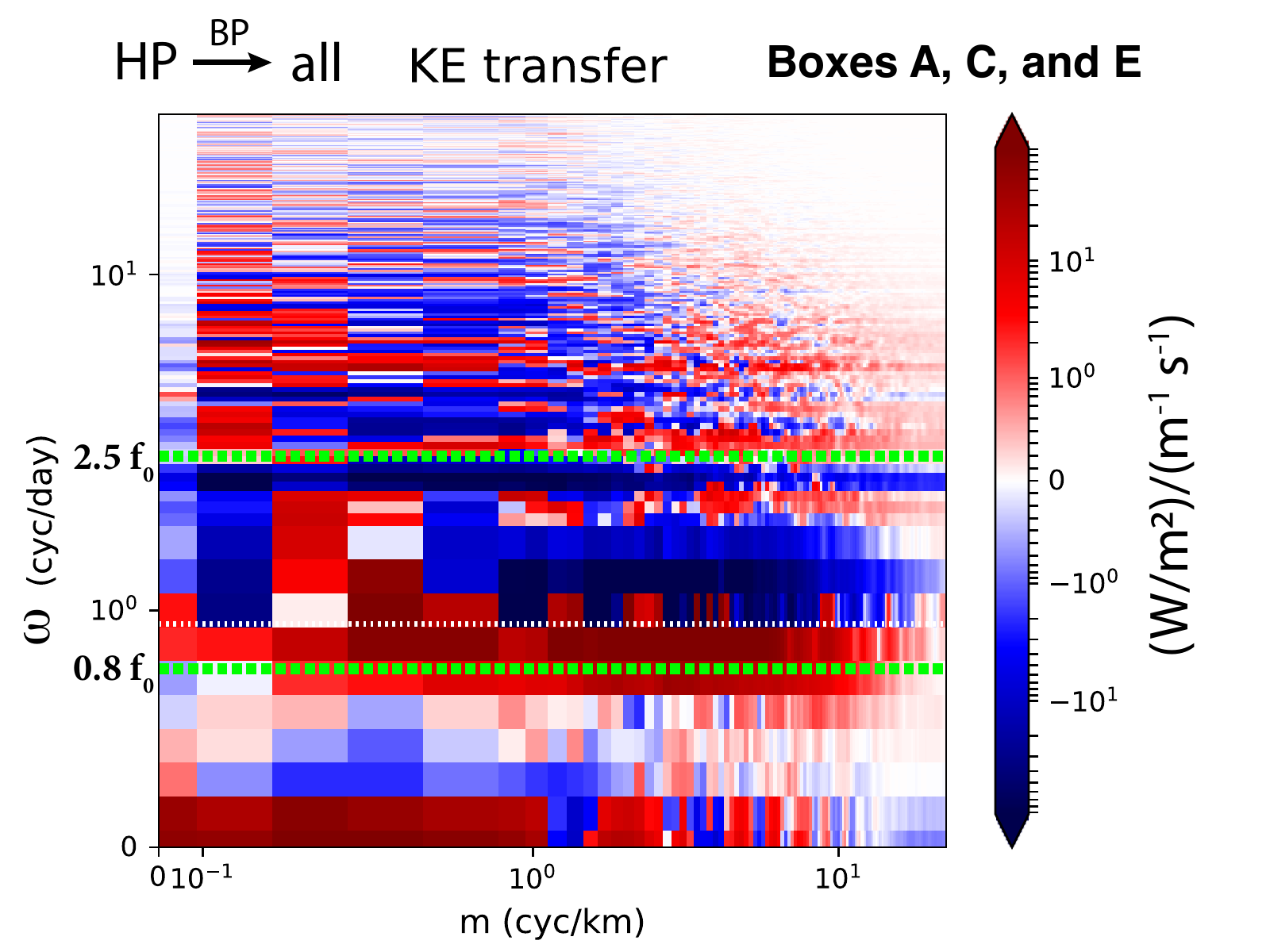}
\centering \caption{$\mathcal{T}_{HP\xrightarrow{BP}HP} \left(m,\omega\right)$ in Boxes A, C, and E.}
\label{fig:10c}
\end{subfigure}
\begin{subfigure}{0.45\textwidth}
\centering
\includegraphics[width=\textwidth]{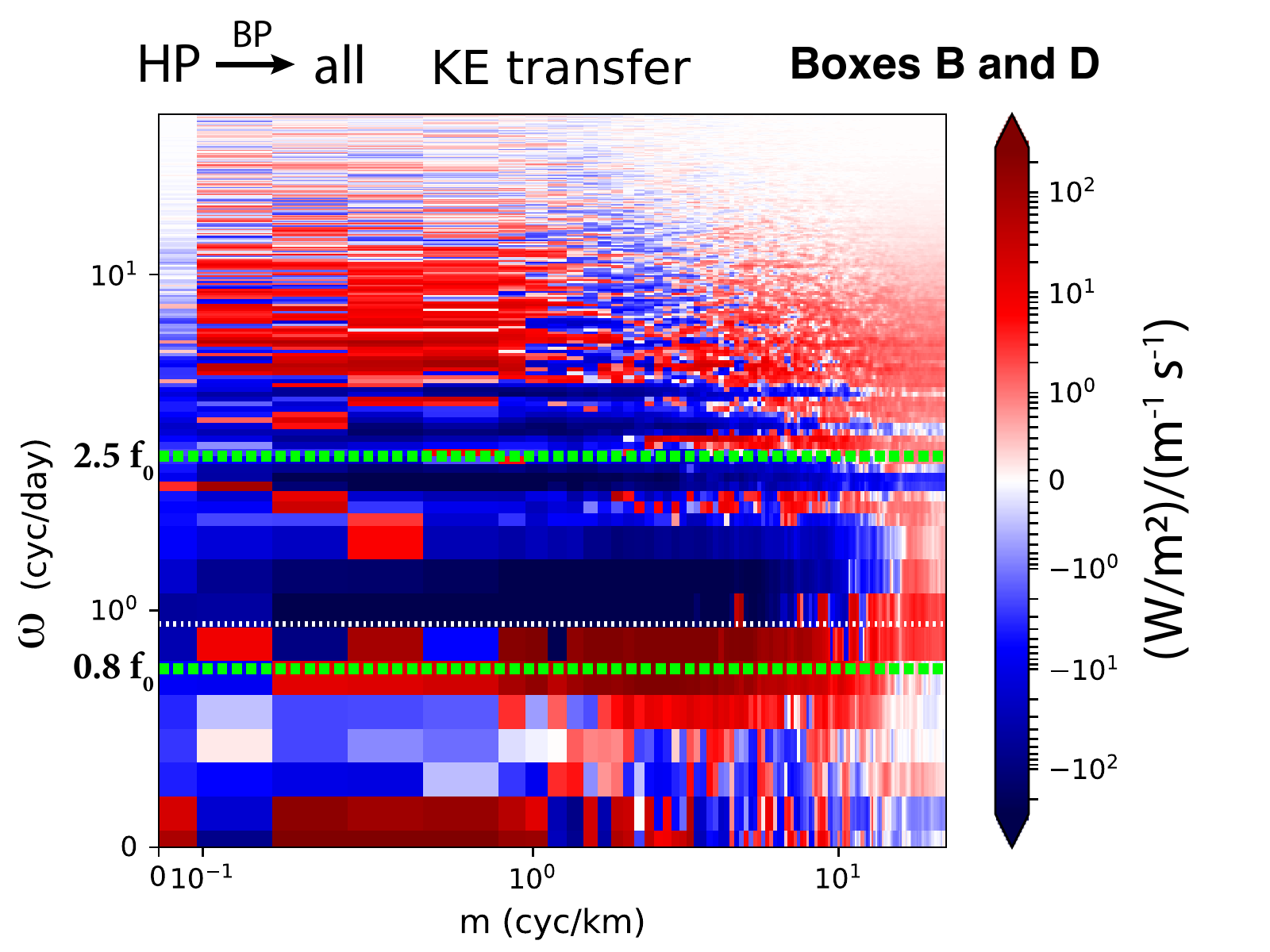}
\centering \caption{$\mathcal{T}_{HP\xrightarrow{BP}HP} \left(m,\omega\right)$ in Boxes B and D. }
\label{fig:10d}
\end{subfigure}

\begin{subfigure}{0.45\textwidth}
\includegraphics[width=\textwidth]{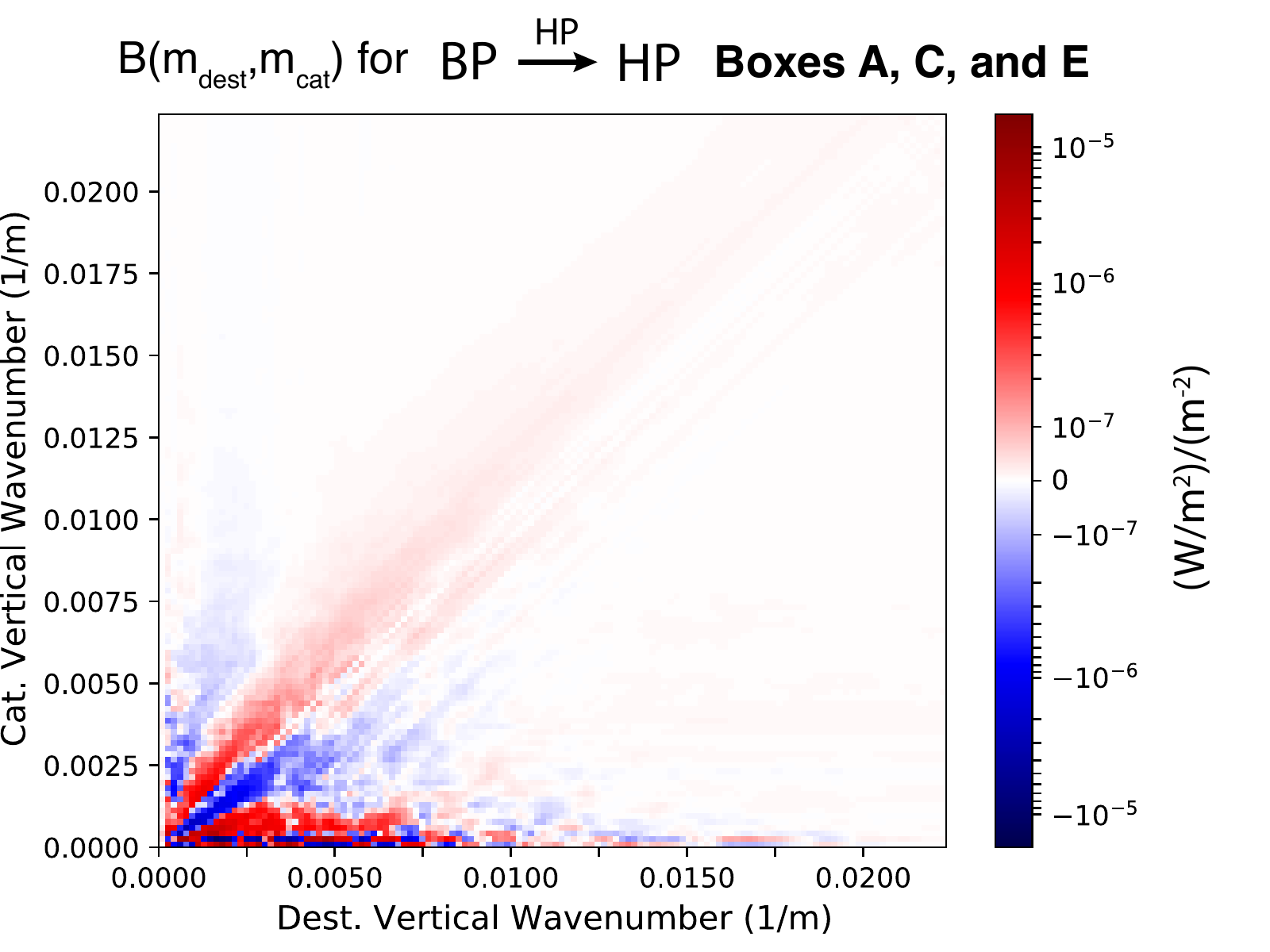}
\centering \caption{Bispectrum in Boxes A, C, and E.}
\label{fig:10a}
\end{subfigure}
\begin{subfigure}{0.45\textwidth}
\centering
\includegraphics[width=\textwidth]{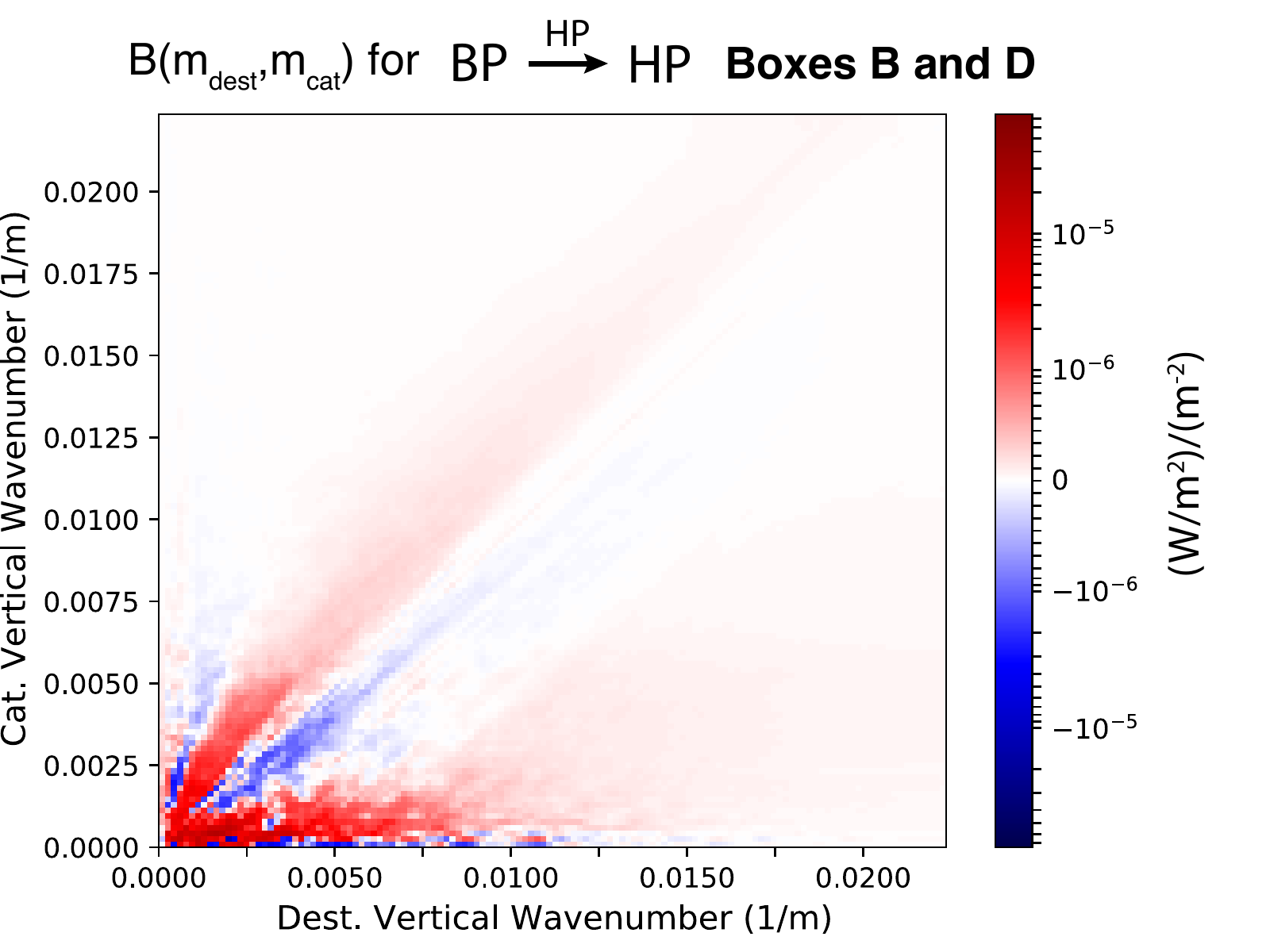}
\centering \caption{Bispectrum in Boxes B and D.}
\label{fig:10b}
\end{subfigure}
\centering \caption{Regional comparison of m-$\omega$ spectra (top row, as in Fig. \ref{fig:m_omega}) and  $\mathcal{B}_{\textrm{ID}_{\textrm{comp}}}\left(m_{dest},m_{cat}\right)$ (bottom row, as in Fig. \ref{fig:sp12_both}).  Note that $\mathcal{B}_{\textrm{ID}_{\textrm{comp}}}$ should capture the spectral transfer from $\mathcal{M}_{\textrm{SP}}$ as well as $\mathcal{M}_{\textrm{ID}_{\textrm{comp}}}$ (refer to discussion in section \ref{sec:methods}\ref{sec:methods:budgets}\ref{sec:flux}).  Boxes B and D were selected for separate analysis because that is where most of the forward frequency cascade observable in the five boxes occurs.  We hypothesized that $\mathcal{M}_{\textrm{ID}_{\textrm{comp}}}$ would be greater in these regions to conserve wave action.  $\mathcal{M}_{\textrm{SP}}$ and $\mathcal{M}_{\textrm{ID}_{\textrm{comp}}}$ can both be seen to be slightly larger in these boxes.}
\label{fig:10}
\end{figure*}

\cite{sun12}'s assumption that the horizontal-gradient contribution to the bispectra vanishes is also verified explicitly by comparing the bispectra with and without the horizontal-gradient contribution in Figs. \ref{fig:sp12} and \ref{fig:sp12_both}, respectively.  We also compute the vertical- and horizontal-gradient spectral fluxes of $\Pi_{\textrm{ID}_{\textrm{diff}}}$ in Fig. \ref{fig:vert_horz}.  It is apparent that the assumption of small horizontal-gradient contribution is not valid in the present model output. Thus, care should be taken to devise alternative means to measure bispectra from observational data.  A more realistic option may be to compare observational data of the vertical component to that of regional model output, such as this one, in an attempt to infer general features of the entire bispectra from the vertical component.  This also suggests that we should be wary of attempts to interpret the bispectra features of \cite{sun12} in the first place.  Nonetheless, the general features of Fig. \ref{fig:sp12_both} are reflected in Fig. \ref{fig:sp12}. The inclusion of the horizontal gradients mostly shifts the signal in the positive direction.  Thus, with the aid of the computational results in Fig. \ref{fig:10}, the features evident in the bispectra of \cite{sun12} can still be interpreted.  

\cite{sun12} found a positive signal at $m_{cat} \ll m_{dest}$; this behavior is not expected from $\mathcal{M}_{\textrm{ID}}$.  Rather, a signal at $m_{cat} \approx m_{dest}$, which would be consistent with $\mathcal{M}_{\textrm{ID}}$, was absent or very weak and not bicoherent.   These features of \cite{sun12}'s results are generally consistent with the bispectra computed from the present model output in Fig. \ref{fig:sp12}, which also shows a strong positive signal at $m_{cat} \ll m_{dest}$, similar to in \cite{sun12}.  This signal indicates a transfer of energy from high vertical-wavenumber near-inertial and tidal waves to high vertical-wavenumber supertidal IWs via scattering off a low-vertical wavenumber supertidal IWs.  As this signal has been reproduced in the present numerical simulation, it will be referred to as a distinct mechanism, $\mathcal{M}_{\textrm{SP}}$, in subsequent discussion.  

The plots in Fig. \ref{fig:sp12} show a larger range of catalyst modes and a smaller range of destination modes than those in \cite{sun12}, thereby making the aspect ratio of $m_{cat} \approx m_{dest}$ easily identifiable as the diagonal.  The bispectra of \cite{sun12} are presented on a linear scale which makes it only easy to discern about one order of magnitude in their signal, possibly reflecting a observational noise floor.  The present numerical results appear to have a lower noise floor (although we did not compute a bicoherence analysis.)  The bispectra are presented on a log-scale colorbar to reveal the weaker signal.  The far left of the left panel of \cite{sun12}'s figure 16 shows alternating-signed bands near the origin.  In Fig. \ref{fig:sp12} of this paper, a similar signal can be seen to extend weakly along the diagonal, $m_{cat} \approx m_{dest}$.  It is not clear exactly how this particular pattern arises, but it should include  $\mathcal{M}_{\textrm{ID}_{\textrm{comp}}}$. 

As previously discussed, a signal at $m_{cat} \approx m_{dest}$ corresponding to $\mathcal{M}_{\textrm{ID}_{\textrm{comp}}}$ would be absent if $\mathcal{M}_{\textrm{ID}_{\textrm{diff}}}$ is not diffusing energy forward in frequency space in the supertidal band.  We found significant variation among the subregions in terms of whether a forward frequency cascade is visible in the frequency-wavenumber spectra of $\mathcal{T}_{\textrm{ID}_{\textrm{diff}}}$.  A forward frequency transfer is most apparent in boxes B and D, those with the roughest topography, while none is apparent in the other boxes (see the top row of Fig. \ref{fig:10}).  Therefore, compensating energy is expected to be positive in the bispectra of boxes B and D for $m_{cat} \approx m_{dest}$ but not for the other boxes.  The bottom panes of Fig. \ref{fig:10} show that the compensating energy is more positive in boxes B and D, but that this occurs at both $m_{cat} \approx m_{dest}$ and the strong signal at $m_{cat} \ll m_{dest}$.  The strong signal at $m_{cat} \ll m_{dest}$ reflects a transfer of energy from large vertical wavenumbers at near-inertial and tidal frequencies (BP) into supertidal modes at similar vertical wavenumbers (HP) via scattering off of catalyst modes of low vertical wavenumber in the supertidal band (HP).  Perhaps such interactions help to compensate energy and conserve wave action in the supertidal band through a mechanism that is of higher order (and involving more than one triadic scattering process) than through the $\mathcal{T}_{\textrm{ID}_{\textrm{comp}}}$ part of $\mathcal{M}_{\textrm{ID}}$.  Otherwise, based on the variation by location of the frequency cascade associated with $\mathcal{M}_{\textrm{ID}}$, the absence of a signal at $m_{cat} \approx m_{dest}$ in \cite{sun12} would be consistent with a weak or neutral frequency cascade. 

\section{Conclusions}
\label{sec:conclusion}

With the increasing availability of high-resolution IW-permitting numerical models, wave-turbulence phenomenologists and theorists have acquired a validation tool capable of filling in gaps that are not easily accessible from observational data.  At the same time, IW-permitting numerical models can enable observations, constrained by budgetary and other limitations from fully sampling complex four-dimensional IW fields, to be more meaningful by serving as a basis of comparison. Another useful tool for these purposes is the unsymmetrized scattering triad, which readily serves as a point of comparison with observations of triads \citep[such as in ][]{sun12}) and also allows for the construction of bispectra that isolate, based on physical reasoning, some of the resonant nonlinear mechanisms, such as induced diffusion and local interactions, that govern the IW cascade.  

As a starting point, integrated vertical KE spectral budgets were computed for decomposition of both nonlinear advective scattering and dissipation.  Advection and dissipation are found to be nearly in balance at high vertical wavenumbers, although an inertial range in which forcing and dissipation are absent was not present.  Further results indicate that the extent of nonlinearity in such an IW continuum is highly dependent on the horizontal resolution of the model. In coarse 2km resolutions (common in the highest-resolution global models), the IW continuum is approximately (generalized) quasilinear around lower frequency ($\omega<2.5 f_0$) background waves and eddies, meaning very little energy is transferred within the supertidal IW continuum.  At such resolutions, the vast majority of energy in the IW continuum has been transferred to a high-frequency mode from a low-frequency mode through a single nonlinear interaction rather than a series of cascade-like processes (such as through $\mathcal{M}_{\textrm{ID}}$ or $\mathcal{M}_{\textrm{LI}}$).  Thus, fully nonlinear aspects of the IW continuum are largely absent in global models and at best can be parameterized based on available quasilinear flow information.  

A central motivation of this work is to better understand nonlinear IW scattering mechanisms in a realistic ocean model.   On one hand, recent theoretical advances in WTT pertaining to $\mathcal{M}_{\textrm{ID}}$ and $\mathcal{M}_{\textrm{LI}}$ in \cite{dematteis22} can be tested as well as decomposed using unsymmetrized spectral transfer triads.  On the other hand, mechanisms that involve the eddy field or that do not fit neatly into existing frameworks can be measured, with the efficient computation of bispectra in position space being particularly instrumental for the latter activity. We find that $\mathcal{M}_{\textrm{ID}}$ is associated with a small forward frequency cascade, which is strongest over rough topography.  This stands in contrast to the theoretical framework of \cite{mccomas81a} who predicted an inverse frequency cascade associated with $\mathcal{M}_{\textrm{ID}}$.  We also test \cite{dematteis22}'s prediction that $\mathcal{M}_{\textrm{LI}}$ (as opposed to $\mathcal{M}_{\textrm{ID}}$) will be the dominant nonlinear mechanism in the IW continuum.  We find that $\mathcal{M}_{\textrm{LI}}$ increases with horizontal resolution. However, even at the highest resolutions used here, $\mathcal{M}_{\textrm{LI}}$ fluxes do not exceed 10\% of the $\mathcal{M}_{\textrm{ID}}$ spectral flux and much of that is only spectrally local in frequency and not wavenumber. In addition, spectral fluxes are highly spatially inhomogeneous, and are largest within the subregion studied that had rough topography at depth.  

The method of decomposing spectral fluxes into "catalyst", "source", and "destination" modes enabled the separation of two distinct exchanges within $\mathcal{M}_{\textrm{ID}}$: (1) $\mathcal{M}_{\textrm{ID}_{\textrm{diff}}}$, the supertidal energy diffusion and (2) $\mathcal{M}_{\textrm{ID}_{\textrm{comp}}}$, energy compensation from the near-inertial and tidal frequencies.  Partial spectral flux budgets reveal that $\Pi^>_{\textrm{ID}_{\textrm{comp}}}$ is larger than $\Pi^>_{\textrm{ID}_{\textrm{diff}}}$, particularly at low resolutions.  However, this decomposition is limited in that (as implemented here) it only constrains the frequencies rather than the wavenumbers to be consistent with these two parts of $\mathcal{M}_{\textrm{ID}}$ (see Fig. \ref{fig:mechanisms}).   Bispectra of these spectral transfers in vertical wavenumber space can further constrain the mechanisms that contribute to this flux decomposition.  Within $\Pi^>_{\textrm{ID}_{\textrm{comp}}}$, a strong nonlinear energy exchange (termed $\Pi^>_{\textrm{SP}}$, reflecting the mechanisms $\mathcal{M}_{\textrm{SP}}$) is identified both here and in the observational results of \cite{sun12}.  This exchange moves KE from near-inertial and tidal frequencies of high vertical wavenumber into supertidal IWs of similar vertical wavenumber via scattering off of supertidal IWs that have lower vertical wavenumbers.  Being an energy exchange from BP to HP, $\mathcal{M}_{\textrm{SP}}$ is a key component of the generalized quasilinear dynamics that govern the IW continuum at low resolutions studied here.  $\mathcal{M}_{\textrm{ID}}$ does not fit into existing frameworks of nonlinear energy exchange among IWs, but perhaps it is real and important in the ocean given its presence in both \cite{sun12} and this paper.  

Additionally, the vertical spectral KE flux of a mechanism analogous to $\mathcal{M}_{\textrm{ID}_{\textrm{diff}}}$ in which eddy fields catalyze supertidal energy diffusion (termed $\Pi^>_{\textrm{ID}_{\textrm{eddy}}}$) is measured.   $\Pi^>_{\textrm{ID}_{\textrm{eddy}}}$ is found to be much larger than $\Pi^>_{\textrm{ID}_{\textrm{diff}}}$ at lower resolutions and comparable to $\Pi^>_{\textrm{ID}_{\textrm{diff}}}$ only in the highest vertical- and horizontal- resolution simulation.  

In summary, the present findings identify eddy-induced IW KE diffusion ($\Pi^>_{\textrm{ID}_{\textrm{eddy}}}$) as an important mechanism in shaping the IW spectrum in this simulation, while, contrary to expectations, $\mathcal{M}_{\textrm{LI}}$ is not.  $\mathcal{M}_{\textrm{SP}}$ (observed in \cite{sun12}) is also found to be important in energizing the IW continuum in the present numerical simulation.  Taken as a whole, the present results suggest internal wave interactions give rise to the GM spectrum in a way that differs markedly from the idealized picture in many previous studies.  This may be a reflection of the limited resolution or representation of IW dynamics.  However, these findings can serve as a starting point for studying a more complete spectral budget of IW interactions that can be improved upon over time. 

A natural next step will be to direct attention to dissipation mechanisms \citep[building on ][]{thakur22} and parameterizations of the IW continuum along with the spectral transfer decompositions presented in this paper.  In particular, we will examine vertical and horizontal spatial distributions.  This strategy will hopefully reveal different modeling strategies for quasilinear and fully nonlinear mechanisms that, in the real ocean, would move energy beyond the grid scale of a numerical model.  Looking forward, we anticipate high resolution regional models, such as this one, becoming increasingly important in probing details of the IW cascade and potentially developing model parameterizations for such details.

\section{Acknowledgements}
Joseph Skitka and Brian Arbic acknowledge support from Office of Naval Research grant N00014-19-1-2712.  RT and BKA acknowledge support from National Science Foundation grant OCE-1851164 and NASA grant 80NSSC20K1135.  

\section{Data Availability}
Data and data-processing scripts are publicly available online through Harvard Dataverse at https://doi.org/10.7910/DVN/FSMM1O \citep{skitka22}.

\appendix

\section{KE transfer within unsymmetrized advective triads}
\label{app:triads}

\begin{figure}[]
\centering
\includegraphics[width=0.48 \textwidth]{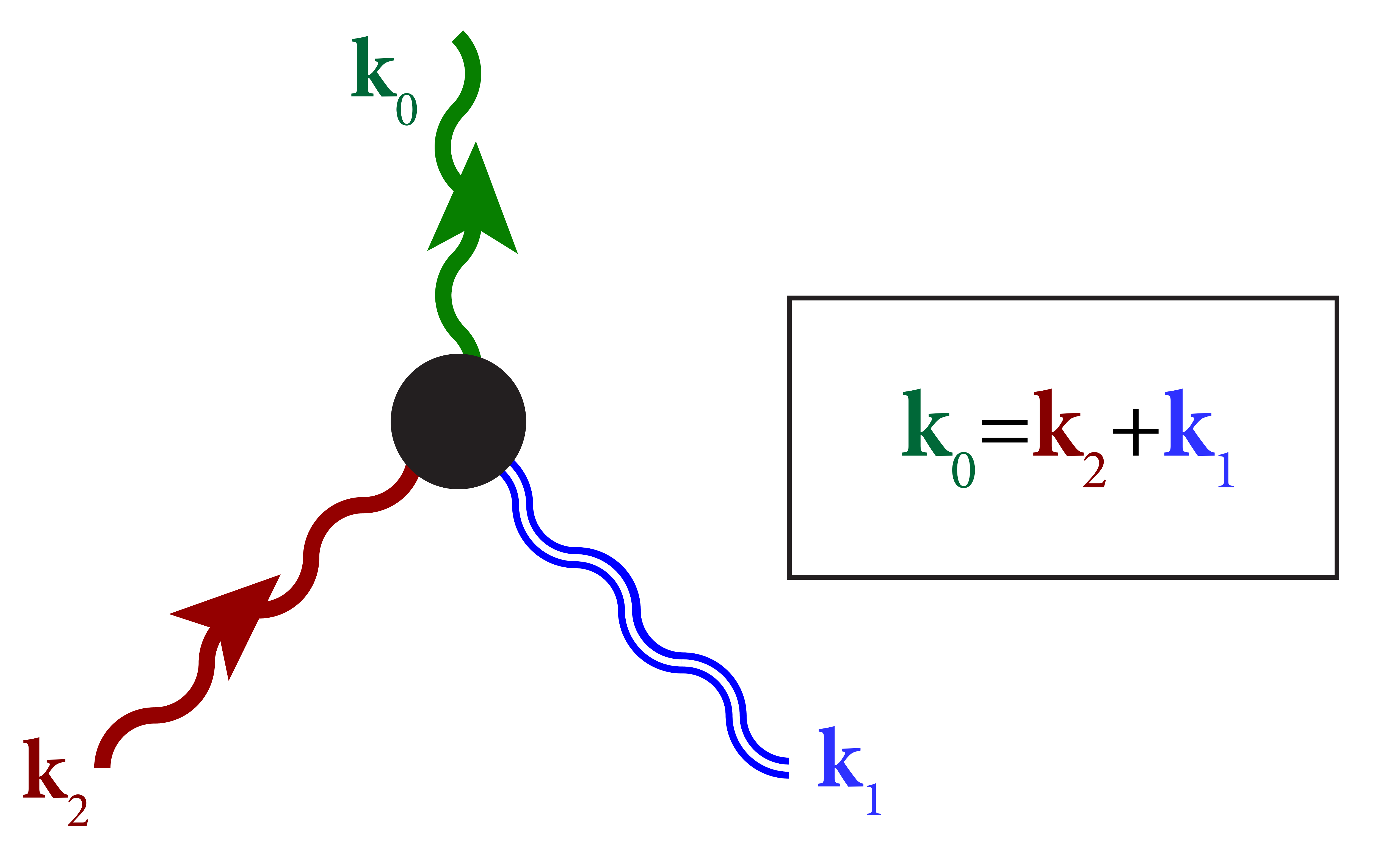}
\caption{\small An advective scattering triad from equation \ref{eq:nl_forwardscatter}.  The arrows indicate the direction of kinetic energy flow.  Here, $\mathbf{k}_0$ refers to the ``destination'' mode of the KE, $\mathbf{k}_2$ is the ``source'' mode of the kinetic energy, and $\mathbf{k}_1$ is the ``catalyst'' mode that neither contributes nor receives KE.  This diagram represents the change in energy in the destination mode, $\mathbf{k}_2$, only, so pairs of triads are required to represent an exchange of energy between modes, as displayed in Fig. \ref{fig:triad_energy}.}
\label{fig:triads}
\end{figure}

\begin{figure}[]
\centering
\includegraphics[width=0.48 \textwidth]{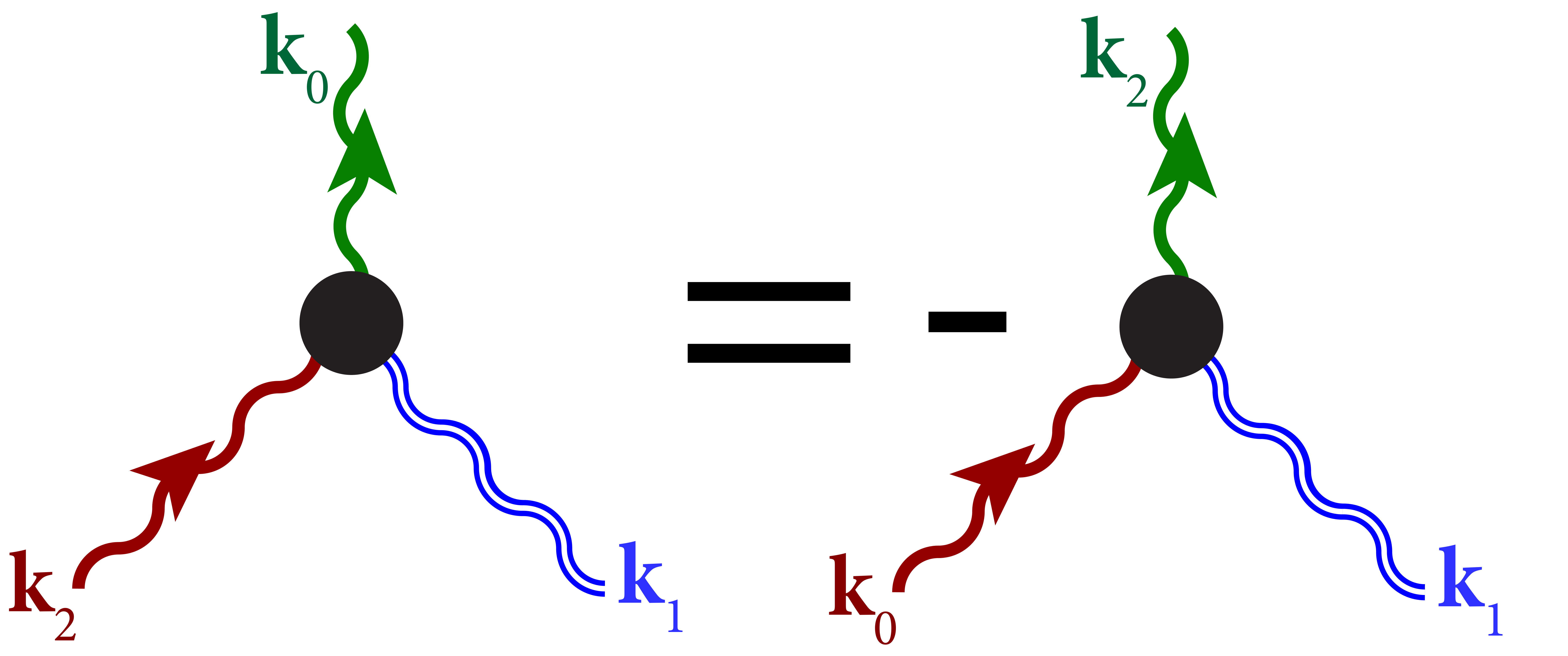}
\caption{\small Energy conservation is expressed through a pair of oppositely signed triads corresponding to the removal of energy from one mode and the injection of energy from another.}
\label{fig:triad_energy}
\end{figure}

\begin{figure}[]
\centering
\includegraphics[width=0.48 \textwidth]{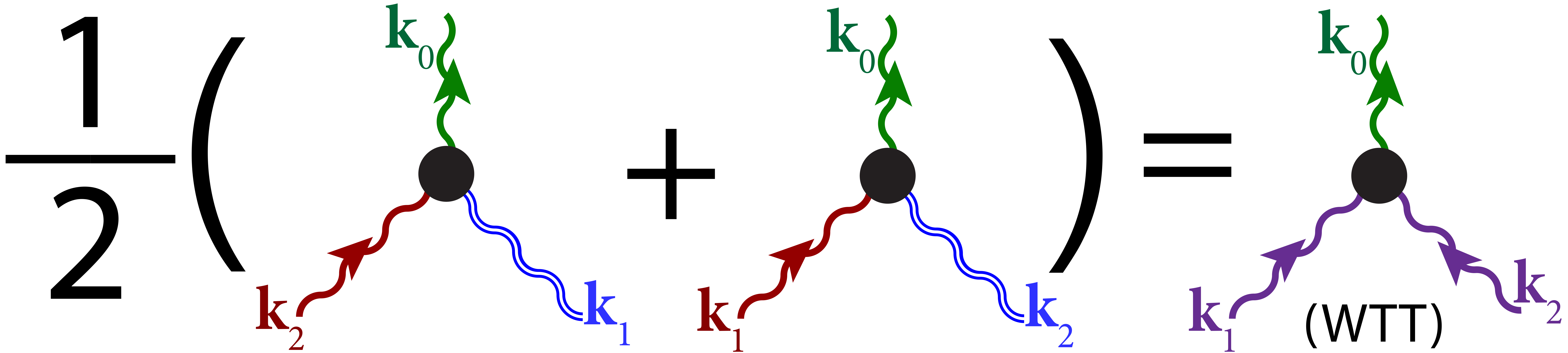}
\caption{\small In Wave Turbulence Theory (WTT), scattering coefficients are implicitly symmetrized between the catalyst and source terms. Thus the resulting triads (which also include the field amplitudes here) can be depicted graphically as in this figure.  The correspondence is not an exact representation of a particular WTT triad but rather a demonstration of the mode-to-mode energy exchange information loss in WTT that occurs with symmetrization.  }
\label{fig:wtt}
\end{figure}

The exchange of energy among modes of waves and eddies of different wavenumber in incompressible flow due to the advective nonlinearity can be clearly understood in terms of triads of waves corresponding to source, catalyst, and destination modes of the kinetic energy.  $\partial_t \boldsymbol{\varv}$ is taken to denote the velocity field evolution strictly due to the advective term:
\begin{align}
\partial_t \varv_i &= - \varv_j \partial x_j \varv_i \\
&= -\partial x_j \left( \varv_j \varv_i \right) + \varv_i \partial x_j \varv_j \\
&= -\partial x_j \left( \varv_j \varv_i \right) \label{eq:ff}
\end{align}
Here, $\varv$ is the 3D velocity field and subscripts are taken to imply a sum over dimensions.  Equation \ref{eq:ff} follows from the divergenceless condition, $\partial x_j \varv_j=0$.  Start by considering the change in KE in the horizontal velocity field from the advective term:
\begin{equation}
  \frac{1}{2} \partial_t \left(\tilde{\varv}_i \left(\mathbf{k}_0 \right) \tilde{\varv}_i^* \left(\mathbf{k}_0  \right) \right) = \textrm{Re}\left(\tilde{\varv}_i \left(\mathbf{k}_0 \right)  \partial_t\left(\tilde{\varv}_i^* \left(\mathbf{k}_0 \right) \right)\right)
\end{equation}
Here, $\mathbf{k}$ is the 3D wavenumber (although solely horizontal or vertical may be used, the latter of which is done in the main part of this paper.)  This can be expanded in Fourier space:
\begin{equation}
  \varv_i^* \left(\mathbf{k}_0  \right) = \int \varv_i^*\left(\mathbf{x}\right) e^{-2\pi \imag \mathbf{x}\cdot\left(-\mathbf{k}_0\right)}d^3\mathbf{x}_0 \label{eq:spectral_transfer}
\end{equation}
Here, integrals are assumed to be from $-\infty$ to $\infty$ and each integral symbol represents a triple integral in 3D space.  Also, note $\varv_i^*\left(\mathbf{k}\right) = \varv_i\left(-\mathbf{k}\right)$.  Then, we can show:
\begin{align}
\frac{1}{2} \partial_t &\left(\tilde{\varv}_i \left(\mathbf{k}_0 \right) \tilde{\varv}_i^* \left(\mathbf{k}_0  \right) \right) \nonumber \\
&= \textrm{Re}\left(\tilde{\varv}_i \left(\mathbf{k}_0 \right)  \partial_t\left(\tilde{\varv}_i^* \left(\mathbf{k}_0 \right) \right)\right) \\
&= \textrm{Re}\left(     \tilde{\varv}_i \left(\mathbf{k}_0 \right)      \int      - \partial_{x_j}\left(     \varv_j\left(\mathbf{x}\right)     \varv_i\left(\mathbf{x}\right)     \right)      e^{2\pi \imag \mathbf{x}\cdot\mathbf{k}_0}     d^3\mathbf{x}\right) \\
&= 2 \pi \textrm{Re}\left(    \imag \int           \mathunderline{green}{\left(\mathbf{k}_0\cdot\boldsymbol{\hat{\Jmath}}\right)}        \mathunderline{green}{\tilde{\varv}_i \left(\mathbf{k}_0 \right)}      \mathunderline{blue}{\tilde{\varv}_j^*\left(\mathbf{k}_0-\mathbf{k}_2\right)}     \mathunderline{red}{\tilde{\varv}_i^*\left(\mathbf{k}_2\right)}         \; d^3\mathbf{k}_2      \right)  \label{eq:nl_forwardscatter} 
\end{align}
The integrand of this equation can be symbolically represented as an interacting triad of modes as in Fig. \ref{fig:triads}.  In equation \ref{eq:nl_forwardscatter}, the colored underlines correspond to the different arms of the scattering diagram in Figs. \ref{fig:triads}-\ref{fig:wtt}.
We also assume that $\boldsymbol{\nabla} \cdot \boldsymbol{\varv} = 0$ here.  Invoking this condition:
\begin{align}
0 &= \partial x_j \varv_j \left(\mathbf{x}\right) \\
&= \int \partial x_j \tilde{\varv}_j\left(\mathbf{k}\right) e^{-2\pi \imag \mathbf{x}\cdot \mathbf{k} } d^3\mathbf{x}_0 \\
&= - \int 2 \pi \imag \left(\mathbf{k}\cdot\boldsymbol{\hat{\Jmath}}\right) \tilde{\varv}_j \left(\mathbf{k}\right) d^3\mathbf{x}_0 \\
&= \left(\mathbf{k}\cdot\boldsymbol{\hat{\Jmath}}\right) \tilde{\varv}_j \left(\mathbf{k}\right) 
\end{align}
Each of the triads in equation \ref{eq:nl_forwardscatter} represents the change in energy in just one mode, $\mathbf{k}_0$.  In order to represent an exchange of energy from one mode to another, we must consider pairs of triads.  Because the advective nonlinearity conserves energy in an incompressible flow, the simplest scenario is for the value of the diagram (representing the integral quantity of kinetic-energy change of the mode $\mathbf{k}_0$) to be opposite under the exchange of $\mathbf{k}_0 \leftrightarrow \mathbf{k}_2$ and that this holds for all $d^3\mathbf{k}_2$ and $d^3\mathbf{k}_0$ in their respective integrands.  To verify this, write the flux of the exchanged triad:
\begin{align}
\frac{1}{2} \partial_t &\left(\tilde{\varv}_i \left(\mathbf{k}_2 \right) \tilde{\varv}_i^* \left(\mathbf{k}_2  \right) \right) \nonumber \\
&= 2 \pi \textrm{Re}\left(    \imag \int           \left(\mathbf{k}_2\cdot\boldsymbol{\hat{\Jmath}}\right)         \tilde{\varv}_i \left(\mathbf{k}_2 \right)      \tilde{\varv}_j^*\left(\mathbf{k}_2-\mathbf{k}_0\right)      \tilde{\varv}_i^*\left(\mathbf{k}_0\right)         \; d^3\mathbf{k}_0      \right) \\
&= 2 \pi \textrm{Re}\left(    \imag \int           \left(\mathbf{k}_2\cdot\boldsymbol{\hat{\Jmath}}\right)         \tilde{\varv}_i \left(\mathbf{k}_2 \right)      \tilde{\varv}_j\left(\mathbf{k}_0-\mathbf{k}_2\right)      \tilde{\varv}_i^*\left(\mathbf{k}_0\right)         \; d^3\mathbf{k}_0      \right)   \label{eq:nl_backscatter} 
\end{align}
From the divergenceless condition:
\begin{equation}
\left(\mathbf{k}_0 \cdot\boldsymbol{\hat{\Jmath}} \right) \tilde{\varv}_j\left(\mathbf{k}_0 - \mathbf{k}_2\right) = \left(\mathbf{k}_2 \cdot\boldsymbol{\hat{\Jmath}} \right) \tilde{\varv}_j\left(\mathbf{k}_0 - \mathbf{k}_2\right) 
\end{equation}
this can be substituted this back into equation \ref{eq:nl_backscatter}
\begin{align}
\frac{1}{2} \partial_t & \left(\tilde{\varv}_i \left(\mathbf{k}_2 \right) \tilde{\varv}_i^* \left(\mathbf{k}_2  \right) \right) = \nonumber \\
& 2 \pi \; \textrm{Re}\left(    \imag \int           \left(\mathbf{k}_0\cdot\boldsymbol{\hat{\Jmath}}\right)         \tilde{\varv}_i \left(\mathbf{k}_2 \right)      \tilde{\varv}_j\left(\mathbf{k}_0-\mathbf{k}_2\right)      \tilde{\varv}_i^*\left(\mathbf{k}_0\right)         \; d^3\mathbf{k}_0      \right) 
\end{align}
Swapping the $\tilde{\varv}_i$ and noting that the real part of a complex number is invariant under conjugation
\begin{align}
\frac{1}{2} \partial_t &\left(\tilde{v}_i \left(\mathbf{k}_2 \right) \tilde{\varv}_i^* \left(\mathbf{k}_2  \right) \right) = \nonumber \\
&- 2 \pi \; \textrm{Re}\left(    \imag \int           \left(\mathbf{k}_0\cdot\boldsymbol{\hat{\Jmath}}\right)         \tilde{\varv}_i \left(\mathbf{k}_0 \right)      \tilde{\varv}_j^* \left(\mathbf{k}_0-\mathbf{k}_2\right)      \tilde{\varv}_i^*\left(\mathbf{k}_2\right)         \; d^3\mathbf{k}_0      \right) \label{eq1}
\end{align}
and comparing this with \ref{eq:nl_forwardscatter}, we can see that energy is indeed conserved, leaving one mode at the same rate it enters another through nonlinear scattering.  This is represented graphically in Fig. \ref{fig:triad_energy}.  We stress that for the basic advection term $\partial_t \varv_i = - \varv_j \partial x_j \varv_i$, the energy comes entirely from the $\varv_i$-field on the right-hand side rather than from the $\varv_j$-field, as one would expect using physical intuition. 

Wave turbulence theory, the framework used by many theorists to understand resonant nonlinear interactions of IWs, uses a symmetrized version of the scattering amplitudes that is most clearly illustrated with triads in Fig. \ref{fig:wtt}.  This symmetrization is evident in equation 15 of \cite{lvov10} in which all of the equations are invariant under an exchange of $\mathbf{p}_1$ and $\mathbf{p}_2$.  A similar symmetrization is also apparent in older eddy turbulence theory, for example, in equation 2.2 of \cite{kraichnan71}; the practice likely goes back much further.

\section{Extension to Spectral Frequency Transfers}
\label{app:omega}

The spatial spectra transfers discussed in appendix \ref{app:triads} are often extended to include frequency space without a formal justification or physical interpretation \citep{arbic14,muller15}.  A more formal derivation and physical interpretation of this decomposition draws on \cite{morten15}.  Following \cite{morten15}, imagine the kinetic energy (KE) averaged over a window in time of duration $T$ centered at time $\tau$.  
\begin{align}
& \left<KE\right> = \frac{1}{2 T V} \int_{\tau-\frac{T}{2}}^{\tau-\frac{T}{2}} \int_V  \varv_i\left(\mathbf{x},t\right) \varv_i\left(\mathbf{x},t\right) \; d^3\mathbf{x} \; dt \quad \textrm{,}
\end{align}
where the angle brackets indicate an average over the spatial domain and the time window, $V$ is the domain volume.  This can be expressed as an integral over the spectral decomposition of the velocity field in both wavenumber and frequency:
\begin{align}
& \left<KE\right> = \frac{1}{2} \iint \textrm{Re}\left(\hat{\tilde{\varv}}_i\left(\mathbf{k}_0,\omega_0,\tau \right) \hat{\tilde{\varv}}_i^*\left(\mathbf{k}_0,\omega_0,\tau \right)\right) \; d^3\mathbf{k} \; d\omega \quad \textrm{,}
\end{align}
where the hat indicates the spectral component in frequency space defined over a window of width $T$ centered at $t=\tau$, the latter of which is considered a free variable.  The KE can be written in terms of components as:
\begin{align}
& \left<KE\right> \left(\mathbf{k}_0,\omega_0,\tau \right)= \frac{1}{2} \textrm{Re}\left(\hat{\tilde{\varv}}_i\left(\mathbf{k}_0,\omega_0, \tau \right) \hat{\tilde{\varv}}_i^*\left(\mathbf{k}_0,\omega_0, \tau \right)\right) \quad \textrm{.}
\end{align}
The change in the KE in the time window centered on $\tau$ contributed from $\hat{\tilde{\varv}}_i\left(\mathbf{k}_0,\omega_0 \right)$ is then
\begin{align}
\partial_\tau & \left<KE\right>\left(\mathbf{k}_0,\omega_0,\tau \right) = \textrm{Re}\left(\hat{\tilde{\varv}}_i\left(\mathbf{k}_0,\omega_0,\tau\right)\partial_\tau \hat{\tilde{\varv}}_i\left(\mathbf{k}_0,\omega,\tau\right)\right) \quad \textrm{.}
\end{align}
A key result of \cite{morten15} is that (subject to certain constraints on how the velocity field is detrended and tapered), the above is equivalent to:
\begin{align}
\partial_\tau & \left<KE\right>\left(\mathbf{k}_0,\omega_0,\tau \right) = \textrm{Re}\left(\hat{\tilde{\varv}}_i\left(\mathbf{k}_0,\omega_0,\tau \right) \reallywidehat{\partial_t \tilde{\varv}_i^*}\left(\mathbf{k}_0,\omega_0,\tau \right)\right)  \quad \textrm{.} \label{eq:morten}
\end{align}
This can then be reworked
\begin{align}
\partial_\tau & \left<KE\right>\left(\mathbf{k}_0,\omega_0,\tau \right) = \nonumber \\
&= \textrm{Re}\left(\tilde{v}_i \left(\mathbf{k}_0, \omega_0  \right) \int \partial_t \left(\tilde{\varv}_i\left(\mathbf{k}_0,t \right) \right) e^{-2 \pi \imag \omega t} dt  \right) \\
&= \scriptstyle \; \textrm{Re}\left(     \tilde{\varv}_i \left(\mathbf{k}_0, \omega_0 \right)      \int      - \partial_{x_j}\left(     \varv_j\left(\mathbf{x},t\right)     \varv_i\left(\mathbf{x},t\right)     \right)      e^{2\pi \imag \left(\mathbf{k}_0\cdot\mathbf{x} - \omega_0 t\right) }   d^3\mathbf{x} \, dt \right) \nonumber \\
& \qquad + \quad \textrm{O.T.} 
\end{align}
where the flux-form of the advective term is written out, O.T. refers to the other terms in the momentum equations, and time is integrated from $t=\tau-\frac{T}{2}$ to $t=\tau+\frac{T}{2}$.   Dropping the other terms, one can follow the derivation of KE energy flow from the source to the destination field  presented in appendix \ref{app:triads} with frequencies included and arrive at
\begin{align}
\partial_\tau & KE\left(\mathbf{k}_0,\omega_0,\tau \right) = \nonumber \\
&= 2 \pi \textrm{Re}\left(    \imag \int           \left(\mathbf{k}_0\cdot\boldsymbol{\hat{\Jmath}}\right)         \tilde{\varv}_i \left(\mathbf{k}_0, \omega_0 \right)   \tilde{\varv}_j^*\left(\mathbf{k}_0-\mathbf{k}_2, \omega_0 - \omega_2\right)   \right.      \nonumber \\
& \qquad \qquad  \left. \tilde{\varv}_i^*\left(\mathbf{k}_2, \omega_2\right)         \; d^3\mathbf{k}_2   \, d\omega_2  \vphantom{\int}    \right) \quad \textrm{.} \label{eq:freq_sym}
\end{align}
Similar to equation \ref{eq:nl_forwardscatter}, equation \ref{eq:freq_sym} changes sign under the exchange of $\left\{\mathbf{k}_0,\omega_0\right\} \leftrightarrow \left\{\mathbf{k}_2,\omega_2\right\}$. \\

Thus, the extension of spectral fluxes and transfers to frequency space as used by \cite{arbic14} and \cite{muller15} as well as the extension of directional triadic interactions in equations \ref{eq:T}, \ref{eq:adv_flx_sources}, \ref{eq:bispectra_cat}, etc. can be interpreted thanks to \cite{morten15} as the exchange of energy from a wave of one frequency to another as the sample window over which the frequencies are computed is moved forward in time.

\section{Local Spectral Budget}
\label{app:lsb}

The local spectral KE budget, $\mathcal{T}\left(\mathbf{k}\right)$, is the evolution of the kinetic energy coming from the spectral components of the velocity fields $\tilde{\varv}_i\left(\mathbf{k}\right)$:
\begin{align}
\mathcal{T}\left(\mathbf{k},t\right) &= \partial_t KE \left(\mathbf{k},t\right) \\
&= \textrm{Re}\left(\tilde{\varv}_i{\left(\mathbf{k},t\right)} \; \widetilde{\partial_t \varv^*_i}\left(\mathbf{k},t\right)\right) \quad \label{eq:total_transfer} \textrm{.}
\end{align}
 Equation \ref{eq:total_transfer} can then be decomposed into the contributions to the time derivative of the velocity fields coming from the various terms in the spectral momentum equations (which are arrived at by taking the Fourier transform of equation \ref{eq:horz_eq}):
\begin{align}
\mathcal{T}_{term}\left(\mathbf{k},t\right) &= \textrm{Re}\left(\tilde{\varv}_i\left(\mathbf{k},t\right) \; \widetilde{\partial_t \varv^*_{i,term}}\left(\mathbf{k},t\right)\right) \label{eq:loc_spec_budg} \quad \textrm{.}
\end{align}
Note that equations \ref{eq:spectral_transfer}-\ref{eq:nl_forwardscatter} refer to the local spectral budget for the advective term, which is referred to as the spectral transfer.  Equation \ref{eq:loc_spec_budg} can be computed in position space by writing $\tilde{\varv}_i\left(\mathbf{k},t\right)$ as a sinusoidal function:
\begin{align}
\mathcal{T}_{term}\left(\mathbf{k},t\right) &= \left<\varv_i\left(\mathbf{x},\mathbf{k},t\right) \; \partial_t \varv_{i,term} \left(\mathbf{x},t\right)\right> \label{eq:T_k} \quad \textrm{,}
\end{align}
where the angle brackets are a spatial average over the domain that is used to define the spectral components, the time derivative of the velocity field is not decomposed into spectral components, and 
\begin{align}
\varv_i\left(\mathbf{x},\mathbf{k},t\right) =& \int \tilde{\varv}_i\left(\mathbf{k},t\right) \delta\left(|\mathbf{k}|-|\mathbf{k}'|\right)e^{2 \pi \imag (\mathbf{k}' \cdot \mathbf{x})} d\mathbf{k}' \\
\tilde{\varv}_i \left(\mathbf{k},t\right) &= \int \varv_i\left(\mathbf{x},t\right) e^{-2 \pi \imag \left(\mathbf{k} \cdot \mathbf{x} \right)} \; d^3\mathbf{x}  \quad \textrm{.}
\end{align}
Here, $\varv_i\left(\mathbf{x},\mathbf{k},t\right)$ is the real-valued spatial distribution of the velocity field with wavenumber $\mathbf{k}$.  

The extension of equation \ref{eq:T_k} to include frequency as well as wavenumber given in equation \ref{eq:T} can be derived by starting with equation \ref{eq:morten}.  To do this, we must now define the local spectral budget as the change in KE from a given mode, $\left\{\mathbf{k}_0,\omega_0\right\}$, defined over a window in time of duration $T$ centered at time $t=\tau$ that moves forward with time, as described in appendix  \ref{app:omega}.  
\begin{align}
\mathcal{T}\left(\mathbf{k},\omega,\tau\right) &= \partial_{\tau} KE \left(\mathbf{k},\omega,\tau\right) \label{eq:total_omega_transfer0}\\
&= \textrm{Re}\left(\hat{\tilde{\varv}}_i{\left(\mathbf{k},\omega\tau\right)} \; \reallywidehat{\widetilde{\partial_t \varv^*_i}}\left(\mathbf{k},\omega,\tau\right)\right) \quad \label{eq:total_omega_transfer} \textrm{.}
\end{align}
Here, the tilde-hat indicates Fourier amplitudes in both frequency and wavenumber and equation \ref{eq:total_omega_transfer} follows from equation \ref{eq:total_omega_transfer0} by equation \ref{eq:morten}.  Then, the velocity evolution can be decomposed into the budget contributions from individual terms and expressed as an integral of a sinusoid over position and time:
\begin{align}
\mathcal{T}_{term}\left(\mathbf{k},\omega,\tau\right) &= \left[\left(\varv_i\left(\mathbf{k},\mathbf{x},\omega,t,\tau\right) \; \partial_t \varv_{i,term}\left(\mathbf{x},t\right) \right)\right] \label{eq:T_omega} \quad \textrm{,}
\end{align}
where the square brackets are a space-time average over the domain and time window that is used to define the spectral components and 
\begin{align}
\varv_i\left(\mathbf{k},\mathbf{x},\omega,t,\tau\right) =& \iint^{\tau+\frac{T}{2}}_{\tau-\frac{T}{2}} \hat{\tilde{\varv}}_i\left(\mathbf{k},\omega,\tau\right) \delta\left(|\mathbf{k}|-|\mathbf{k}'|\right)\delta\left(|\omega|-|\omega'|\right) \nonumber \\
& \qquad \qquad \quad e^{2 \pi \imag (\mathbf{k}' \cdot \mathbf{x} - \omega' t)} d\mathbf{k}' d\omega' \\
\hat{\tilde{\varv}}_i \left(\mathbf{k},\omega,\tau\right) &= \iint^{\tau+\frac{T}{2}}_{\tau-\frac{T}{2}} \varv_i\left(\mathbf{x},t\right) e^{-2 \pi \imag \left(\mathbf{k} \cdot \mathbf{x} - \omega t \right)} \; d^3\mathbf{x} dt  \quad \textrm{.}
\end{align}
Here, $\varv_i\left(\mathbf{k},\omega,\tau\right)$ is the real-valued spatial distribution of the velocity field with wavenumber $\mathbf{k}$, frequency $\omega$ and computed over the time window centered at $\tau$. \\

\section{Divergence correction}
\label{app:div_corr}

To handle the horizontal boundaries in computing spectral transfers and fluxes from the advective term, the flow field is assumed to be periodic and the resulting divergence is explicitly removed:
\begin{align}
  \Pi_{\textrm{LP} \xrightarrow{\textrm{BP}} \textrm{HP}}^>\left(m,\omega\right) &= - \left<u_i^{>, \textrm{HP}}{\left(\omega\right)} \; \left(\varv_j \partial x_j u_i^{<} \right)\right> \nonumber \\
  & + \frac{1}{2} \left<u_i^{>, \textrm{HP}}{\left(\omega\right)} u_i^{<} \; \left(\partial x_j \varv_j \right)\right> \quad \textrm{.}
\end{align}
This procedure forces the spectral flux to conserve energy within the domain (up to discretization and interpolation error) and has the added benefit of correctly handling the advective flux associated with the free surface of the flow.  The procedure does introduce error from the sharp gradients imposed at the periodic boundaries, but this error is deemed to be a worthwhile sacrifice in order to conserve energy under the advective operator.

\section{Budget Term Diagnosis Implementation}
\label{app:terms}
Below are details of the numerical methods used to diagnose various terms in the spectral budgets used in this paper.  Also, see \cite{mitgcm} and \cite{nelson20} for further information on the solver details and simulation settings.
\begin{itemize}
\item Advection ($\mathsf{A}$) is a flux-form centered second-order operator.  In our analysis, we compute this term exactly as in the model, except that it is computed on a vertically uniform grid.  Experiments suggest that this procedure yields better energy-conservation than interpolation after the velocity tendencies are computed.  All other terms are computed on the nonuniform grid and then interpolated. Horizontal boundary conditions are handled by assuming periodicity and explicitly removing the energy sources/sinks associated with the resulting unphysical (three-dimensional) flow divergence, as described in appendix \ref{app:div_corr}.
\item Advection of kinetic energy into the simulation domain from the boundaries (BCs) is computed explicitly at the surfaces of the region being analyzed.  This is done with the same flux-form discretization as for the previously mentioned volume advection operator.
\item KPP background and shear ($\mathsf{D}_{\textrm{KPP BG}}$ and $\mathsf{D}_{\textrm{KPP Shear}}$) are computed using the same discretizations as in the model below 20 meters depth, using a 30-meter one-sided Tukey taper from 20 to 50 meters.  The taper excludes the vast majority of the diagnosed mixed layer, which always lies within the upper 50 m during the week-long model output period across the domain.   The mixed-layer dissipation ($\mathsf{D}_{\textrm{KPP ML}}$) is entirely omitted in the hope t   hat it will prove to be unimportant in the general balance of the IW field and any other conclusions drawn from this work.  The findings of this work generally support this assumption.  All other terms are computed throughout the entirety of the water column.  
\item The Leith scheme ($\mathsf{D}_{\textrm{Leith}}$) is computed using the same discretizations as in the model.  
\item Quadratic bottom boundary layer drag ($\mathsf{D}_{\textrm{QBD}}$) is computed exactly as it is in the model.  This term implicitly handles the no-slip bottom boundary condition.
\item A no-slip side boundary condition is computed exactly as it is computed in the model (BCs).  This contribution was found to be negligible and is therefore omitted from the results.
\item Bottom scattering and the pressure boundary condition (i.e. the no-normal flow condition, BCs) are handled implicitly in the other terms, such as advection.  
\item The pressure term ($\mathsf{P}$) mediates the linear transfer of potential energy into kinetic energy via the hydrostatic pressure but will not transfer kinetic energy among scales away from the boundary conditions.  This potential-energy-to-kinetic-energy conversion was computed through the hydrostatic pressure in a manner that is a good approximation of how it is represented in the model.  The conversion turns out to not have a significant impact for higher vertical wavenumbers and is not included in any budget plots.
\item The Coriolis force ($\mathsf{C}$) rotates the modes containing KE but, as a linear term, does not transfer it among scales.  Thus the Coriolis term is not computed for spectral budgets in this paper.  The contribution to the drag from the scattering of the Coriolis tendency off of the bottom topography is not computed but has been determined to be very small in a different test case.
\item In some figures, the total energy dissipation is given. This total dissipation is the sum of dissipations associated with the KPP background and shear terms, the Leith scheme, and quadratic bottom drag, but omits the KPP mixed-layer term.
\item Finally, a residual is computed as the sum of the aforementioned terms. The residual includes the combined interpolation and discretization errors of each term, any omitted forcing, the KPP mixed-layer dissipation, the small contributions of intentionally neglected terms that were previously mentioned, as well as energy buildup over time.
\end{itemize}
The KE budget terms are computed across the entirety of the water column (with the exception of KPP) in part because the flow is vertically inhomogeneous.  In terms of spectral analysis, the flow resembles a transient signal in position space, and it is appropriate to use a rectangular taper function (i.e. no taper function) to capture parts of the signal at the top and bottom of the domain.  Put another way, no taper is needed for this analysis because the analysis does not compute the spectrum from a sample of an infinite dimension.   Rather, the spectrum is computed on the entire domain on which the signal is defined. The exception to this, as previously mentioned, is KPP’s interior dissipation, which tapered from 20m to 50m while KPP's mixed-layer dissipation is simply omitted.  Also note that a Tukey taper is applied in time.


\bibliographystyle{ametsoc2014}

\end{document}